\documentclass[12pt]{article}
\usepackage{amsmath}
\usepackage{amssymb}
\usepackage{graphicx,bbm,mathrsfs}
\usepackage{color}

\newcommand{\lsim}{\raisebox{-0.13cm}{~\shortstack{$<$ \\[-0.07cm] $\sim$}}~} 
\newcommand{\gsim}{\raisebox{-0.13cm}{~\shortstack{$>$ \\[-0.07cm] $\sim$}}~} 
\newcommand{\beq}{\begin{eqnarray}} 
\newcommand{\eeq}{\end{eqnarray}} 
\newcommand{\tb}{\tan\beta} 
\newcommand{\ee}{e^+ e^-} 
\usepackage{listings}
\usepackage{caption}
\usepackage{cite}

\newcommand{\bea}{\begin{align}}
\newcommand{\eea}{\end{align}}
\newcommand{\nbea}{\begin{align*}}
\newcommand{\neea}{\end{align*}}
\newcommand{\nbeq}{\begin{equation*}}
\newcommand{\neeq}{\end{equation*}}
\newcommand{\bear}{\begin{eqnarray}}  
\newcommand{\eear}{\end{eqnarray}}

 \newcommand{\comment}[1]{}
\usepackage{array}
\newcolumntype{M}[1]{>{\centering\arraybackslash}m{#1}}
\newcolumntype{N}{@{}m{0pt}@{}}

\numberwithin{equation}{section}

\begin{document}

\begin{flushright}\small{CERN-TH/2016-008, KCL-PH-TH/2016-01 \\ 
LCTS/2016-01, LPT-Orsay--16--02} \end{flushright} 

\vspace*{5mm} 

\begin{center}


 
\mbox{\large\bf Future Collider Signatures of the Possible 750 GeV State}

 

\vspace*{8mm}

{\sc Abdelhak~Djouadi$^{1,2}$, John Ellis$^{2,3}$,  Rohini Godbole$^4$} 

\vspace{2mm}

and {\sc J\'er\'emie Quevillon$^3$ } 

\vspace{6mm}

{\small 
$^1$ Laboratoire de Physique Th\'eorique,  CNRS and Universit\'e Paris-Sud, \\  
B\^at. 210, F--91405 Orsay Cedex, France \\
\vspace{2mm}
$^2$ Theory Department, CERN, CH 1211 Geneva 23, Switzerland \\
\vspace{2mm}
$^3$ Theoretical Particle Physics \& Cosmology Group, Department of Physics, \\ 
King's College,  Strand, London WC2R 2LS, United Kingdom\\
\vspace{2mm}
$^4$ Center for High Energy Physics, Indian Institute of Science, \\
Bangalore 560 012, India
}
\end{center}

\vspace*{8mm}

\begin{abstract}

If the recent indications of a possible state $\Phi$ with  mass $\sim 750$~GeV decaying into 
two photons reported by ATLAS and CMS in LHC collisions at 13~TeV were to become confirmed, the prospects for future collider physics at the LHC and beyond would be affected radically, as we explore in this paper. Even minimal scenarios for the $\Phi$ resonance and its $\gamma \gamma$ decays require additional particles with masses $\gtrsim \frac12 m_\Phi$.
We consider here two benchmark scenarios that exemplify the range of possibilities: 
one in which $\Phi$ is a singlet scalar or pseudoscalar boson whose production and $\gamma \gamma$ decays are due to loops of coloured and charged fermions, and another benchmark scenario in which $\Phi$ is a superposition of (nearly) degenerate CP--even and CP--odd Higgs bosons in a (possibly supersymmetric) two--Higgs doublet model also with additional fermions to account for the $\gamma \gamma$ decay rate. We explore the implications of these benchmark scenarios for the production of $\Phi$ and its new partners at colliders in future runs of the LHC and beyond, at higher-energy $pp$ colliders and at $\ee$ and $\gamma \gamma$ colliders, with emphasis on the bosonic partners expected in the doublet scenario and the fermionic partners expected in both scenarios. 

\end{abstract}
\vspace*{1cm}
\begin{flushleft}\small{January 2016} \end{flushleft} 

\newpage

\section{Introduction}

The world of particle physics has been set alight by the reports from the ATLAS and CMS Collaborations in LHC  collisions at 13~TeV of hints of a possible state, that we denote  $\Phi$, with mass $\sim 750$~GeV decaying into two photons \cite{yearend,ATLAS-diphoton,CMS-diphoton}, echoing the discovery of the 125~GeV Higgs boson~\cite{SMH-discovery1}. The product of the cross section and branching ratio for $\Phi \to \gamma \gamma$ decay hinted by the data is $\sim 6$~fb, with the ATLAS data hinting that it may have a significant total decay width. If these hints are confirmed, a changed and much brighter light will be cast on the future of particle physics, because the putative $\Phi$ particle must be accompanied by additional massive particles.

The Landau-Yang theorem \cite{Landau-Yang} tells us that $\Phi$ cannot have spin one,
and spins zero and two are the most plausible options. Since a graviton--like spin--2 particle would in principle have similar decays into dileptons, dijets and dibosons that have not been observed by the experiments \cite{LHC-others},  
we focus here on the more likely possibility of spin zero. Gauge invariance requires the $\Phi \gamma \gamma$ coupling to have dimension $\ge 5$, and hence be induced by additional physics with a mass scale $\gtrsim \frac12 m_\Phi$. 
Historical precedent ($\pi^0, H \to \gamma \gamma$) \cite{history,venerable} and many models suggest that the $\Phi \gamma \gamma$ coupling is induced by anomalous loops of massive charged fermions and/or bosons whose form factors vanish if their masses are much smaller  than $M_\Phi$. In addition, the absence of a strong $\Phi$ signal in LHC collisions at 8~TeV motivates gluon-gluon fusion as the $\Phi$ production mechanism, presumably mediated by massive coloured fermions and/or bosons. The null results of LHC searches for coloured fermions require them to have masses $\gtrsim M_\Phi$, whereas new uncoloured ones might weigh $\sim \frac12 M_\Phi$.

These arguments apply whatever the electroweak isospin assignment of the possible $\Phi$ particle. If it is a
singlet, it need not be accompanied by any bosonic partner particles. However, if it is a non-singlet, it must also be accompanied by bosonic isospin partners that would be nearly degenerate with $\Phi$
in many scenarios, since $M_\Phi$ is larger than the electroweak scale. The minimal example of a non-singlet scenario for $\Phi$ is an electroweak doublet, as in a 2-Higgs-doublet model (2HDM), e.g., in the context of supersymmetry.

The necessary existence of such additional massive fermions and/or bosons would yield exciting new perspectives for
future high-energy collider physics, if the existence of the $\Phi$ particle is confirmed. In the absence of such confirmation, some might consider the exploration of these perspectives to be premature, but we consider a preliminary discussion to be appropriate and interesting, in view of the active studies of the physics of the high-luminosity LHC (HL-LHC) \cite{HL-ATLAS,HL-CMS}, possible future $\ee$ colliders (ILC \cite{ee-ILC}, CLIC \cite{ee-CLIC}, FCC-ee \cite{Fcc-ee}, 
CEPC \cite{CEPC-ee}) and their eventual $\gamma \gamma$ options \cite{ginzburg,gamma-gamma},
and possible future higher-energy proton-proton colliders (higher-energy LHC (HE-LHC) 
\cite{LHC-33}, SPPC \cite{SPPC} and FCC-hh \cite{FCC-hh, 100TeV}.

In this paper we explore three aspects of the possible future $pp$ and $\ee$ collider physics of the $\Phi$ resonance and its putative partners. We consider single $\Phi$ production, production of the fermions that are postulated to mediate its $\gamma \gamma$ decays and its production via gluon-gluon fusion, production of $\Phi$ in association with fermionic mediators, and the phenomenology of possible bosonic partners. For definiteness, we focus on two alternative benchmark scenarios that  illustrate the range of different possibilities for the possible $\Phi$ particle,  in which it is either an isospin singlet state~\cite{EEQSY,othersinglet} or an isospin doublet in a 2HDM~\cite{Phi-ADM,otherdoublet}. Both scenarios have rich fermionic phenomenology, and the doublet scenario also has rich bosonic phenomenology.

Section 2 of this paper discusses the singlet benchmark scenario, paying particular attention to its production
at the LHC and possible future higher-energy $pp$ colliders, as well as in $\gamma \gamma$ fusion at
a high--energy  $\ee$ collider. Section~3 of this paper discusses the 2HDM benchmark scenario, in which the $\Phi$
signal is a combination of scalar and pseudoscalar Higgs bosons $H, A$, and discusses their single, pair
and associated production in $pp$, $\ee$ and $\gamma \gamma$ collisions, with some comments on
production in $\mu^+ \mu^-$ collisions. The capabilities of $pp$ and $\ee$ colliders to observe directly 
the massive fermions postulated to mediate $\Phi$ production by gluon-gluon fusion and $\Phi \to \gamma \gamma$
decay are discussed in Section~4, including the possibilities for pair and single production in association with
Standard Model fermions and vector bosons. Finally, Section~5 summarizes our conclusions.

\section{Singlet Scenario}

\subsection{Singlet models of the $\Phi \to \gamma \gamma$ signal}

We consider in this Section the minimal scenario in which the $\Phi$ resonance is an isospin singlet scalar state\footnote{As mentioned before, we ignore the unlikely possibility of spin--2. There are also possibilities to circumvent the Landau-Yan theorem for spin--one particles \cite{evading-LY}  but we will ignore them here.}, unaccompanied by
bosonic isospin partners. As already mentioned, the $\Phi \gamma \gamma$ and $\Phi gg$ couplings could be induced by new, massive particles that could be either fermions or bosons, with spins 0 and 1 being possibilities for the latter. We concentrate here on the fermionic option, since loops of scalar bosons make smaller contributions to the anomalous loop amplitudes than do fermions of the same mass and, mindful of William of Occam, we avoid enlarging the gauge group with new, massive gauge bosons. 
In view of the stringent constraints on massive chiral fermions~\cite{PDG}, we assume that the new fermions are vector-like. 

The couplings of a generic scalar $S$ and  pseudoscalar $P$ state to pairs of photons and gluons are described via dimension-five operators in an effective field theory
\beq
{\cal L}_{\rm eff}^S &=& \frac{e}{v } c_{S\gamma\gamma} \, S F_{\mu\nu} F^{\mu\nu} 
                 + \frac{g_s}{v}       c_{Sgg} \, S G_{\mu\nu}  G^{\mu\nu} \nonumber \\
{\cal L}_{\rm eff}^P &=& \frac{e}{v } c_{P\gamma\gamma} \, P F_{\mu\nu} \tilde F^{\mu\nu} 
                 + \frac{g_s}{v}   c_{Pgg}  \, P G_{\mu\nu} \tilde G^{\mu\nu}
\label{eq:lag-eff}
\eeq
with $F_{\mu \nu}= (\partial_\mu A_\nu - \partial_\nu A_\mu)$ the field strength of the
electromagnetic field, $\tilde{F}_{\mu\nu}=\epsilon_{\mu \nu \rho \sigma} F^{\rho \sigma}$ 
and likewise for the SU(3) gauge fields, where $v\approx 246$ GeV is the standard Higgs vacuum expectation value.  Within this effective theory, the partial widths of the $\Phi=S/P$ particle decays into two gluons and two photons are given by
\beq 
\Gamma (\Phi \to \gamma \gamma) = c_{\Phi \gamma\gamma}^2 \frac{\alpha}{v^2} M_\Phi^3    \, , \ \ \  
\Gamma (\Phi \to g g) = c_{\Phi gg}^2 \frac{ 8 \alpha_s}{ v^2} M_\Phi^3\, .
\eeq
If the  gluonic decay is dominant, one would have the following branching ratio for the photonic 
decay:
\beq 
{\rm BR}(\Phi  \to \gamma \gamma) = \frac{ \Gamma (\Phi \to \gamma \gamma) }
{\Gamma (\Phi \to \gamma \gamma) + \Gamma (\Phi \to g g) } \approx 
\frac{ \Gamma (\Phi \to \gamma \gamma) }{ \Gamma (\Phi \to g g)  } \approx 
\frac{c_{\Phi \gamma\gamma}^2  }{ c_{\Phi gg}^2 } \frac{ \alpha}{8 \alpha_ s }  \, ,
\eeq
leading to  ${\rm BR}(\Phi  \to \gamma \gamma) \approx 10^{-2}$ if $c_{\Phi 
\gamma\gamma} \approx  c_{\Phi gg}$. In general, decays into $WW,ZZ$ 
and $Z\gamma$ final states also occur through similar effective couplings. 
Writing  the Lagrangian (\ref{eq:lag-eff}) in terms of the ${\rm SU(2)_L \times U(1)_Y}$
fields $\vec W_\mu$ and $B_\mu$ rather than the electromagnetic field $A_\mu$, one
obtains for the coupling constants of the electroweak gauge bosons, where in the scalar case we define the coefficients $c_1 \equiv c_{\Phi BB}$ and $c_2 \equiv c_{\Phi \vec W \vec W}$ and $s_W^2= 1- c^2_W \equiv \sin^2\theta_W$:
\beq
c_{\Phi \gamma \gamma}= c_1 c^2_W + c_2 s^2_W \, ,  \, 
c_{\Phi ZZ}=c_1 s^2_W + c_2 c^2_W \, , \, 
c_{\Phi Z \gamma}=s_W c_W (c_2-c_1) \, , \, 
c_{\Phi WW}= c_2~ \, ,
\label{eq:lag-eff2}
\eeq
and in the pseudoscalar case we denote the corresponding coefficients  by ${\tilde c_{1, 2}}$.

Turning to the new fermionic content and  following Ref.~~\cite{EEQSY}, we consider here four models for the massive vector-like fermions:\vspace*{-3mm} 


\begin{description}

\item[$\bullet$] {\bf Model 1}: A single vector-like pair of charge 2/3 quarks, $T_{R,L}$.\vspace*{-3mm} 

\item[$\bullet$] {\bf Model 2}: A vector-like doublet of charge 2/3 and charge - 1/3 quarks, $(U, D)_{R, L}$.\vspace*{-3mm}  

\item[$\bullet$] {\bf Model 3}: A vector-like generation of quarks, including charge 2/3 and -1/3 singlets, \\ \hspace*{1.2cm} $(U, D)_{R, L}, T_{R,L}, B_{R,L}$.\vspace*{-3mm}  

\item[$\bullet$] {\bf Model 4}: A vector-like generation of heavy fermions including leptons and quarks, \\ \hspace*{1.2cm} $(U, D)_{R, L}, T_{R,L}, B_{R,L}, (L^1, L^2)_{R, L}, E_{R, L}$.\vspace*{-3mm}
\end{description}

For simplicity, we consider the case where the mixing between the $\Phi$ state and the Standard Model Higgs boson is negligible, and favour the possibility that mixing between the new heavy fermions and their Standard Model counterparts is also small. 

\begin{figure}[!h]
\vspace*{0.7cm}
\centerline{\hspace*{-7.cm} \vspace{-0.5cm} \includegraphics[scale=0.57]{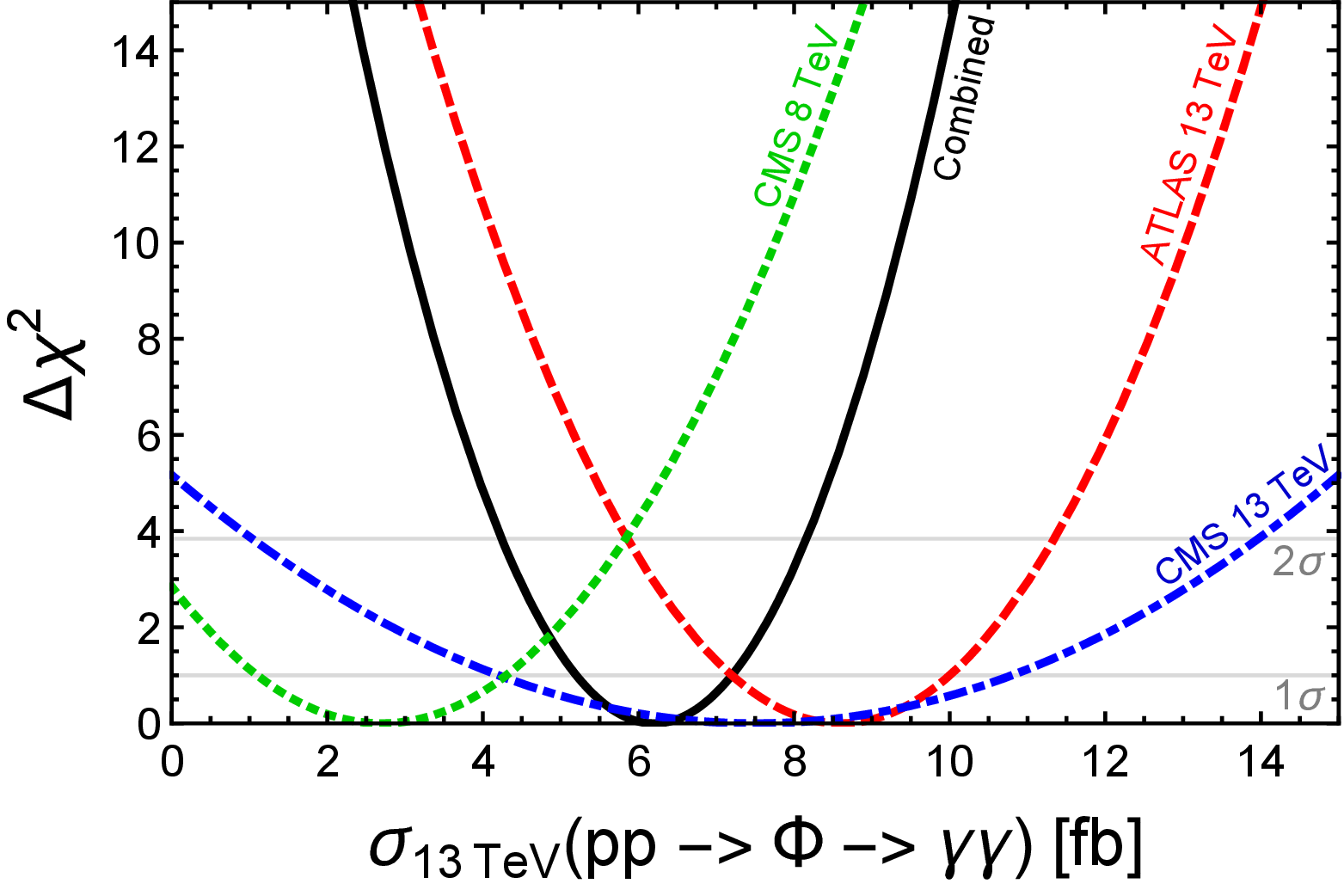}}
 \vspace*{-6.3cm} \centerline{\hspace*{+9cm} \includegraphics[scale=0.33]{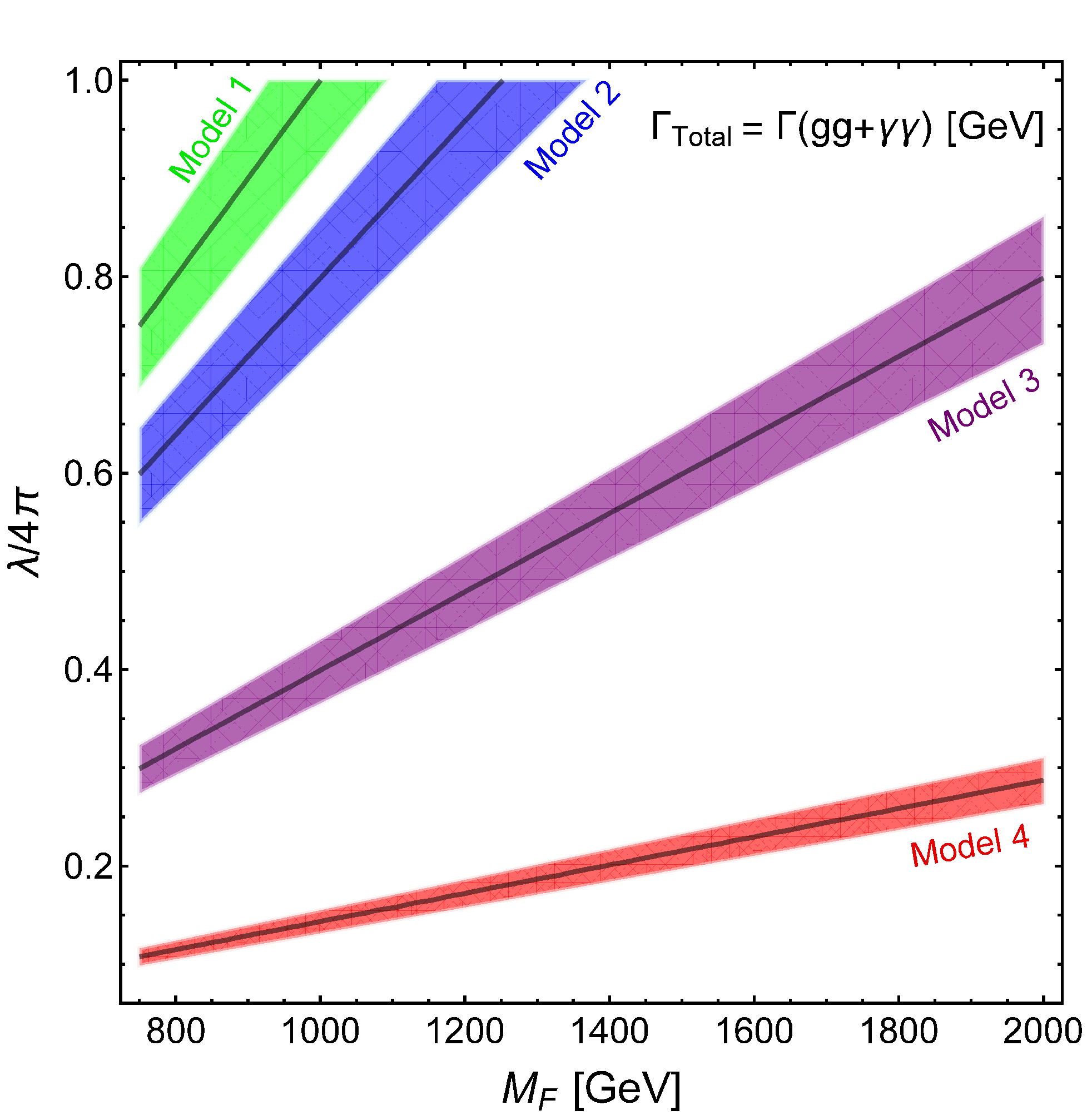}}
\vspace*{0.1cm}
\caption{\it Left panel: A compilation of the possible strengths of the 750 GeV  $\Phi$
resonance signal found by CMS in Run~1 at 8~TeV (green dashed line), CMS at 13~TeV (blue dashed line), ATLAS at 13~TeV (dashed red line) and their combination (black solid line) assuming a $gg$ fusion production mechanism. Right panel: Values of the vector-like fermion mass $m_F$ and coupling $\lambda$ (both assumed to be universal) required in singlet Models 1, 2, 3 and 4 to accommodate the possible $gg\to \Phi \to \gamma \gamma$ signal reported by CMS and ATLAS~\protect\cite{yearend,ATLAS-diphoton,CMS-diphoton}. The black lines are for the central value of the cross section, $6$~fb, and the coloured bands represent the $1$-$\sigma$ uncertainties. Plots adapted from~\protect\cite{EEQSY}.}
\label{fig:xsec}
\vspace*{-.2mm}
\end{figure}

The left panel of Fig.~\ref{fig:xsec} reproduces Fig.~1 of Ref.~\cite{EEQSY}, and shows the
possible strengths of the $\Phi (750~{\rm GeV})$ signal found by the CMS collaboration in Run~1 at 8~TeV (green dashed line) and at 13~TeV (blue dashed line), the ATLAS signal at 13~TeV (dashed red line) and their combination (black solid line). The figure is made assuming a gluon fusion production mechanism, $gg\to \Phi$, and formally, this combination yields $\sigma (pp \to \Phi \to \gamma \gamma) = 6 \pm 2$~fb at 13~TeV.
The right panel of Fig.~\ref{fig:xsec} is a simplified version of Fig.~4 of~Ref.~\cite{EEQSY}, and is obtained assuming that total width of the resonance  is such that $\Gamma (\Phi) \approx \Gamma (\Phi \to gg)$. It shows that reproducing the $\sim 6$~fb $gg\to \Phi \to \gamma \gamma$ signal reported by the  ATLAS and CMS collaborations requires a relatively large value  of the fermion-antifermion-$\Phi$ coupling $\lambda$. For simplicity, both the masses $m_F$ and the couplings $\lambda$ were assumed in~Ref.~\cite{EEQSY} and here to be universal.  If one requires $\lambda^2/4 \pi^2 \le 1$ so as to remain with a perturbative r\'egime, corresponding to $\lambda /4 \pi \le 1/2$,  one finds no solutions in Model 1, 
whereas Model 2 would require $m_F \lesssim 800$~GeV even after allowing for the uncertainties. On the other hand, Model 3 would be consistent with perturbativity for $m_F \lesssim 1.4$~TeV and Model 4 could accommodate $m_F \lesssim 3$~TeV~\footnote{We do not 
discuss in this Section the case where $\Gamma (\Phi) \approx 45$~GeV as hinted by ATLAS, which would require  non-perturbative values of $\lambda$ in all the models studied~\cite{EEQSY}. However, as discussed in Section~3, this value of $\Gamma (\Phi)$ could be accommodated within two--doublet models.}. 

As will be discussed in Section~4, the LHC has good prospects to explore all the range of 
the fermion masses $m_F$ allowed by perturbativity in Model 3, whereas a higher-energy collider may be required to  explore fully the range of $m_F$ allowed in Model 4.

In the following we discuss the production of the $\Phi$ resonance first at $pp$ colliders,  
focusing on the gluon fusion mechanism~\footnote{Additional processes likes Higgs--strahlung
$q\bar q\to \Phi W, \Phi Z$ and vector boson fusion $qq \to \Phi qq$ can occur but will have smaller cross sections and will be discussed only in the context of $\ee$ collisions.}  $gg\to \Phi$, and then at high energy electron-positron colliders in both the $\ee$ and $\gamma\gamma$ modes. 

\subsection{$\Phi$ production in pp collisions}

The dominant gluon-gluon fusion mechanism for $\Phi$ production in these singlet models in $pp$ collisions has the following
leading-order partonic subprocess cross section $\sigma(gg\to \Phi)$, which is proportional to the 
$\Phi \to gg$ partial width:
\begin{eqnarray}
\sigma(pp \rightarrow \Phi) = \frac{1}{M_\Phi s} C_{gg} \Gamma(\Phi \rightarrow gg) \, :~
C_{gg} =  \frac{\pi^2}{8} \int_{M_\Phi^2/s}^{1} \frac{dx}{x} g(x)g(\frac{M_\Phi^2}{s x} ) \, ,
\end{eqnarray}
where $g(x)$ is the gluon distribution inside the proton at a suitable factorization scale $\mu_F$. Since we assume here that $\Phi$ is an isospin singlet, production in association with a Standard Model vector boson, $W^\pm$ or $Z$, is much smaller, and production in association with a ${\bar t} t$ pair or a vector-like fermion pair is also relatively small (see Section 2.4). 

\begin{figure}[!h]
\vspace*{-.01cm}
\centerline{\hspace*{-15mm} \includegraphics[scale=0.77]{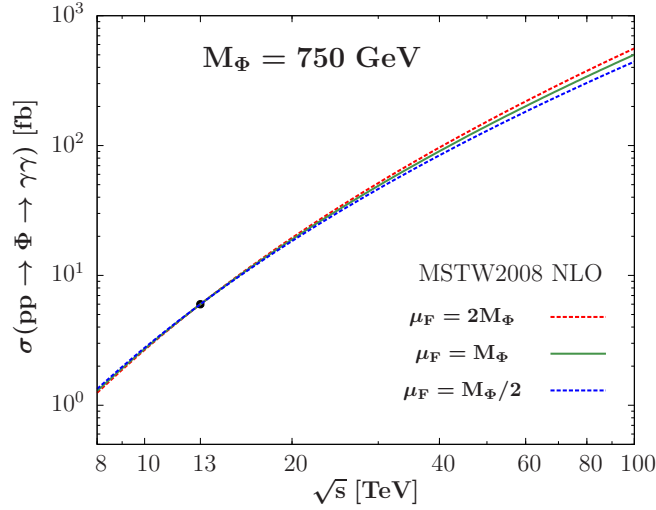}}
\vspace*{-.2cm}
\caption{\it Cross section for producing a singlet $\Phi$ boson with mass $750$~GeV at a $pp$ collider as a function of the centre-of-mass energy from $\sqrt s=8$ TeV to $100$ TeV, assuming $gg$ fusion with a cross section of $6$~fb at $13$~TeV. The extrapolation to other energies uses the MSTW2008 NLO parton distributions~\protect\cite{MSTW} and the central value of the factorization scale $\mu_F= M_\Phi = 750$~GeV (solid green line), compared with the choices $\mu_F = 2 M_\Phi$ (red dotted line) and $\mu_F = \frac12 M_\Phi$ (blue dotted line).}
\label{fig:ggPhi}
\vspace*{-2mm}
\end{figure}

The $gg \to \Phi \to \gamma \gamma$ production cross section times branching ratio 
at different $pp$ centre-of-mass energies can be obtained directly from the estimated rate $\sigma \times {\rm BR} \simeq 6 \pm 2$~fb at a centre-of-mass energy of 13~TeV, simply by rescaling the gluon-gluon luminosity function as shown in Fig.~\ref{fig:ggPhi}. 
In this figure we use MSTW2008 NLO parton distributions with various choices of the factorization scale~\cite{MSTW}: the central value $\mu_F= M_\Phi = 750$~GeV (solid green line) and the choices $\mu_F = 2 M_\Phi$ (red dotted line) and $\mu_F = M_\Phi/2$ (blue dotted line). We see that the cross section grows by a modest factor $\simeq 1.2$ from 13 to 14~TeV, but by larger factors $\sim 10 (24) (57) (84)$ at $\sqrt s= 33 (50) (80) (100)$ TeV
which correspond to the energies mooted for the HE-LHC~\cite{LHC-33}, SPPC~\cite{SPPC} and FCC-hh~\cite{FCC-hh}. The uncertainty associated with the variation in $\mu_F$ is $\sim 20$\%
at 100~TeV, and we find an additional uncertainty of $\sim 30$\% associated with different choices of parton distributions that are recommended by the LHC Higgs working group \cite{PDF4LHC}.

It is possible in each of the Models 1 to 4 above to calculate the ratios of rates for decays into Standard Model vector bosons via anomalous triangle diagrams, as shown in Table~\ref{tab:Gamma}, which is adapted from Ref.~\cite{EEQSY}. As discussed there, there are interesting prospects for observing some of these decays,  in particular the $\Phi \to Z \gamma$ and $\Phi \to W^+ W^-$ decays in Model 2~\footnote{In view of the uncertainties and the small  branching ratios of the $W/Z$ bosons into leptons that are easier to search for, 
we do not regard this model as being excluded by searches for these modes.}. 
It was assumed in Ref.~\cite{EEQSY} that all the heavy fermions are degenerate. However,
this might not be the case and, in particular, it is natural to consider the possibility that the heavy vector-like leptons $L$ are much lighter than the quarks. 

\begin{figure}[!h]
\vspace*{-2.3cm}
\centerline{ \hspace*{-2.3cm}
\includegraphics[scale=0.77]{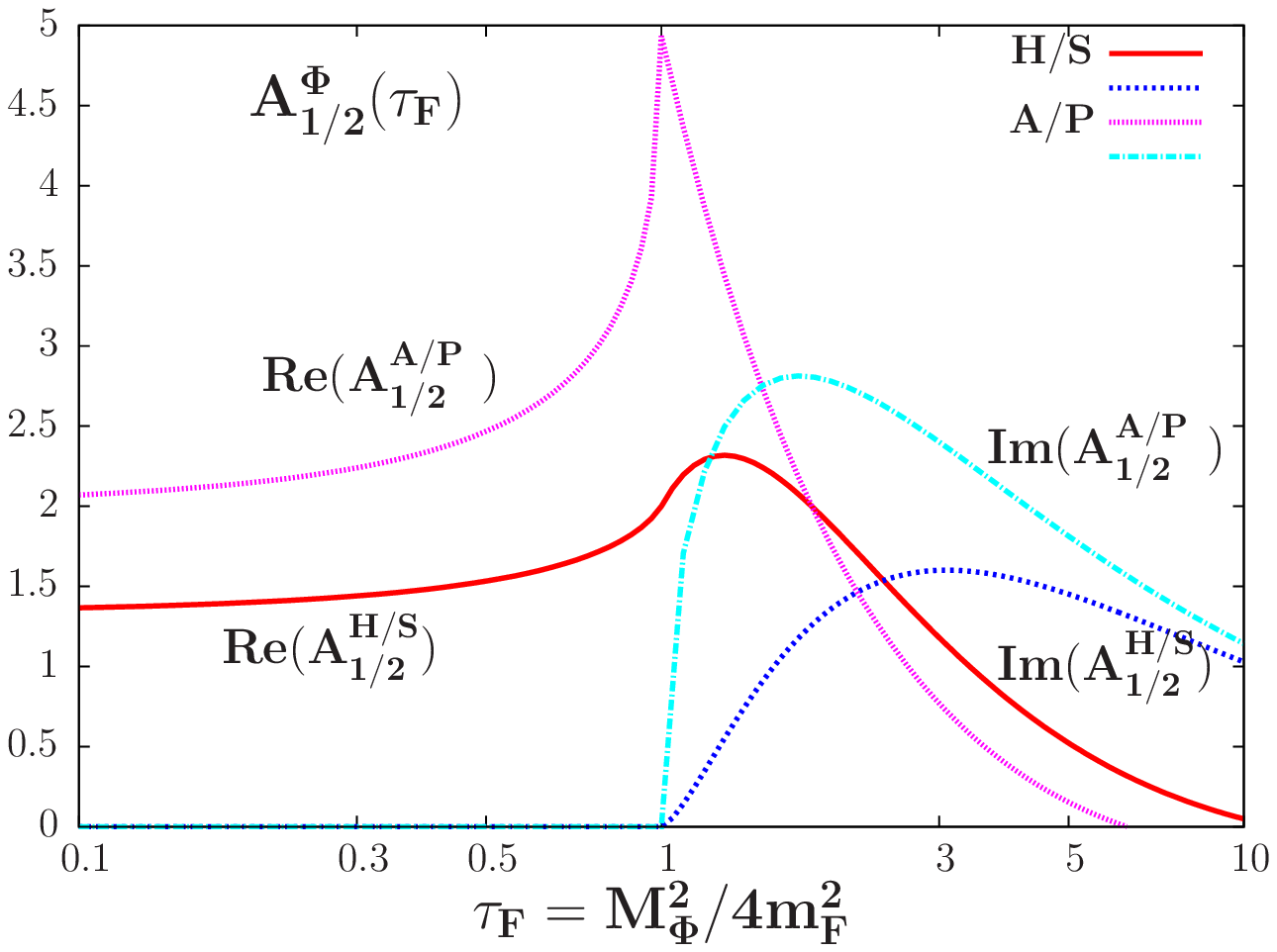} }
\vspace*{-12.9cm}
\caption{\it
The real and imaginary parts of the form factors $A^\Phi_{1/2}$ with fermion loops in the case of CP--even $H/S$ and CP--odd $A/P$ states as functions of the variable $\tau_F=M_\Phi^2/4m_F^2$.} 
\label{fig:enhancement}
\end{figure}

In general, the $\Phi \to gg$ and $\Phi \to \gamma\gamma$ partial decay widths, assuming that only heavy fermions are running in the loops,  are  given by~\cite{venerable,Review1} 
\begin{eqnarray}
\Gamma(\Phi  \to gg) & = &  \frac{G_\mu\alpha_s^2 M_\Phi^3}
{64\sqrt{2}\pi^3} \bigg| \sum_Q  \hat g_{\Phi QQ} A_{1/2}^\Phi  (\tau_Q) 
\bigg|^2 \, , \nonumber \\
\Gamma(\Phi  \to \gamma\gamma) & = &  \frac{G_\mu\alpha^2 M_\Phi^3}
{128\sqrt{2}\pi^3} \bigg| \sum_F  \hat g_{\Phi FF} N_c e_F^2 A_{1/2}^\Phi  (\tau_F) 
\bigg|^2 \, ,
\label{eq:Gammagg}
\end{eqnarray}
with $N_c$ a color factor, $e_F$ the electric charge of the fermions $F$, and  $g_{\Phi FF}$
the Yukawa coupling normalised to its Standard Model value, $\hat g_{\Phi FF}^{\rm SM} = m_F/v$. The partial widths are the same in the scalar and pseudoscalar cases, apart from the form factors $A_{1/2}^\Phi (\tau_F)$ that characterize the loop contributions of spin--$\frac12$ fermions as functions of the variable $\tau_F =M_\Phi/4m_F^2$, which depend on the parity of the spin-zero state. They are given by 
\begin{eqnarray}
A_{1/2}^{H/S}  & = & 2 \left[  \tau_{F} +( \tau_{F} -1) f(\tau_{F})\right]  \tau_{F}^{-2} \, , \\
A_{1/2}^{A/P}  & = & 2 \tau_{F}^{-1} f(\tau_{F}) \, ,
\label{eq:ASAP}
\end{eqnarray}
for the scalar/CP-even ($S/H$) and pseudoscalar/CP-odd ($P/A$) cases, respectively, where
\begin{eqnarray}
f(\tau_{F})=\left\{ \begin{array}{ll}  \displaystyle
\arcsin^2\frac{1}{\sqrt{\tau_{F}}} & {\rm for} \; \tau_{F} \geq 1 \, , \\
\displaystyle -\frac{1}{4}\left[ \log\frac{1+\sqrt{1-\tau_{F}}}
{1-\sqrt{1-\tau_{F}}}-i\pi \right]^2 \hspace{0.5cm} & {\rm for} \; \tau_{F}<1 \, .
\end{array} \right.
\label{eq:formfactors}
\end{eqnarray}
The real and imaginary parts of the  form factors for the different  $H/S$ and  $A/P$ 
CP cases are shown in Fig.~\ref{fig:enhancement} as functions of the reduced variable $\tau_F$.

When the fermion mass in the loop is much larger than the mass $M_\Phi$, namely in the limit $m_F \to \infty$, one obtains $A^{S}_{1/2}\! =\! \! \frac43$ and $A^P_{1/2}\!=\!2$ for the real parts of the form factors, and in the opposite limit, $m_F \to 0$, one has $A_{1/2}^{\Phi} \to 0$. For $M_\Phi \le 2 m_F$ ($\tau_F \le 1$), so that $\Phi \to {\bar F} F$ decays are forbidden, the maximal values of the form factors are attained when $\tau_F=1$, i.e., just at the $\Phi \to {\bar F}F$ threshold. In this case, one has the real parts Re($A^S_{1/2}) \approx 2 $ and Re($A^P_{1/2}) \approx \frac12 \pi^2 \approx 5$, and Im($A^{\Phi}_{1/2}$) = 0. In the case of the top quark contributions, when $M_\Phi=750$ GeV, one has $\tau_t \approx 4$, the form factors have both real and imaginary parts, and $| A^P_{1/2}/A_{1/2}^S|^2  \approx 2$. 

We have included in Table~\ref{tab:Gamma} predictions in Model 4 for the ratios of scalar 
and pseudoscalar diboson decay rates if $m_L = 400$~GeV and the vector-like quark
masses are much larger than $M_\Phi$.  We see that in both these low-lepton-mass cases, the $\gamma \gamma$ decay rate is enhanced relative to all the other diboson decay rates, as compared with the case where the vector-like quark and lepton masses are the same.

\begin{table}[!]
\vspace*{3mm}
\centering
\renewcommand{\arraystretch}{1.24}
\begin{tabular}{| c | c | c | c | c | c |}
\hline
\textbf{Model} & Masses & $\frac{\Gamma(\Phi\to gg)}{\Gamma(\Phi\to \gamma \gamma)}$ & $\frac{\Gamma(\Phi\to Z \gamma)}{\Gamma(\Phi\to \gamma \gamma)}$ & $\frac{\Gamma(\Phi\to Z Z)}{\Gamma(\Phi\to \gamma \gamma)}$ & $\frac{\Gamma(\Phi\to W^\pm W^\mp)}{\Gamma(\Phi\to \gamma \gamma)}$ \\
\hline
\textbf{1} & all $\gg m_\Phi$ & 180 & 1.2 & 0.090 & 0\\
\hline
\textbf{2} & all $\gg m_\Phi$ & 460 & 10 & 9.1 & 61\\
\hline
\textbf{3} & all $\gg m_\Phi$ & 460 & 1.1 & 2.8 & 15\\
\hline
& all $\gg m_\Phi$ & 180 & 0.46 & 2.1 & 11\\
\textbf{4} & S: $m_L = 400$~GeV & 140 & 0.10 & 1.4 & 6.6 \\
& P: $m_L = 400$~GeV & 110 & 0.12 & 1.5 & 6.9 \\
\hline
\end{tabular}
\caption{\it Ratios of $\Phi$ decay rates for the singlet models under consideration,
where we have used $\alpha_s (m_X) \simeq 0.092$. Extended version of Table~6 in Ref.~\protect\cite{EEQSY}.}
\label{tab:Gamma}
\vspace*{-4mm}
\end{table}

\subsection{$\Phi$ production in $\gamma \gamma$ collisions}

Since this state has been observed in the diphoton channel at the LHC at 13~TeV, it is natural to discuss $\Phi$ production via $\gamma \gamma$ collisions; see also 
Ref.~\cite{previousgammagamma}. Many aspects of a possible $\gamma \gamma$ collider associated with a parent linear $e^{+} e^{-}$ collider have been discussed quite extensively, see, e.g., Ref.~\cite{gamma-gamma}, starting from the original idea~\cite{ginzburg}. A $\gamma\gamma$ collider  can be constructed using Compton 
back-scattering from a laser beam via the processes~\cite{ginzburg,grs} 
\begin{eqnarray}
e^-(\lambda_{e^-}) \ \gamma(\lambda_{l_1}) \rightarrow e^- \ \gamma(\lambda_1) \, , \;
e^+(\lambda_{e^+}) \ \gamma(\lambda_{l_2}) \rightarrow e^+ \ \gamma(\lambda_2) \, ,
\end{eqnarray}
The back-scattered laser photons then carry a large fraction of the parent $e^{+}/e^{-}$ energy. Their energy spectrum and polarization depend on the helicities of the lasers $\lambda_{l_{1}}, \lambda_{l_2}$ and of the leptons $\lambda_{e^{+}},\lambda_{e^{-}}$, as well as on the laser energy.  The virtue of such a collider is that it provides a direct and accurate probe of the $\gamma \gamma$ coupling of a  diphoton resonance. Moreover, it offers an unique opportunity to study the CP properties of such resonances. 

For the production cross section, one has in general
\begin{eqnarray}
\sigma(\lambda_{e^+},\lambda_{e^-},\lambda_{l_{1}},\lambda_{l_{2}},E_{b}) = \int dx_1 dx_2 L_{\gamma\gamma}(\lambda_{e^+},\lambda_{e^-},\lambda_{l_1},\lambda_{l_2},x_1,x_2)\, \hat \sigma(\lambda_1,\lambda_2, 2 E_{b} \sqrt{x_1x_2})  ,~~ 
\end{eqnarray}
where $L_{\gamma\gamma}(\lambda_{e^+},\lambda_{e^-},\lambda_{l_1},\lambda_{l_2},x_{1},x_{2})$ 
is the luminosity function for polarizations $\lambda_1(\lambda_{e^-},\lambda_{l_{1}},$ $x_1)$  and $\lambda_2(\lambda_{e^+},\lambda_{l_{2}},x_2)$ of the colliding photons.  $\hat \sigma(\lambda_1,\lambda_2, 2E_{b}\sqrt{x_1x_2})$ is the cross section for the process under consideration, $\gamma \gamma \rightarrow \Phi \rightarrow X$  in this case.  The invariant mass of the two-photon system is given by $W = \sqrt{\hat s} =  2 E_{b} \sqrt{x_{1} x_{2}}$, where $x_{1}, x_{2}$ are the fractions of the beam energy $E_{b}$ carried by the two back-scattered photons.  The cross section for  $\Phi$ production via $\gamma \gamma$ fusion is given by
\begin{eqnarray}
\hat {\sigma} (W,\lambda_{1},\lambda_{2})\! =\! 8 \pi  {\frac{\Gamma (\Phi \rightarrow \gamma \gamma) \Gamma (\Phi \rightarrow X)}{(W^{2} - M_{\Phi}^{2})^{2} + M_{\Phi}^{2} \Gamma_{\Phi}^{2}}} (1 + \lambda_{1} \lambda_{2}) \, ,
\label{gmgmcsec}
\end{eqnarray}
where $W$ is the centre-of-mass energy of the $\gamma \gamma$ system. The factor of $(1 + \lambda_{1} \lambda_{2})$ projects out the $J_{Z} = 0$ component of the cross section, thereby maximizing the scalar resonance contribution relative to the continuum backgrounds.

We recall that in this singlet $\Phi$ resonance scenario, the total $\Phi$ decay width may be dominated by $\Phi \to gg$, in which case it would be much narrower than the experimental resolution in any measurable final state~\cite{EEQSY}. Accordingly, in this subsection we treat $\Phi$ in the narrow-width approximation. The value of $\Gamma(\Phi  \to \gamma \gamma)$ may be calculated directly from the cross section for $gg \to \Phi \to \gamma \gamma$ inferred from the LHC measurements, if $\Phi \to gg$ is indeed the dominant decay mode as would be the case if mixing between the heavy and Standard Model fermions is negligible as we assume. In this case, 
$\sigma (pp \to \Phi \to \gamma \gamma) \propto 
\Gamma(\Phi \to \gamma \gamma)$ and the value $\sigma (pp \to \Phi \to \gamma \gamma) \sim 6$~fb indicated by the ATLAS and CMS collaborations   would correspond to $\Gamma(\Phi \to \gamma \gamma) \sim 1$~MeV. We note that this should be regarded as a {\it lower limit} on $\Gamma (\Phi \to \gamma \gamma)$, which would be enhanced by a factor $\Gamma (\Phi \to {\rm all})/\Gamma (\Phi \to gg)$ if $\Phi \to gg$ is not the dominant decay mode.

\begin{figure}[!h]
\vspace*{-2.3cm}
\centerline{\hspace*{-5mm} \includegraphics[scale=0.75]{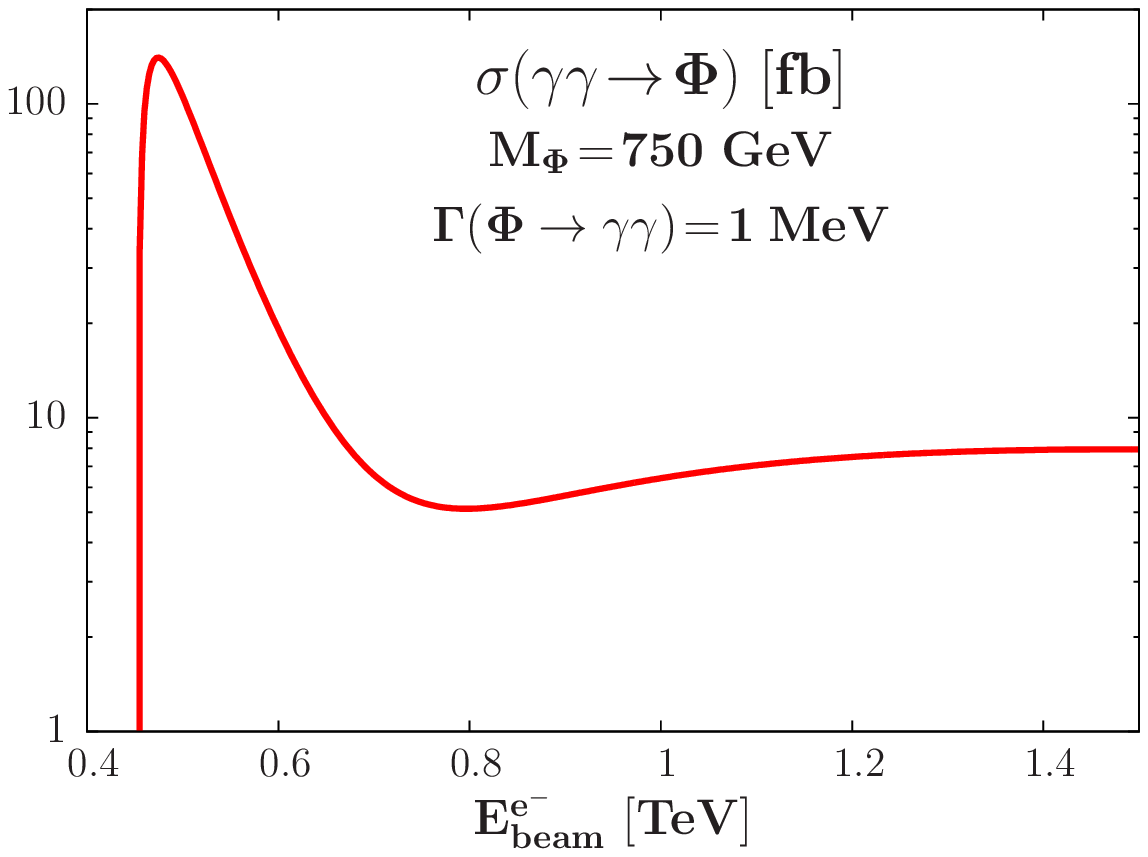}}
\vspace*{-13.4cm}
\caption{\it Cross section for producing a singlet $\Phi$ boson with mass $750$~GeV  via $\gamma \gamma$ fusion at an $\ee$ collider as a function of the $\ee$ centre-of-mass energy in the range from $\sqrt s=0.8$ TeV to $3$ TeV. The $\Phi \to \gamma \gamma$ partial width is assumed to be $1$ MeV as can be inferred from $\sigma( gg\to \Phi ) \approx 6$ fb at $\sqrt s=13$~TeV when the decay $\Phi \to gg$ is dominant.}
\label{fig:gammagammaPhi}
\vspace*{-3mm}
\end{figure}

The value of $\Gamma (\Phi \to \gamma \gamma)$ inferred from the LHC data motivates the option of a $\gamma\gamma$ collider discussed above. In the narrow-width approximation and assuming that $\Phi \rightarrow gg$ dominates $\Gamma_{\Phi}$  we obtain for the $gg$ final state the following expression for $\hat{\sigma} (\sqrt{\hat s})$, where $\hat s= x_1 x_2 s$ with $\sqrt{s}$ the centre-of-mass energy of the $e^{+}e^{-}$ machine
\beq
\hat \sigma (\sqrt{\hat s})  = {\frac {8 \pi^2}{M_\Phi}} \Gamma (\Phi \rightarrow \gamma \gamma) \delta (M_\Phi^2 - s x_1 x_2) (1 + \lambda_1 \lambda_2) \, ,
\eeq
The dependence of the energies and the polarizations of the back-scattered photons, i.e.,  $(E_{b}x_{1}, \lambda_{1})$ and  $(E_{b} x_{2}, \lambda_{2})$,  on the electron and positron beam energy $E_{b}$ as well as on  the frequency and the polarization of the laser, has been  computed in Ref.~\cite{ginzburg}. The results are that the spectrum peaks in the region of  high photon energy for $\lambda_{e} \lambda_{l} = -1$. If further one chooses  the laser energy  $\omega_{0}$ such that $x=4E_b~\omega_0/ {m_e^2} = 4.8$,  the two-photon luminosity is peaked at $z = 0.5 \times W/E_{b} = 0.8$. The mean helicity of the back-scattered photons depends on their energy. For the choice $\lambda_{e} \lambda_{l} = -1$ and $x = 4.8$, in the region of high energy for the  back-scattered photon where the spectrum is peaked, the  back-scattered photon also carries the polarisation of the parent electron/positron beam. Thus, choosing  $\lambda_{e^{-}} = \lambda_{e^{+}}$ ensures that the dominant photon helicities are the  same,  which in turn maximizes the Higgs signal relative to the QED background, leading to a luminosity $L_{\gamma \gamma} \equiv L_{\gamma \gamma} (\lambda_{e^-},x_{1},x_{2})$. The relevant expressions used for $L_{\gamma\gamma}(\lambda_{e^-},x_1,x_2)$ as well as those for $\lambda_1(\lambda_{e^-},x_1),\lambda_2(\lambda_{e^+},x_2)$ are taken from Ref.~\cite{ginzburg},
as presented in Ref.~\cite{grs}.

The total cross section for $\gamma \gamma \rightarrow \Phi \rightarrow gg$, where we write down explicitly the expression for $L_{\gamma \gamma}$ for the above choices of helicities,  is then given by
\beq
\sigma \; = \; {\frac {8 \pi^2}{M_\Phi s}}~~\Gamma (\Phi \rightarrow \gamma \gamma)~ 
\int_{x_1^m}^{x_1^M} {\frac{1}{x_1}} f(x_1) f(M^2_\Phi/s/y_1) \left(1 + \lambda_1(x_{1},\lambda_{e^{-}}) \lambda_2(x_{2},\lambda_{e^{+}})\right), 
\eeq
where $f(x_{i})$ denotes the probability that the backscattered photon carries a fraction $x_{i}$ of the beam energy for the chosen laser and lepton helicities, with
\beq
x_1^m = {\frac{M_\Phi^2}{s}} {\frac{(1 + x_c)}{x_c}},\,\, x_1^M = {\frac{x_c}{1+x_c}} ~~\mathrm {with ~} x_{c} = 4.8\, .
\eeq
Because of this cutoff on the fraction of the energy of the $e^{-}/e^{+}$ beam carried by the photon,  one needs a minimum energy $E_b= 453$ GeV to produce the 750 GeV resonance.

Our results for the cross section for $\Phi$ production via $\gamma \gamma$ collisions at
different $e^{+} e^{-}$ collision centre-of-mass energies are presented in  Fig.~\ref{fig:gammagammaPhi}. The above-mentioned choices of laser energy and the helicities of $e^{-}, e^{+}$ as well as those of the lasers $l_{1},l_{2}$, are used in our numerical calculations, ensuring that the $J_{Z } =0$ contribution is dominant for the production of the scalar resonance. Our results include thus the folding of the expected helicities of the backscattered photons  with the cross section. We see that the $\Phi$ production cross section is maximized for an $e^{+} e^{-}$ centre-of-mass energy $\sim 950$~GeV.

\subsection{$\Phi$ production in $\mathbf{e^+ e^-}$ collisions}

As the $\Phi$ state has the loop-induced couplings to electroweak gauge bosons given in eqs.~(\ref{eq:lag-eff})  and (\ref{eq:lag-eff2}), it can be produced in the same processes as the Standard Model--like Higgs boson,  namely the $WW$ and $ZZ$ fusion processes $\ee \to \Phi \nu \bar \nu$ and $\ee \to \Phi  \ee$ and the Higgsstrahlung process $\ee \to \Phi Z$. We also consider the companion process $\ee \to \Phi \gamma$, which occurs via the 
$\Phi \gamma\gamma$ and $\Phi Z\gamma$ couplings that are generated through the same loops 
as the $\Phi ZZ$ coupling. The couplings used in the discussion are\footnote{In principle, 
the $\Phi VV^*$ induced couplings should be damped by the virtuality of the 
off-shell gauge bosons, in much the same way as in pion scattering where the quadratic pion scalar radius plays an important role; see for instance Ref.~\cite{pion-FF}. Nevertheless, in the approximation that we are using in our exploratory work, we ignore these corrections and consider only the ``point--like" coupling below.} 
\beq
S\, V^\mu (p_1)V^\nu (p_2) &: &e/(v s_W) (p_1 \cdot p_2 g^{\mu \nu} - p_1^\mu p_2^\nu) 
c_{\Phi VV}  \nonumber \\
P\, V^\mu (p_1)  V^\nu (p_2) &: &e/(v s_W) ( i \epsilon_{\mu \nu \rho \sigma } p_1^\rho p_2^\sigma)  \tilde c_{\Phi VV} 
\eeq 
Neglecting the small standard--like contribution~\footnote{The full differential cross section including a Standard Model--like contribution  as well as the new contributions and their possible interferences can be found in~Ref.~\cite{Fernand}.}, the total cross section reads \cite{Fernand}
\beq
\sigma(\ee \to Z \Phi) = \frac{ 2\pi \alpha^2}{s } \, \lambda^{1/2} \, 
\left[ \left( 1+ \frac16 \frac{\lambda}{z}  \right) (D_+^2+D_-^2) 
+ \frac16 \frac{\lambda}{z}   ( \tilde D_+^2+\tilde D_-^2) \right] \, ,
\eeq
with $z=M_Z^2/s$, and $\lambda^{1/2}$ the usual two--particle phase--space function defined by $\lambda^{1/2} = \sqrt { (1-M_\Phi^2/s-M_Z^2/s)^2-4M_\Phi^2M_Z^2/s^2} \to  1-M_\Phi^2/s$ in the limit $M_Z \ll \sqrt s$. The scalar contributions $D_\pm$ are given in terms of the scalar coefficients $c_1$ and $c_2$ and reduced propagator, $P_Z=1/(1-z)$ by
\beq 
D_+ &=&  c_2 (1- P_Z) - c_1( 1+ P_Z s_W^2/c_W^2) \, , \nonumber \\ 
D_- &=&  c_2 [ 1 +  P_Z (1- 2s_W^2)/(2s_W^2) ] -c_1 [ 1 +  P_Z (1- 2c_W^2)/(2c_W^2) ] \, ,
\label{eq:eeXZ}
\eeq
and  the pseudoscalar contributions $\tilde D_\pm$ are given by similar expressions in terms of the corresponding pseudoscalar coefficients $\tilde c_1$ and $\tilde c_2$. 

The  process $\ee \to \Phi \gamma$ proceeds through the $s$--channel  exchange of the $Z$ boson and the photon via, respectively, the $\Phi Z\gamma$ and $\Phi\gamma\gamma$ induced couplings. Neglecting the small Standard Model--like loop-induced contribution~\cite{ee-Hgamma}, the cross section is given by \cite{Fernand}  
\beq
\sigma(\ee \to  \Phi\gamma ) = \frac{\pi \alpha^2}{3} \, \frac{ \lambda^{3/2} }{M_Z^2
c_W^2 s_W^2}  \, \left[  (D_1+\tilde D_1) + (D_2+\tilde D_2) + (D_3 +\tilde D_3) \right] \, ,
\label{eq:eeXp}
\eeq
with 
\beq 
D_1 &=&  c_2^2 [ 2s_W^4 + P_Z (1- 4s_W^2)s_W^2 +P_Z^2 (1/4-s_W^2 +2 s_W^4) ] \, , \nonumber \\
D_2 &=&  c_1^2 [ 2c_W^4 - P_Z (1- 4s_W^2)c_W^2 +P_Z^2 (1/4-s_W^2 +2 s_W^4) ] \, , \nonumber \\
D_3 &=&  c_1 c_2[4s_W^2c_W^2 +P_Z (1- 4s_W^2)(1- 2s_W^2)  -2 P_Z^2 (1/4-s_W^2 +2 s_W^4) ] \, ,
\eeq
and similarly for the CP--odd $\tilde D_i$ contributions. One should note that in the CP--odd case both the $\ee \to Z \Phi$ and $\ee \to \gamma\Phi$ cross sections behave like $ \sigma \propto \lambda^{3/2}$ near the kinematical threshold, and hence are strongly suppressed there, and have an angular distribution that follow the $1+ \cos^2\theta $ law~\cite{Bargeretal}. These features also hold for a CP--even state in the cases of both the $\ee \to \Phi \gamma$ process eq.~(\ref{eq:eeXp}) and also in the  $\ee \to \Phi Z$ process at high enough energies when $M_Z \ll \sqrt s$.  

\begin{figure}[!h]
\vspace*{-2.5cm}
\centerline{\hspace*{-1.2cm} 
\includegraphics[scale=0.8]{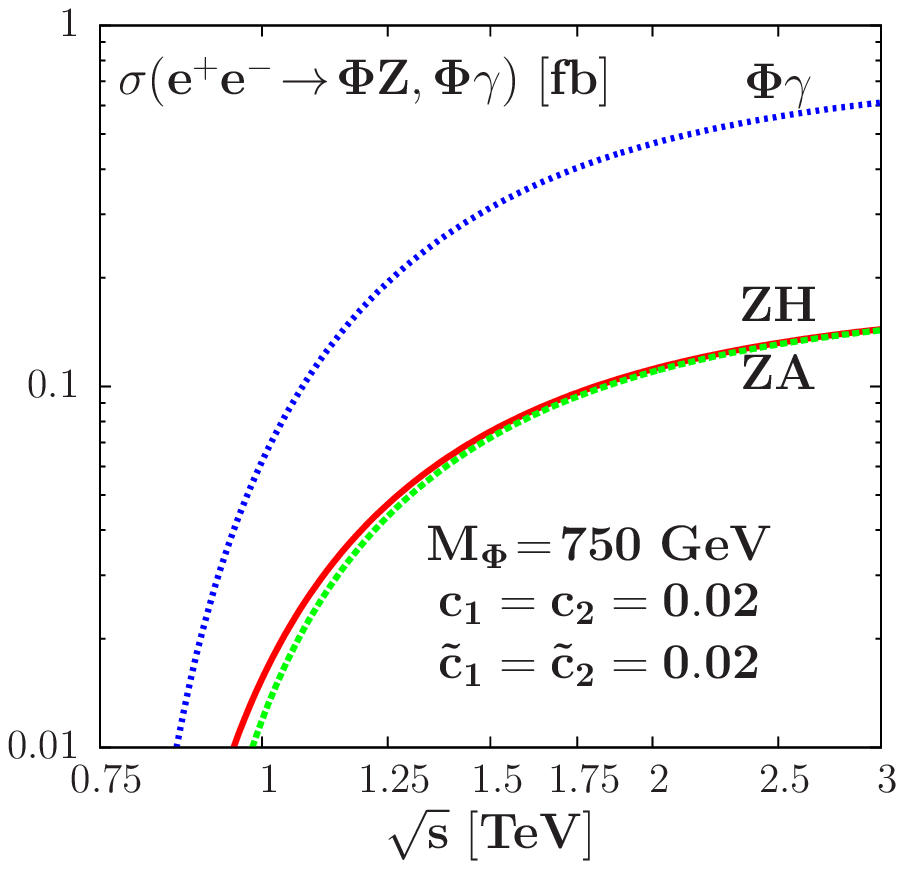}\hspace*{-9cm} 
\includegraphics[scale=0.8]{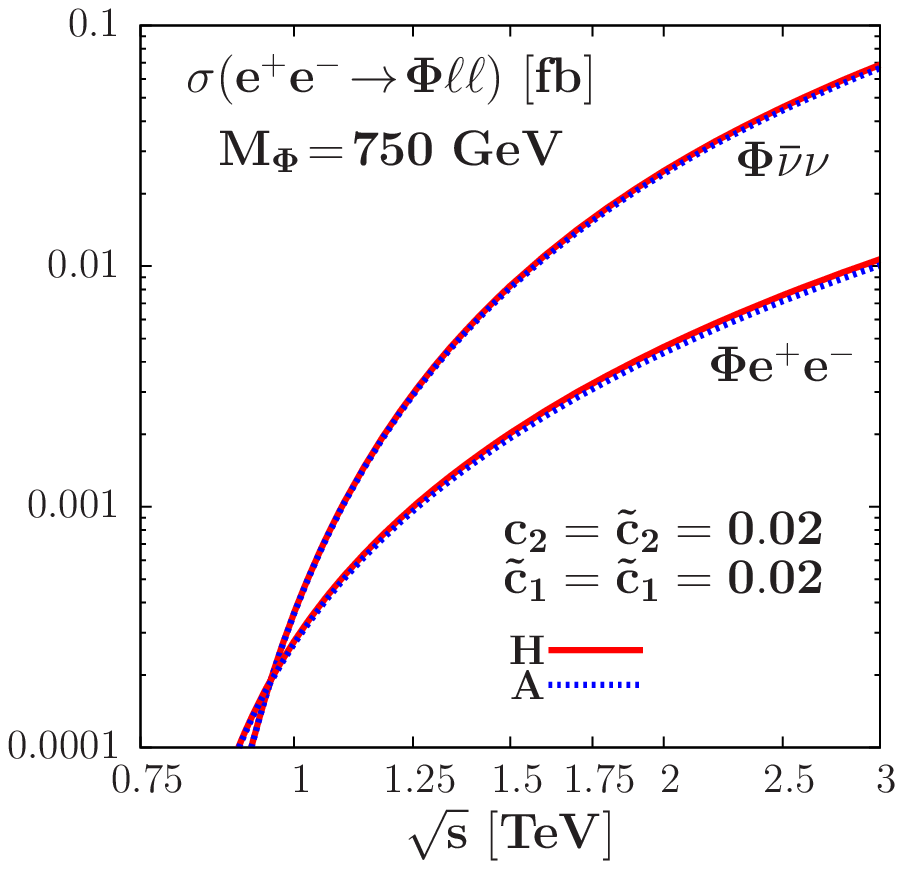}
}
\vspace*{-14.2cm}
\caption{\it
Cross sections in $\ee$ collisions for producing a singlet scalar or pseudoscalar state with $M_\Phi=750$ GeV state  as functions of the energy $\sqrt s$, for induced couplings to electroweak gauge bosons, $c_1=c_2=0.02= \tilde c_1= \tilde c_2$.  Left panel: Higgsstrahlung $\ee\to \Phi Z$ and associated production with a photon $\ee\to \Phi \gamma$. Right panel: the $WW$ fusion $\ee \to \Phi \nu \bar \nu$ and $ZZ$ fusion $\ee \to \Phi \ee$ processes.} 
\label{Fig:ee-Phi-sing}
\vspace*{-2mm}
\end{figure}

The production cross sections for the two processes $\ee \to Z \Phi$ and $\ee \to \gamma\Phi$
are shown in the left panel of Fig.~\ref{Fig:ee-Phi-sing} as functions of the centre-of-mass energy, where loop-induced couplings $c_1=c_2=0.02= \tilde c_1= \tilde c_2$ have been assumed in both the scalar and pseudoscalar cases (these values yield a partial decay width 
$\Gamma(\Phi \to \gamma\gamma) \approx$ a few MeV).  As can be seen, for such couplings, the cross sections are small but not negligible. They are approximately (exactly) the same for the CP--even and CP--odd scalar particles sufficiently above the $M_\Phi+M_Z$ threshold (for any $M_\Phi$) in  the $Z\Phi$ $(\gamma \Phi)$ case and, for the chosen $c_{1,2}, \tilde c_{1,2}$ values, they are a factor of four larger in $\ee \to \Phi \gamma$ than in the $\ee \to Z\Phi$ processes. The most important message of the figure is that,  contrary to the $\ee \to ZH$ cross section with standard--like Higgs couplings which drops like $1/s$, the cross sections with the anomalous induced couplings increase with energy (at least at the level of
approximation used here; see e.g. Ref.~\cite{pion-FF}). Hence, the highest energies are favored and rates at the 1 fb level can be generated in the chosen example for couplings.  

The other processes for the production of a scalar or a pseudoscalar resonance in $\ee$ collisions are due to vector boson fusion, $\ee \to V^* V^* \ell \bar \ell \to \Phi \ell \bar \ell$, which leads to the $\Phi \nu \bar \nu $~ and $\Phi \ee$ final states in the $WW$ and $ZZ$ fusion modes, respectively. In the Standard Model, the spin-summed and -averaged amplitude squared of the $e^-(k_1) + e^+(k_2) \to \nu (p_1) + \bar \nu  (p_2) + \Phi (p_3)$ process for $WW$ fusion is given by \cite{ep-VBF1}
\begin{eqnarray}
|{\cal M}|^2_{\rm SM}   =  \sigma_0 [ 4 M_W^4\, u_1t_2 ] ~{\rm with}~  \sigma_0= \frac{4\pi^3\alpha^3}{s_W^6 M_W^2} \frac{1}{ (t_1 - M_W^2)^2 \, (u_2 - M_W^2)^2} \, ,
\eeq
where the variables are defined as $s = (k_1 + k_2)^2, s' = (p_1 + p_2)^2,  t_1= (k_1 - p_1)^2, u_1  = (k_1 - p_2)^2, t_2 =(k_2 - p_1)^2$ and $u_2 = (k_2 - p_2)^2$. In the case 
of the scalar resonances $S$ and $P$ the amplitudes-squared would become \cite{ep-VBF1} 
\beq 
|{\cal M}|^2_{\rm S}  &= & \sigma_0 g_{SWW}^2 \left[ t_1 u_2 ( u_1^2  + t_2^2 + t_1  u_2- 2 s s' ) + (s s' - t_2 u_1)^2 \right] \, , \nonumber \\
|{\cal M}|^2_{\rm P} &= & \sigma_0 g_{PWW}^2 \left[ t_1 u_2 ( u_1^2   + t_2^2 - t_1  u_2+ 2 s s' ) - (s s' - t_2 u_1)^2 \right] \, .
\eeq
A similar expression can be obtained for the $ZZ$ fusion process $\ee \to \Phi \ee$ and for
equal $g_{ \Phi ZZ} = g_{WW \Phi}$ induced couplings, but the cross section is about a factor of ten smaller compared to $\sigma( \ee \to \Phi \nu \bar \nu)$, as a result of the smaller $Z\ee $ couplings compared to the $We\nu$ couplings. \comment{Using a model file for the program {\tt Madgraph} \cite{Madgraph} that was developed to study vector boson fusion at the LHC \cite{pp-VBF2}, we have calculated the cross section for the $WW$ and $ZZ$ fusion processes.} We have calculated the cross section for the $WW$ and $ZZ$ fusion processes using the calculations that were developed to study vector boson fusion for anomalous vertices  at the $e^{+} e^{-}$ colliders ~\cite{ee-VBF3} and the LHC \cite{pp-VBF2}.   The cross sections are  shown in 
Fig.~\ref{Fig:ee-Phi-sing} (right) as a function of $\sqrt s$ again for $c_2= \tilde c_2=0.02$.
Here again they are the same for CP--even and CP--odd particles. The $WW$ fusion cross section is comparable to that of $\ee \to Z\Phi$  and that of $ZZ$ an order of
magnitude smaller. Hence, even if the couplings of the $\Phi$ resonances to $\gammaû\gamma, \gamma Z, ZZ$ and $WW$ states are loop-induced, the production rates are not negligible  
at high--energy and high--luminosity $\ee$ colliders.

\section{Benchmark Two-Higgs-Doublet Models}

\subsection{Properties of the scalar resonances}

\subsubsection{Review of models and couplings}

In this Section, we discuss a second possibility \cite{Phi-ADM}: namely that the observed scalar $\Phi$ resonance is the heavier CP--even $H$ state and/or the CP--odd $A$ state of a two Higgs doublet model (2HDM)~\cite{2HDM} as realised, for instance, in the Minimal Supersymmetric extension of the Standard Model (MSSM) \cite{HHG,Review2}.   We start by reviewing the CP--conserving 2HDM and, more precisely,  a special MSSM scenario called  the $h$MSSM \cite{hMSSM,hMSSM-fully}, which will be the basic framework for our second benchmark scenario for the $\Phi$ resonance.  

The scalar potential of this model, in terms of the two Higgs doublet fields $\Phi_1$ and $\Phi_2$, is described by three mass parameters and five quartic couplings and is given by
\cite{2HDM}
\begin{eqnarray}
 V &=m_{11}^2\Phi_1^\dagger\Phi_1^{\phantom{\dagger}} +m_{22}^2\Phi_2^\dagger\Phi_2^{\phantom{\dagger}} -m_{12}^2 ( \Phi_1^\dagger\Phi_2^{\phantom{\dagger}} +\Phi_2^\dagger\Phi_1^{\phantom{\dagger}}) 
+\tfrac12 \lambda_1(\Phi_1^\dagger\Phi_1^{\phantom{\dagger}})^2
   +\tfrac12 \lambda_2(\Phi_2^\dagger\Phi_2^{\phantom{\dagger}})^2
 \nonumber \\  &\phantom{{}={}}
  +\lambda_3(\Phi_1^\dagger\Phi_1^{\phantom{\dagger}})
            (\Phi_2^\dagger\Phi_2^{\phantom{\dagger}})
  +\lambda_4(\Phi_1^\dagger\Phi_2^{\phantom{\dagger}})
            (\Phi_2^\dagger\Phi_1^{\phantom{\dagger}})
  +\tfrac12 \lambda_5[ (\Phi_1^\dagger\Phi_2^{\phantom{\dagger}})^2
                      +(\Phi_2^\dagger\Phi_1^{\phantom{\dagger}})^2] \, . \label{eq:pot}
\end{eqnarray}
The model contains two CP--even neutral Higgs bosons $h$ and $H$, a CP--odd neutral boson $A$ and two charged $H^\pm$ bosons, whose masses $M_h, M_H, M_A$ and $M_{H^\pm}$  are free parameters. We assume that the lighter CP--even $h$ boson is the light Higgs state with a mass of $M_h=125$ GeV  that was discovered at the LHC in 2012 \cite{SMH-discovery1}. Three other parameters characterize the model: the mixing angle $\beta$ with $\tb = v_2/v_1$, where $v_1$ and $v_2$ are the vacuum expectation values of the neutral components of the fields $\Phi_1$ and $\Phi_2$, with $\sqrt {v_1^2+v_2^2} = v = {\rm 246~GeV}$, the angle $\alpha$ that diagonalises the  mass matrix of the two CP--even $h$ and $H$ bosons, and another mass parameter $m_{12}$ that enters only in the quartic couplings among the Higgs bosons, which is not relevant for our analysis. 
 
In this parametrisation, the neutral CP--even $h$ and $H$ bosons share the coupling of
the Standard Model Higgs particle to the massive gauge bosons $V=W,Z$ and one has, at tree level, the following couplings normalised relative to those of the standard Higgs
\begin{eqnarray}
\hat g_{hVV}=  \sin(\beta-\alpha) \ , ~~~
\hat g_{HVV}=  \cos(\beta-\alpha) \, ,
\end{eqnarray}
while, as a consequence of CP invariance,  the CP--odd $A$ does not couple to vector bosons, $\hat g_{AVV}=0$. There are also couplings between two Higgs and a vector boson which, up to a normalization factor, are complementary to the ones above. For instance, one has
\begin{eqnarray}
\hat g_{hAZ}= \hat g_{h H^\pm W}=  \cos(\beta-\alpha) \ , ~~~ 
\hat g_{HAZ}= \hat g_{H H^\pm W} =  \sin(\beta-\alpha) \, .
\end{eqnarray}
For completeness, additional couplings of the charged Higgs boson will be needed in our discussion: they do not depend on any extra parameter and one has, for instance, 
$\hat g_{A H^\pm W}= 1$ and $g_{H^+ H^- Z} = -e \cos2\theta_W /(\sin\theta_W \cos\theta_W)$.

The interactions of the Higgs states with fermions are more model--dependent, and there are two major options that are discussed in the literature; see again Ref.~\cite{2HDM}. 
In Type-II 2HDMs, the field $\Phi_1$ generates the masses of down--type quarks and charged leptons, while $\Phi_2$ generates the masses of up--type quarks, whereas in Type-I 2HDMs
the field $\Phi_2$ couples to both up-- and down--type fermions. The couplings of the neutral 
Higgs bosons to gauge bosons and fermions in the two models are summarized in Table~\ref{Tab:cpg-2HDM}.  (The couplings of the charged Higgs to  fermions follow those of the CP--odd 
Higgs state.)

\begin{table}[!h]
\vspace*{2mm}
\begin{center}
\renewcommand{\arraystretch}{1.2}
\begin{tabular}{|c|c|c|c|c|c|c|} \hline
\ \ $\Phi$ \ \  &\multicolumn{2}{c|}{$\hat g_{\Phi \bar{u}u}$}&  
                  \multicolumn{2}{c|}{$\hat g_{\Phi \bar{d}d}$}&  
$\hat g_{ \Phi VV} $ \\ \hline
& Type I & Type II & Type I & Type II & Type I/II \\   \hline
$h$  &  $\; \cos\alpha/\sin\beta       \; $ 
     &  $\; \cos\alpha/\sin\beta       \; $  
     &  $ \; \cos\alpha/\sin\beta \; $  
     &  $ \; -\sin\alpha/\cos\beta \; $  
     &  $ \; \sin(\beta-\alpha) \; $  \\
$H$  &  $\; \sin\alpha/\sin\beta \; $  
     &  $\; \sin\alpha/\sin\beta \; $  
     &  $ \; \sin\alpha/ \sin\beta \; $  
     &  $ \; \cos\alpha/ \cos\beta \; $  
     &  $ \; \cos(\beta-\alpha) \; $  \\
$A$  &  $\; \cot \beta \; $ 
     &  $\; \cot \beta \; $ 
     &  $ \; \cot \beta \; $    
     &  $\; \tan \beta \; $ 
     &  $ \; 0 \; $ \ \\ \hline
\end{tabular}
\end{center}
\vspace{-4mm}
\caption[]{\small The couplings of the  $h,H,A$ states to fermions and gauge bosons in Type-I and -II 2HDMs relative to standard Higgs couplings; the $H^\pm$ couplings to fermions follow those of $A$.} 
\label{Tab:cpg-2HDM}
\vspace{-4mm}
\end{table}

We see that the Higgs couplings to fermions and gauge bosons depend only on the ratio $\tan\beta$ and on the difference $\beta-\alpha$. However, one needs to take into account the fact that the couplings of the light $h$ boson have been measured at the LHC and found to be
Standard Model--like \cite{HiggsCombo}. With this in mind, we set  $\beta-\alpha=\frac{\pi}{2}$, which is called the  alignment limit \cite{alignment}. In this limit, the $h$ couplings to fermions and vector bosons are automatically standard--like, $\hat g_{hVV}= \hat g_{huu}= \hat g_{hdd} \to 1$, while  the couplings of the CP--even $H$ state  reduce exactly to those of the pseudoscalar $A$ boson. In particular, there is no $H$ coupling to vector bosons, $\hat g_{HVV} \to \hat g_{AVV} =0$, and the couplings to up--type fermions are $\hat g_{Huu} = \cot \beta$, while those to down--type fermions are   $\hat g_{Hdd} = \cot \beta$ and $\hat g_{Hdd} = \tan \beta$ in Type-I and -II models, respectively. 

Finally, there are also some triple couplings among the Higgs bosons  that depend in addition
on the parameter $m_{12}$. However, in the alignment limit $\beta-\alpha=\frac\pi2$ the most important ones involving the lighter $h$ boson are simply $\hat \lambda_{hhh} \approx 1$ and $\hat \lambda_{Hhh} \approx 0$.  

In addition to $\tb$,  the other 2HDM parameters are the three Higgs masses $M_H, M_A$ and  $M_{H^\pm}$, which are in principle free. In our scenario we assume that the possible $\Phi$ resonance is a superposition of the $H$ and $A$ states and set $M_H\approx M_A \approx 750$ GeV. This assumption has several motivations.  First, it is a property of the MSSM in the decoupling limit \cite{decoupling} as will be seen  shortly. Then, there is only one hint 
of a peak at the LHC (not two) and having two degenerate states enhances the signal (which is a necessity in the 2HDM). Finally,  a small breaking of the mass degeneracy would yield a larger signal width (as may be favoured by the ATLAS data), as will be seen later.

The charged Higgs boson mass and $\tb$ will thus be the only free parameters, and in most of our discussion we assume $M_{H^\pm}$ to be comparable to the $H/A$ masses: $M_H \approx M_A \approx M_{H^\pm}$, as happens in the MSSM scenario in the decoupling limit $M_A \gg M_Z$
\cite{decoupling}. 

Indeed, the MSSM is essentially a 2HDM of Type II in which supersymmetry imposes strong constraints on the Higgs sector so that only two parameters, generally taken to be $M_A$ and $\tb$, are independent. This remains true also when the important radiative corrections that introduce dependences on many other supersymmetric model parameters \cite{RC-1loop} are incorporated. These corrections shift the value of the lightest $h$ boson mass from the tree--level value, predicted to be $M_h \leq M_Z |\cos2\beta| \leq M_Z$, to the value $M_h=125$ GeV that has been measured experimentally \cite{HiggsCombo}. Assuming a very heavy supersymmetric particle spectrum, as indicated by LHC data \cite{PDG}, and fixing these radiative corrections in terms of $M_h$, one can write the parameters $M_H, M_{H^\pm}$ and $\alpha$ in terms of $M_A,\tb$ and $M_{h}$ in the simple form (writing $c_\beta \equiv \cos \beta$ etc..) 
\begin{eqnarray}
h{\rm MSSM}:~~ 
\begin{array}{ll} 
M_{H} = \sqrt{ \frac{(M_{A}^2+M_{Z}^2-M_{h}^2)(M_{Z}^2 c^{2}_{\beta}+M_{A}^2s^{2}_{\beta}) 
- M_{A}^2 M_{Z}^2 c^{2}_{2\beta} } {M_{Z}^2 c^{2}_{\beta}+M_{A}^2 s^{2}_{\beta} - M_{h}^2}
} & \stackrel{\small M_A \gg M_Z} \longrightarrow M_A \, , \\
M_{H^\pm} = \sqrt { M_A^2 + M_W^2}  & \stackrel{\small M_A \gg M_Z} \longrightarrow M_A \, , \\ \ \ \  \alpha = -\arctan\left(\frac{ (M_{Z}^2+M_{A}^2) c_{\beta}    s_{\beta}} {M_{Z}^2 
c^{2}_{\beta}+M_{A}^2 s^{2}_{\beta} - M_{h}^2}\right) & \stackrel{\small M_A \gg M_Z} \longrightarrow  \beta -\frac12 \pi \, .
\end{array}
\label{hMSSM} 
\end{eqnarray}

This is the so--called $h$MSSM approach \cite{hMSSM,hMSSM-fully}, which has been shown to provide a very good approximation to the MSSM Higgs sector \cite{LHCWG-hMSSM}. 

When  $M_A \! \gg \! M_Z$, one is in the so--called decoupling r\'egime, where one has 
$\alpha \approx \beta- \frac\pi2$ implying that the light $h$ state has almost exactly the 
standard  Higgs couplings, $\hat g_{hVV}= \hat g_{hff}= 1$. The other CP--even 
boson $H$ and the charged bosons $H^\pm$ become heavy and degenerate in mass  with the $A$ state, $M_H \! \approx \! M_{H^\pm} \! \approx \! M_A$, and decouple from the massive gauge bosons. The strengths of the couplings of the $H$ and $A$ states are the same. Thus, in this r\'egime the MSSM Higgs sector looks almost exactly like that of the 2HDM of Type-II in the alignment limit, especially if we use the additional assumption $M_{H^\pm} = M_A$ on the Higgs masses that simplifies further the model.  Hence, our discussion below covers two scenarios: the 2HDM in the alignment limit and the MSSM in the decoupling limit,
augmented by extra vector-like fermions as we discuss in the next subsection.

\subsubsection{Boosting the $\mathbf{\Phi=H/A}$ production rates at the LHC}

Loops of the top quark (and lighter standard fermions) are, by themselves, insufficient to explain the magnitude of the $\Phi = H, A \to \gamma\gamma$ signal hinted by the LHC experiments. How large is the contribution from heavy vector-like fermions~\footnote{The easiest option would have been  the introduction of a fourth generation of fermions but it is completely excluded  by the observation of the light $h$ state with standard --like couplings \cite{fourth}.} that we need to enhance the $\Gamma(\Phi \to \gamma \gamma)$ rate to fit the estimated LHC cross section at 13~TeV, $\sigma(\Phi \to \gamma \gamma) \approx 6$ fb? For $M_H=M_A=750$ GeV and $\tb=1$, one has in principle branching ratios of the order of BR$(A \to \gamma\gamma) \approx 7 \times 10^{-6}$ and BR$(H \to \gamma\gamma) \approx 6 \times 10^{-6}$ for the $A$ and $H$ resonances and total decay widths $\Gamma_{\rm tot}^\Phi \sim \Gamma (\Phi \to t\bar t) \approx 32\;{\rm GeV}$ for the $H$ and $\approx 35\;{\rm GeV}$ for the $A$ boson~\cite{hdecay} which, as discussed previously, are consistent with the  total width of $\sim 45$ GeV favoured by the ATLAS Collaboration~\footnote{In fact, the best fit to the ATLAS value $\Gamma_\Phi \simeq 45$ GeV can be obtained by allowing a 10--15 GeV mass difference between the $H$ and $A$ states. Curiously enough, this is just what happens in the $h$MSSM: with $M_A \approx 750$ GeV, one obtains $M_H \approx 765$ GeV for $\tan\beta \approx 1$ \cite{hMSSM}.}. At $\sqrt s=13$ TeV, the cross sections in the dominant $gg\to \Phi$ processes are  $\sigma (gg\to H)\approx 0.6$~pb and $\sigma (gg\to A) \approx  1.3$~pb for the chosen $M_\Phi$ and $\tb$ values. This leads to the following cross section times branching fraction when the two states are added  (the numbers are for the $h$MSSM),
$\sum_{\Phi =H,A}  \sigma (gg\to \Phi) \times {\rm BR}( \Phi \to \gamma\gamma) \approx 
1.2 \times 10^{-2}  \; {\rm fb}$. 
This is to be compared with the reported  cross section $\sigma (gg\to \Phi) \times {\rm BR}( \Phi \to \gamma\gamma) \approx  6\; {\rm fb}$. We conclude that an enhancement factor $K^{H+A}_{gg \times \gamma\gamma}$ of about 500  is required when the rates for the two resonances $H$ and $A$ are added in order to accommodate the observations. 

Such an enhancement can be obtained by including singly- or doubly-charged vector-like leptons in the $\Phi \gamma \gamma$ loop vertices and/or also some vector--like quarks in both the $\Phi \gamma \gamma$ and $\Phi gg$ loops. The contributions of some representative scenarios with vector-like fermions are illustrated in Fig.~\ref{Fig:Aboost} (left), where the boost factors $K_{gg \times \gamma \gamma}$ obtained as functions of the charged fermion mass (assumed, for simplicity, to be universal) can be compared with the factor $\sim 500$ that is needed to reach the hinted $gg\! \to \! \Phi \! \to \! \gamma \gamma$ rate. 

Three scenarios are considered in the left--hand side of Fig.~\ref{Fig:Aboost}. The first scenario includes three vector-like pairs of left-- and right--handed leptonic doublets (green dotted line), which leads to the presence of six charged leptons. In this case, the enhancement factor can reach the level of $\sim 200$ when the (common) masses of the vector-like leptons are close to the $\frac12 M_\Phi$ threshold for which the form factors $A_{1/2}^\Phi$ are maximal; see Fig.~\ref{fig:enhancement}. A second scenario is when three charged dileptons  $E^{--}$ contribute to the $\Phi \gamma\gamma$ loop (blue dotted line). Because of the higher electric charge, the enhancement factor is larger than in the previous case (a factor of two at the amplitude level). A final scenario is when an entire generation of vector--like quarks and leptons is added to the standard spectrum (red solid line). In this case, as in the singlet scenario discussed in the previous Section, up- and down-type quarks $(U, D)$ contribute to both the $\Phi gg$ and $\Phi  \gamma\gamma$ loops, and the additional vector-like charged leptons $L$ also contribute to the $\Phi  \gamma\gamma$ amplitude.

\begin{figure}[!h]

\vspace*{-2.5cm}
\centerline{\hspace*{-1cm} \includegraphics[scale=0.75]{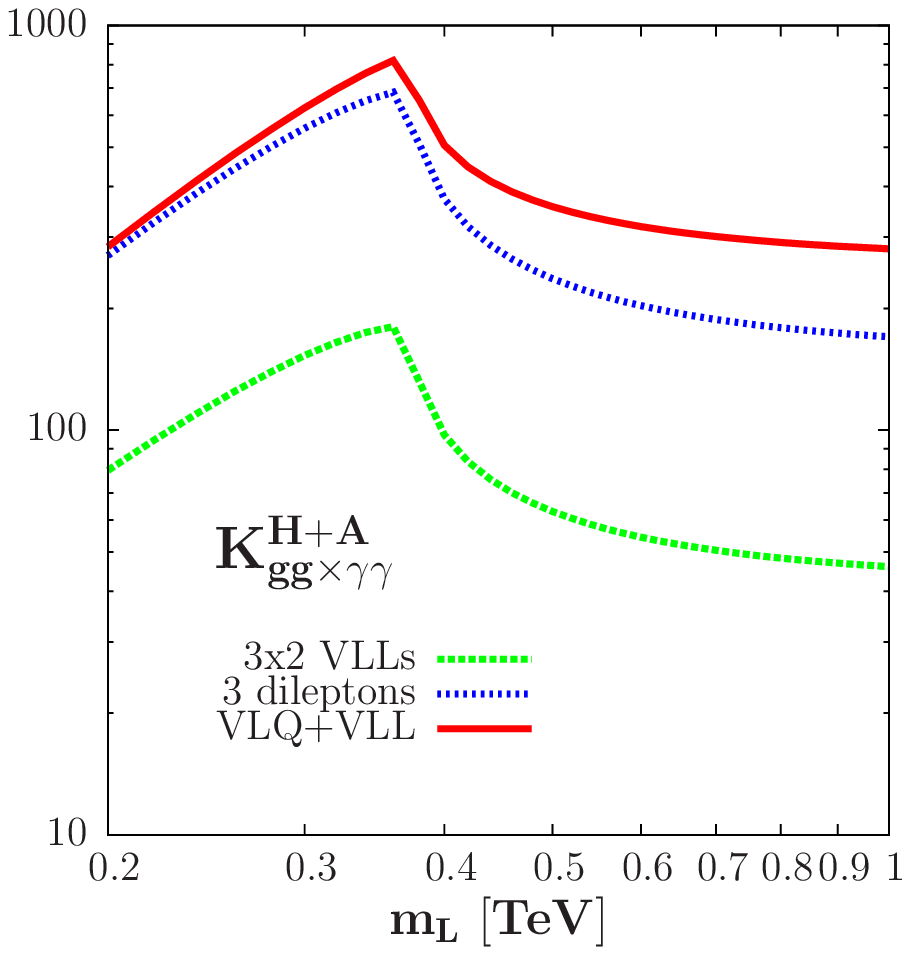}
            \hspace*{-8cm} \includegraphics[scale=0.75]{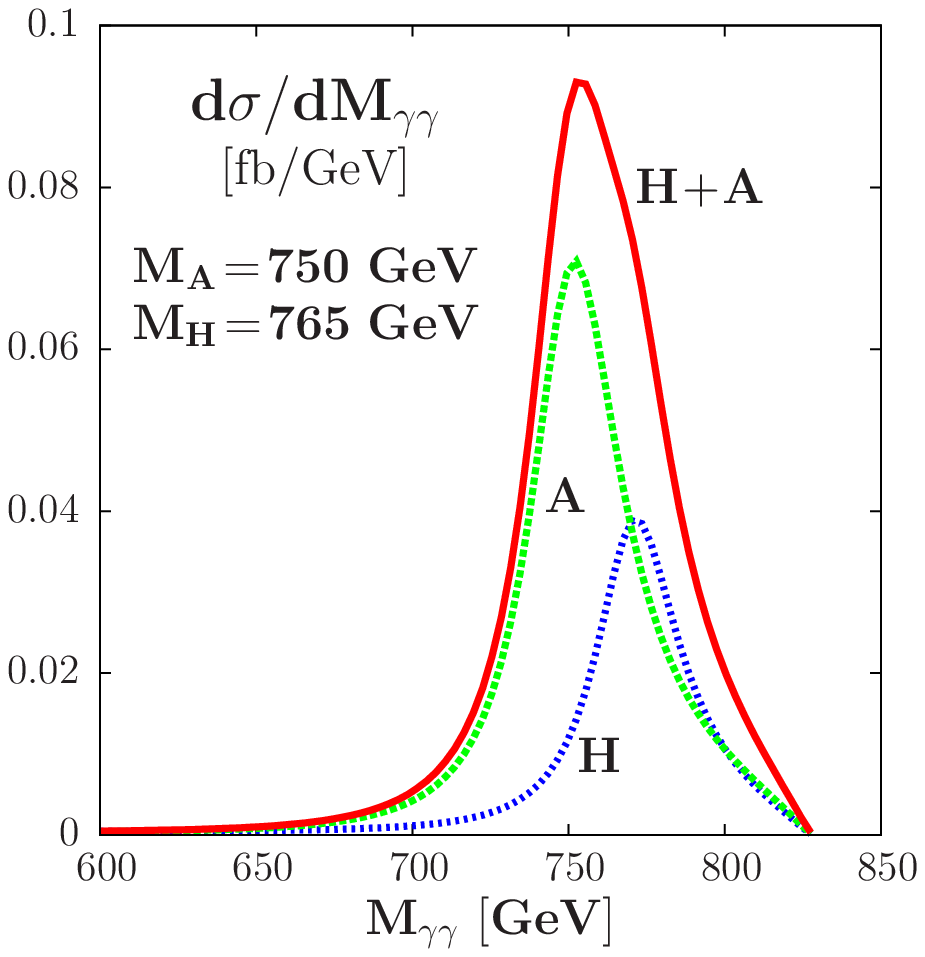}
}
\vspace*{-12.8cm}
\caption{\it Left: The enhancement factors, $K^{H+A}_{gg \times \gamma\gamma}$ as a function of the lepton mass, in scenarios with vector--like fermions, of the prospective $\sigma (gg\to \Phi) \times {\rm BR}( \Phi \to \gamma\gamma)$ signal, compared to the 2HDM case in which only the top quark contribution for $tan\beta =1$ is included. Right: The $\Phi = H, A$ line-shape calculated assuming as in the $h$MSSM $M_A = 750$~GeV, $M_H= 765$~GeV,  $\Gamma_H = 32$~GeV and $\Gamma_A = 35$~GeV for $\tan \beta = 1$, and normalizing the combined signal curve to $\sigma \times {\rm BR}(pp\! \to \! H,A \!  \to \! \gamma \gamma) = 6$~fb.}
\label{Fig:Aboost}
\vspace*{-2mm}
\end{figure}

In this last scenario, in order to suppress the cross section for $\sigma(gg \to \Phi \to t \bar t)$, which would also be enhanced by the presence of the new $U$ and $D$ quarks, we assume that $\tb=3$. In this case, $\Gamma_\Phi \! \approx \! 3$ GeV in the absence of novel $\Phi$ decay modes. A large $\Phi$ width could be recovered, e.g., by allowing invisible $\Phi$ decays into the neutral vector-like leptons~\footnote{This would be the case if, for instance, the partner neutral lepton is the dark matter particle, with interactions mediated by the $\Phi$; see, e.g., Ref.~\cite{Phi-YAD}.}, $\Phi \! \to \! \bar NN$,  with the mass  $m_N$ adjusted in order to obtain the  desired total width. This would also suppress the fraction BR$(\Phi \to t\bar t)$, allowing us to evade even more easily the  constraints from the ATLAS and CMS searches for $t\bar t$ resonances \cite{LHC-ttbar}.

The right--hand side of Fig.~\ref{Fig:Aboost} shows the expected $\Phi = H, A$ 
line-shape, assuming $M_A = 750$~GeV, $M_H - M_A = 15$~GeV as in the $h$MSSM, $\sigma \times {\rm BR}(pp \to A \to \gamma \gamma)/\sigma \times {\rm BR}(pp \to H \to \gamma \gamma)
= 2$, $\Gamma_{\rm tot} (H) = 32$~GeV and $\Gamma_{\rm tot} (A) = 35$~GeV as in the $h$MSSM for $\tan \beta = 1$ (modelled as non-interfering Breit-Wigners), and normalizing the combined signal curve to $\sigma \times {\rm BR}(pp \to \Phi = H,A \to \gamma \gamma) = 6$~fb. The $A$ contribution is the dashed green line, the $H$ contribution is the dotted blue line, and the sum is the solid red line. In principle, this line-shape could be distinguished from a single Breit-Wigner, e.g., by its asymmetry around the peak of the  distribution, though this would require large statistics.

\subsubsection{Decays of the $\mathbf{\Phi=H/A}$ states}

We turn now to the decays of the $\Phi=H/A$ states which, in the two limits of alignment in the general 2HDM and decoupling in the MSSM, are almost the same (this is particularly true as we assume that the supersymmetric particle  spectrum is very heavy and do not enter the
decays either directly or indirectly in the loop-induced modes).  The pattern of these decays  is, to a large extent,  dictated by the value of $\tb$ and the fact that for low values, the top  quark Yukawa coupling $\propto 1/\tb$ is very large, while at high $\tb$ the Higgs couplings to bottom quarks and $\tau$ leptons, $\propto \tb$, are enhanced; see Table \ref{Tab:cpg-2HDM}.
  
We start with the modes by which the $\Phi$ signal has been observed, i.e., the $\Phi \to \gamma\gamma$ mode and the (inverse of the) $\Phi \to gg$ mode. In these two cases, the partial decay widths, assuming that only heavy  fermions are running in the loops~\footnote{There is no $W$ contribution in the $H$ case as the coupling $\hat g_{HVV} \to 0$ is suppressed in these limits and the contribution of the charged  Higgs boson is very small \cite{Phi-ADM}.},  are  given by the same expressions eqs.~(\ref{eq:Gammagg}, \ref{eq:ASAP}, \ref{eq:formfactors}) shown previously~\cite{venerable,Review1}, see also Fig.~\ref{fig:enhancement}.

The other important decays of the $\Phi$ states would be into fermion pairs, with 
partial widths given, in terms of the fermion velocity $\beta_f=(1-4m_f^2/M_{\Phi}^2)^{1/2}$,  by \cite{Review1}
\begin{eqnarray}
\Gamma(\Phi \to f \bar{f} ) =  N_c \frac{ G_\mu m_f^2}{4\sqrt{2} \pi}
\, \hat g_{\Phi ff}^2 \, M_{\Phi} \, \beta^{p_\Phi}_f \, ,
\end{eqnarray}
with  $p_\Phi =3\,(1)$ 
for the CP--even (odd) Higgs boson. In principle, the only relevant decays at low $\tb$ values
are those into $t\bar t$ pairs, while at high $\tb$ values the decays into $b\bar b$ and 
$\tau^+ \tau^-$ pairs are dominant. All other decay modes~\footnote{There is, however, one exception:  the decays $\Phi \to H^\pm W^\mp$ with light charged Higgs bosons. In general,
we assume this channel to be kinematically closed, an assumption that we revisit later.}  
are strongly suppressed in the alignment/decoupling limits of 2HDMs/MSSM. 
In particular, this is the case for the $H\to WW,ZZ$ decays of the CP-even $H$ state
and the $A \to h Z$ decays of the CP--odd $A$ state, which involve the couplings 
$\hat g_{HVV}= \hat g_{ZhA}= \cos(\beta-\alpha) \to 0$. The decay $H\to hh$, 
which involves the trilinear coupling $g_{Hhh}$ that is small or vanishing,  is also suppressed. 

The branching fractions for the main decay modes of the $H/A$ states, namely  $\Phi \to t\bar t, b \bar b, \tau^+ \tau^-$ and the loop-induced decays $\Phi \to \gamma\gamma, gg$ are shown in the left panel of Fig.~\ref{Fig:BR-phi} for $M_H=M_A=750$ GeV as functions of $\tan\beta$. A Type-II 2HDM like the MSSM has been assumed, and the value of $\tb$ is restricted to lie in the range $ \frac13 \lsim \tb \lsim 60$ for which both the top and bottom quark Yukawa couplings, $y_t = m_t/v \tb$ and $y_b= \bar m_b \tb/v$, are perturbative at the weak scale (using the pole $t$--mass $m_t=172$ GeV and the running $b$--mass at the scale of the Higgs mass $ \bar m _b= 3$ GeV \cite{PDG}). 

\begin{figure}[!h]
\vspace*{-2.4cm}
\centerline{\hspace*{-1.5cm} \includegraphics[scale=0.7]{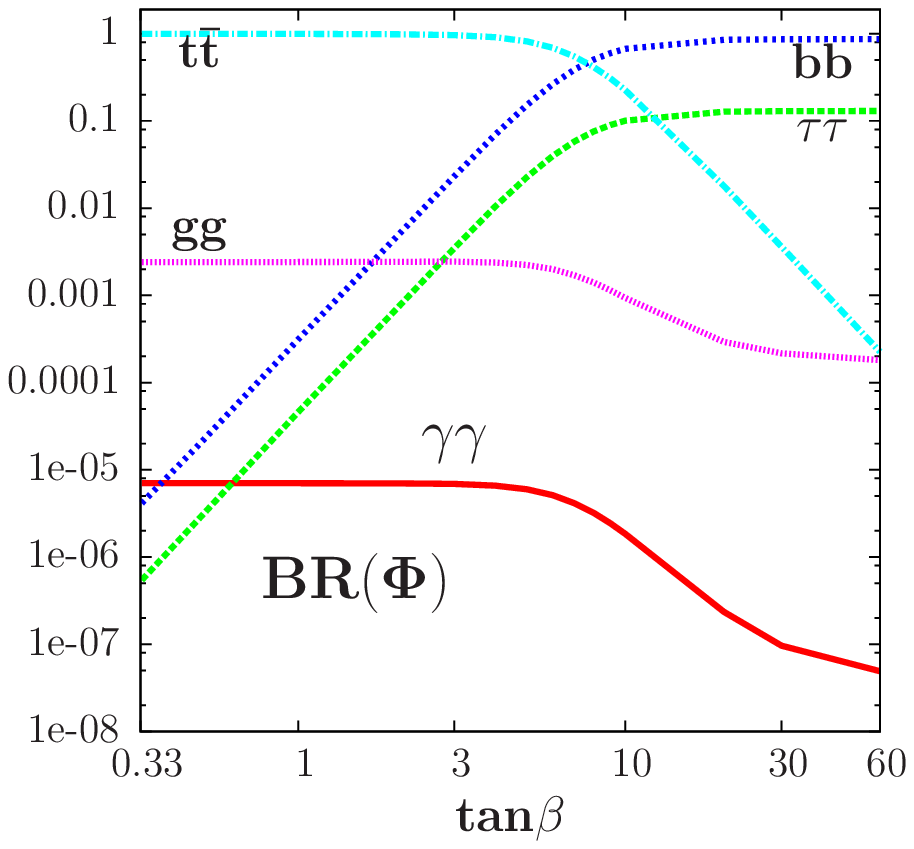}
            \hspace*{-7.2cm} \includegraphics[scale=0.7]{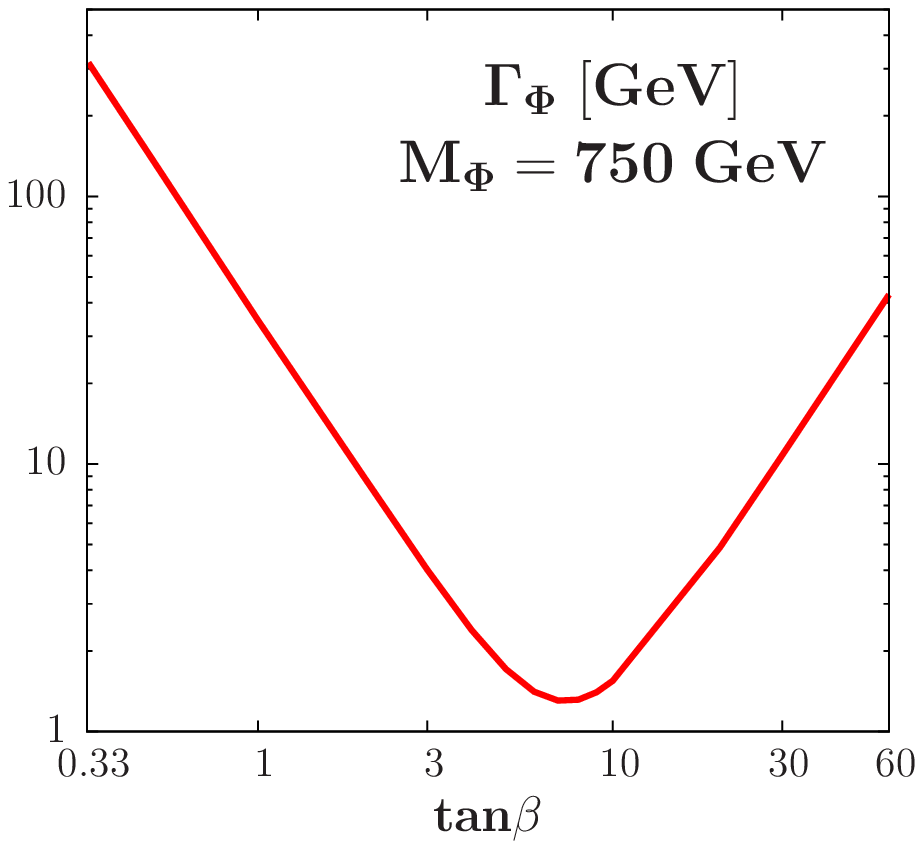}
}
\vspace*{-12.6cm}
\caption{\it The branching ratios for $\Phi = H/A$ decays into various final states
for $M_\Phi=750$ GeV as functions of $\tb$ (left) and the corresponding total decay 
width (right).} 
\label{Fig:BR-phi}
\vspace*{-3mm}
\end{figure}

As expected, the decay $\Phi \to t \bar t$ dominates by far when $\tb \lsim 5$, for which the 
top Yukawa coupling is the largest: $y_t \! \gg \! y_b$. On the other hand, for $\tb \gsim 10$ one has $y_b \gg y_t$ and the decay $\Phi \to b \bar b$ is the dominant one. The branching fractions for decays into $\tau$ pairs is ${\rm BR}(\Phi \to \tau^+\tau^-) \approx 10\%$, 
a simple reflection of the fact that $3 \bar m_b^2/m_\tau^2 \approx 10$, with 3 being the colour factor. For intermediate $\tb$ values, $\tb \approx 5$--10, the suppression of the $\Phi t\bar t$ coupling is already effective, whereas the $\Phi b \bar b$ coupling 
is not yet strongly enhanced, resulting in decay rates into $t\bar t$ and $b \bar b$ that are comparable. The cross-over point is at $\tb \approx \sqrt{m_t / \bar m_b} \approx 7$.

The decays $\Phi \to \gamma \gamma$ have constant branching fractions of the order of 
$10^{-5}$ at low $\tb$ and, starting from $\tb \approx 5$, they decrease with 
increasing $\tan\beta$ reaching BR($\Phi \to \gamma \gamma) \approx 10^{-7}$ for $\tb \approx 30$. The branching ratio for the $\Phi \to gg$ decay is  more than two orders of magnitude higher, ${\cal O}(\frac32 \alpha_s^2/\alpha^2)  \approx 400$. Thus the maximal values of the $\Phi \to \gamma \gamma$ and $\Phi \to gg$ branching ratios, and hence the $\Phi$  cross section at a $pp$ collider, will be obtained at low $\tb$ values. We note also that, for $M_\Phi=750$ GeV, values $\tb \gsim 20$ are excluded by the search for heavy Higgs particles decaying into $\tau^+ \tau^-$ pairs~\cite{hMSSM-fully}~\footnote{The ATLAS and CMS bounds are less restrictive, with only values $\tb \gsim 30$ being excluded for $M_\Phi=750$ GeV~\cite{LHC-tautau}. However, these searches were interpreted in a benchmark scenario in which 
additional decays of the $A/H$ bosons, namely supersymmetric decays into charginos and neutralinos, are present, reducing the interesting $A/H$ branching ratios into $\tau^+\tau^-$ final states.}. 
 
The total decay width of $\Phi = H, A$ is shown in the right panel of Fig.~\ref{Fig:BR-phi} as a function of $\tan\beta$. For the reasons discussed above, the total width is large at low and high $\tb$ values, and values $\tb < 1$ lead to an unacceptably large width for $\Phi$. The width has a minimum of about 1 GeV at the cross-over $\tb$ value, $\tb \approx \sqrt{m_t / \bar m_b} \approx 7$. If the total width  $\Gamma_\Phi \approx 45$ GeV apparently observed by the ATLAS Collaboration \cite{ATLAS-diphoton} is to be attained, values $\tb \approx 1$ or $\tb \approx 60$ would be required, but the high--$\tb$ option is completely excluded by the $H/A \to \tau^+ \tau^-$ searches at the previous LHC run~\cite{LHC-tautau}.

Another argument that disfavours the $\tb < 1$ option is that searches for resonances 
decaying into $t\bar t$ final states have been conducted at $\sqrt s= 8$ TeV with 20 fb$^{-1}$, setting an upper limit on $\sigma ( pp \to X \to t\bar t)$ of about half a pb \cite{LHC-ttbar}, which is attained in our model for the $\Phi$ signal for $\tb \approx 1$.  Any value $\tb < 1$ would lead to a $gg\to \Phi \to t\bar t$ rate that is too high and hence excluded by the ATLAS and CMS searches~\footnote{We note that these searches have in  fact been performed only for electroweak spin--one resonances, like new neutral gauge bosons or electroweak Kaluza--Klein excitations decaying into $t\bar t$ pairs~\cite{LHC-ttbar}. In these cases, the main production channel is $q\bar q$ annihilation and there is no interference with the 
(coloured) QCD $q\bar q \to t\bar t$ continuum background. In our case,  the signal is due to $gg \to \Phi \to t\bar t$, which interferes in a complicated way with the $gg\to t\bar t$ QCD background as discussed in e.g. Ref.~\cite{hMSSM-fully}. A more detailed analysis is thus needed to interpret more accurately the  ATLAS and CMS exclusion limits but, {\it grosso modo}, they should be of the same order as those derived for spin--1 resonances.}.  

Hence, the value $\tb \approx 1$ seems to be optimal for coping with the LHC data
on the $\Phi$ signal, when all constraints from other search channels are fulfilled.
We therefore use the value $\tb=1$ as a benchmark. This choice has the additional
advantage that, in the 2HDM context, the predictions of the Type-I
and -II variants are quite similar, so that our discussion then becomes more general. 

Nevertheless, even for this optimal value of $\tb$, the partial decay widths of the $\Phi$ states into $\gamma\gamma$ and $gg$ are far too small to explain the large cross section for the diphoton signal, as discussed in~\cite{Phi-ADM} and in the singlet model considered previously. However, in our particular 2HDM/MSSM scenario, the presence of heavy vector--like quarks which couple to the $\Phi$ states is in general strongly disfavoured: vector--like quarks would enhance strongly the cross section for $gg \to \Phi$ production and, since the main decay mode is $\Phi \to t \bar t$, the rate for $gg \to \Phi \to t\bar t$ will exceed by far the limit imposed by the null results of searches for $t\bar t$ resonances~\cite{LHC-ttbar} in the previous LHC Run~1. The
mitigating strategy, as was discussed in the previous subsection,  would be to suppress the top quark contribution to the $\Phi gg$ vertex by choosing intermediate values of $\tb$ in an attempt to allow for such vector-like quarks\footnote{Vector--like quarks could be useful in order to explain another excess,
albeit smaller, observed at the LHC: namely a $2\sigma$ deviation of the cross section for
the associated $pp\to t \bar t h$ production process for the standard Higgs boson \cite{HiggsCombo}. This excess can be explained by the presence of vector--like quarks that enhance  the top Yukawa coupling through mixing with no alteration of the $gg \to h$ cross section and the $h\to \gamma \gamma$  decay branching ratio \cite{ttH-VLQ}. Note, however, that care should be taken with the interactions of the new fermions introduced here not to alter  too much the properties of this standard Higgs boson \cite{Phi-ADM}.}.    In this case the total width of $\Phi$, which will be suppressed because it is controlled by the decay $\Phi \to t \bar t$ as the vector-like quarks are supposed to be much heavier than $\frac12 M_\Phi$ \cite{LHC-VLQ},  would be much smaller than the value $\Gamma_\Phi \approx 45$ GeV favored by the ATLAS measurement \cite{ATLAS-diphoton}. This value can be recovered by allowing for decays into the lighter neutral leptons. On the other hand,
charged vector--like leptons enter only in the $\Phi \gamma \gamma$ couplings, so here we consider their presence only. 

Summarizing this 2HDM/MSSM scenario, we have two (near-)degenerate neutral Higgs bosons, one CP--even, $H$, and one CP--odd, $A$, both with masses $M_\Phi=750$ GeV. We assume that $\tb=1$, so that they couple strongly to the top quark, but not to other standard particles. There is also a charged Higgs boson, $H^\pm$,  with a constraint on its mass $M_{H^\pm} 
\gsim  160$ GeV \cite{LHC-MH+}, though we favour $M_{H^\pm} \sim 750$~GeV.   We postulate a number of charged electroweakly-interacting particles with masses not too far from $\frac12 M_\Phi$, that enhance strongly the $\Phi \gamma \gamma$ coupling, and we also have neutral partners, $N$, of these particles that are invisible, which might appear in the decays of the $\Phi $ states.

\subsection{Production of the $\mathbf{\Phi = H, A}$ states at pp colliders}

We come now to the production of the $\Phi$, interpreted in this Section as a pair $H,A$ of Higgs particles, at $pp$ colliders; some generic Feynman diagrams are shown in Fig.~\ref{diag2}.

\begin{figure}[!h]
\vspace*{-1.2cm}
\centerline{\hspace*{.2cm} 
\includegraphics[scale=0.8]{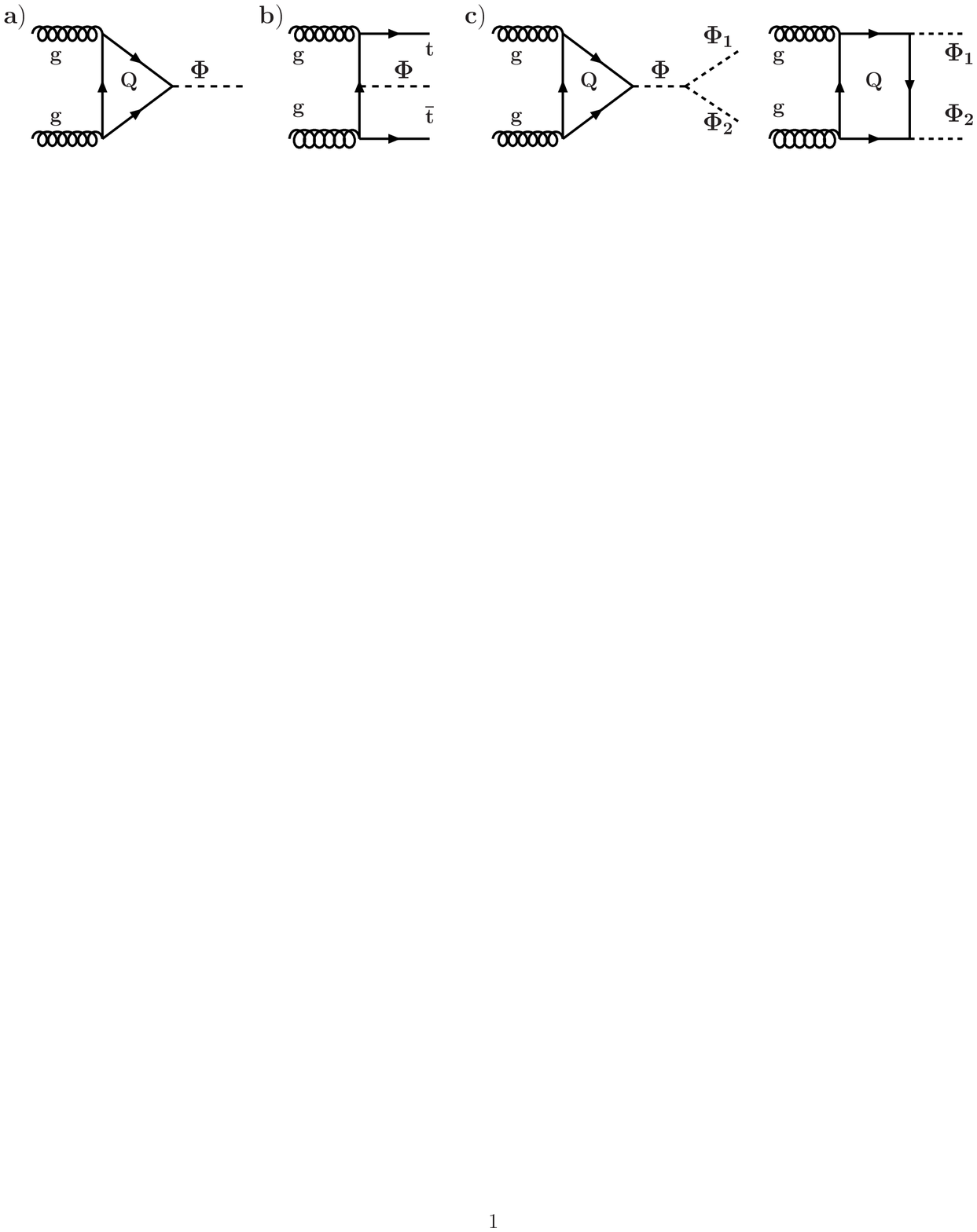} }
\vspace*{-20.2cm}
\caption{\it Generic Feynman diagrams for $\Phi$ production at $pp$ colliders.}
\label{diag2}
\vspace*{-2mm}
\end{figure}

 As already mentioned, the leading production mechanism would be gluon--gluon fusion, $gg\to \Phi$, generated by triangular loops of top quarks, Fig.~\ref{diag2}a, neglecting the possibility of new colored particles such as vector--like quarks. Referring back to eqs.~(\ref{eq:Gammagg}, \ref{eq:ASAP}) and (\ref{eq:formfactors}) for the partial decay width $\Gamma (\Phi \to gg)$ and the form factors $A_{1/2}^{\Phi}$ that govern it, we recall that for the top quark loops with $\tau_t = M_\Phi^2/4m_t^2 \approx 4$ for $M_\Phi=750$ GeV and $m_t=172$ GeV, one has for the form factors $| A^A_{1/2}/A_{1/2}^H|^2  \approx 2$, leading to a characteristic ratio of production cross sections  $\sigma (gg \to A) \approx 2 \sigma(gg\to H)$. 

We have calculated the production cross sections using the program {\tt SUSHI} \cite{SUSHI},
in which important higher-order effects are included, notably the QCD corrections that 
are quite large \cite{ggH-NLO-approx,ggH-NLO,ggH-NNLO,ggH-N3LO} compared to the 
electroweak corrections \cite{LHC-XS}. In principle $\sigma(gg\to \Phi \! +\! X)$  can be evaluated only at next-to-leading-order (NLO) in QCD, since the full top-quark mass corrections are known only at this perturbative order ~\cite{ggH-NLO}: they increase the LO rate by a factor of about 1.7 at LHC energies. The NNLO corrections that are known only in the limit $m_t \gg M_\Phi$ increase the rate by another 30\% \cite{ggH-NNLO} in that case. It has been shown  that at NLO the limit $m_t \gg M_\Phi$, which is not valid in principle, nevertheless provides a good approximation to the exact result provided that the full mass dependence is included in the LO cross sections~\cite{ggH-NLO}. The recently calculated N$^3$LO corrections are small at LHC energies~\cite{ggH-N3LO}, and we can safely neglect them for our purposes. 
 
The production rates for $gg\to H$ and $gg\to A$ at proton colliders are shown in
Fig.~\ref{Fig:main-pp} as a function of the centre-of-mass energy $\sqrt s$ for our 2HDM 
basic inputs $M_\Phi=750$ GeV, $\tan\beta=1$ and the alignment limit, which leads to $\cos(\beta-\alpha)=0$. The MSTW2008 set \cite{MSTW} has been adopted for the parton distribution functions up to NNLO.  At the LHC with $\sqrt s=13$ TeV, the cross sections are of the order of 1 pb, and increase with energy to reach the level of 100 pb at $\sqrt s=100$ TeV. Assuming an accumulated luminosity of  a few ab$^{-1}$, as is expected to be the case at both HL-LHC and FCC-hh/SPPC, one could then collect from $10^6$ to $10^8$ $\Phi$ events at the respective colliders. 
 
\begin{figure}[!h]
\vspace*{-.1cm}
\centerline{\hspace*{-5mm} \includegraphics[scale=0.90]{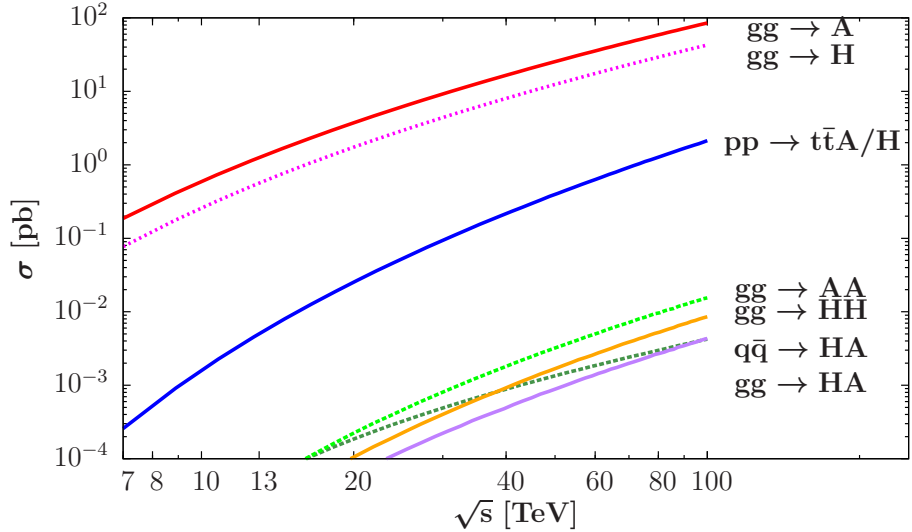}}
\vspace*{-.1cm}
\caption{\it Cross sections for single, associated and pair production of the $\Phi=H,A$ bosons at $pp$ colliders as functions of the c.m. energy from $\sqrt s=7$ TeV to 100 TeV. We assume a common mass $M_{H/A}=750$, $\tan\beta=1$ and
the alignment limit $\cos^2(\beta-\alpha)=0$.}
\label{Fig:main-pp}
\vspace*{-3mm}
\end{figure}

Another important process for the $\Phi$ states is production in association with top quark pairs, $gg \to t\bar t \Phi$ and $ q\bar q \to t\bar t \Phi$, with the first subprocess 
(Fig~\ref{diag2}b) being dominant at high collider energies. In this case,  one is approximately in the chiral limit $M_\Phi \gg m_t$, which leads to almost exactly the 
same cross sections for the CP--even $H$ and CP--odd $A$ states. Because of the reduced phase space,  the production rates are at least two orders of magnitude smaller than in the dominant $gg \to \Phi$ fusion  modes even at $\sqrt s=100$ TeV, as seen in Fig.~\ref{Fig:main-pp}. The rates are nevertheless significant as,  for the planned integrated luminosities,  one can collect more than $10^4$ and $10^6$ events, respectively at HL--LHC and FCC-hh/SPPC, with the $\Phi$ states decaying into $t\bar t$ pairs, resulting in four-top final states.  The processes would give direct measurements of the $t\bar t \Phi$ coupling, a key test of this model and a possible discrimination between the CP--even and odd possibilities \cite{pp-ttH-CP} \footnote{Further, if $\Phi$ indeed has invisible decays into dark matter  particles, then the associated $t \bar t \Phi$ process~\cite{ttH-invis} and  the mono jet channel~\cite{mono jet-invis} are the only ways to search for such a mode at the LHC, as the VBF \cite{VBF-invis} and VH modes~\cite{VH-invis} modes are no longer available in this aligned 2HDM context.}.

We should note that the cross sections for these processes have been obtained using a modified version of the leading-order program {\tt HQQ} \cite{Michael-web} with the renormalisation and factorisation scales fixed to $\mu_0= m_t +\frac12 M_\Phi$, and again using the MSTW2008 set of structure functions. The QCD corrections have been known to NLO  \cite{ttH-NLO} in the case of a CP--even Higgs state for some time  and have been derived more recently for the CP--odd case.  However, at $\sqrt s=14$ TeV,  they lead to a $K$--factor that is of order unity for a  Higgs boson with a mass $\sim 125$~GeV. Considering the process at LO only as is done here should therefore be an adequate  approximation.

There are several processes in which a pair of heavy Higgs bosons is produced,
see, e.g., Ref.~\cite{DoubleH}. First there is $HA$ production that, at moderate energies, proceeds primarily via the $q\bar q\to HA$  process with the $s$--channel exchange of a virtual $Z$ boson that has a maximal coupling to the Higgs pair in the alignment/decoupling limit, $\hat g_{ZHA}=1$. At higher energies, because of the significantly larger gluon luminosity, the dominant production mode becomes $gg \to HA$, which is mediated
almost exclusively by top quark loops. Their contribution comes primarily from box diagrams
in which two Higgs states are emitted from the internal quark lines,
but there are additional ones from the triangular loops that produce an off--shell CP--odd $A$ boson which splits into $HA$ final states: $gg\to A^* \to HA$. The latter process involves the trilinear $HAA$ coupling that is expected to be small in the alignment limit of the 2HDM or the decoupling limit of the MSSM.

One can also produce $HH$ and $AA$ pairs in gluon--fusion processes with contributions from both the box diagrams and triangular loops with intermediate $h,H$ virtual states that then split into to the two $\Phi$ bosons, $gg\to h^*, H^* \to HH$ or $AA$; Fig.~\ref{diag2}.  The cross sections, evaluated at LO using the programs {\tt HPAIR} \cite{Michael-web} and the LO MSTW2008 PDFs, are also shown in Fig.~\ref{Fig:main-pp}c. As can be seen, these cross sections are rather small, barely reaching the 10 fb level even at $\sqrt s=100$ TeV. Their signature, dominated by final states with four top quarks, will be similar to that of the $t\bar t \Phi$ process discussed above. 

We may also consider the possibility of deviating from the alignment or decoupling limit,
with $\hat g_{HVV} = \cos(\beta-\alpha)$ small but non--zero. In this case two interesting
types of process become possible. First, one can produce singly the heavy CP--even $H$ boson in both vector boson fusion $qq\to qq H$ and the Higgs--strahlung process $q\bar q \to VH$, where $V=W,Z$. Especially in the former case, the cross sections would be very large at high energy  if it were not for the $g_{HVV}$ coupling suppression. Associated production of  the light standard--like $h$  state and the pseudoscalar $A$ boson in $q\bar q \to Z^* \to hA$ 
would also be possible for $\hat g_{ZhA} = \cos(\beta-\alpha) \neq 0$. In fact, $h$ production in association with the heavier $A$ and $H$ states is always possible in $gg$ fusion, $gg\to hA$ and $gg\to hH$ through box diagrams involving top quark loops, even in the alignment/decoupling limits. 

We have calculated the cross sections for $qq \to Hqq, q\bar q \to HV$ and $hA$, evaluated at LO only (the QCD corrections in these cases are at the level of 10\% to 30\%  
and can be readily included see e.g.,~Refs.~\cite{LHC-XS,VVH-NLO}). The cross sections 
for $gg\to hH,hA$ are evaluated at NLO, with $K$--factors of the same order as those appearing in single Higgs production in $gg$ fusion, using the programs of Ref.~\cite{Michael-web}. These cross sections are displayed in the left panel of Fig.~\ref{Fig:subl-pp} as functions of the centre-of-mass energy  for $\cos^2(\beta-\alpha)=10^{-2}$. The cross sections for $qq \to Hqq$ and $gg \to hH,hA$ are of the same magnitude and, at $\sqrt s= 100$ TeV, they are quite substantial as they reach the level of 100 fb.  At LHC energies, however, they are two orders of magnitude smaller. The production rates for the $HW$ and $HZ$ Higgs--strahlung processes are very small already for $\cos^2(\beta-\alpha)=1$ and are hence negligible for the more realistic value $\cos^2(\beta-\alpha)=10^{-2}$. 

\begin{figure}[!h]
\vspace*{-9mm}
\centerline{
\includegraphics[width=7.3cm,height=6.6cm]{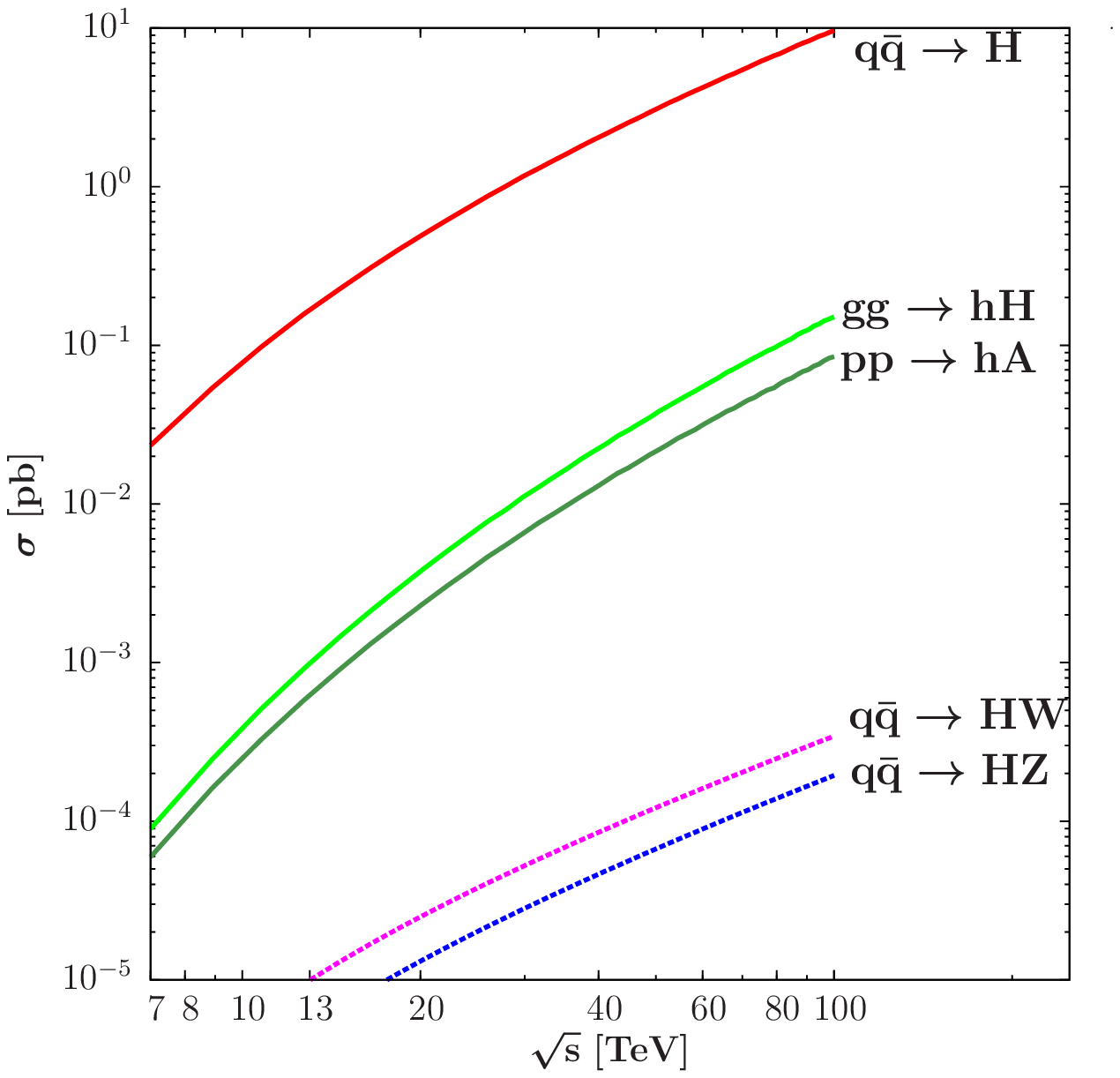}~~~~~~
\includegraphics[width=7.3cm,height=8cm]{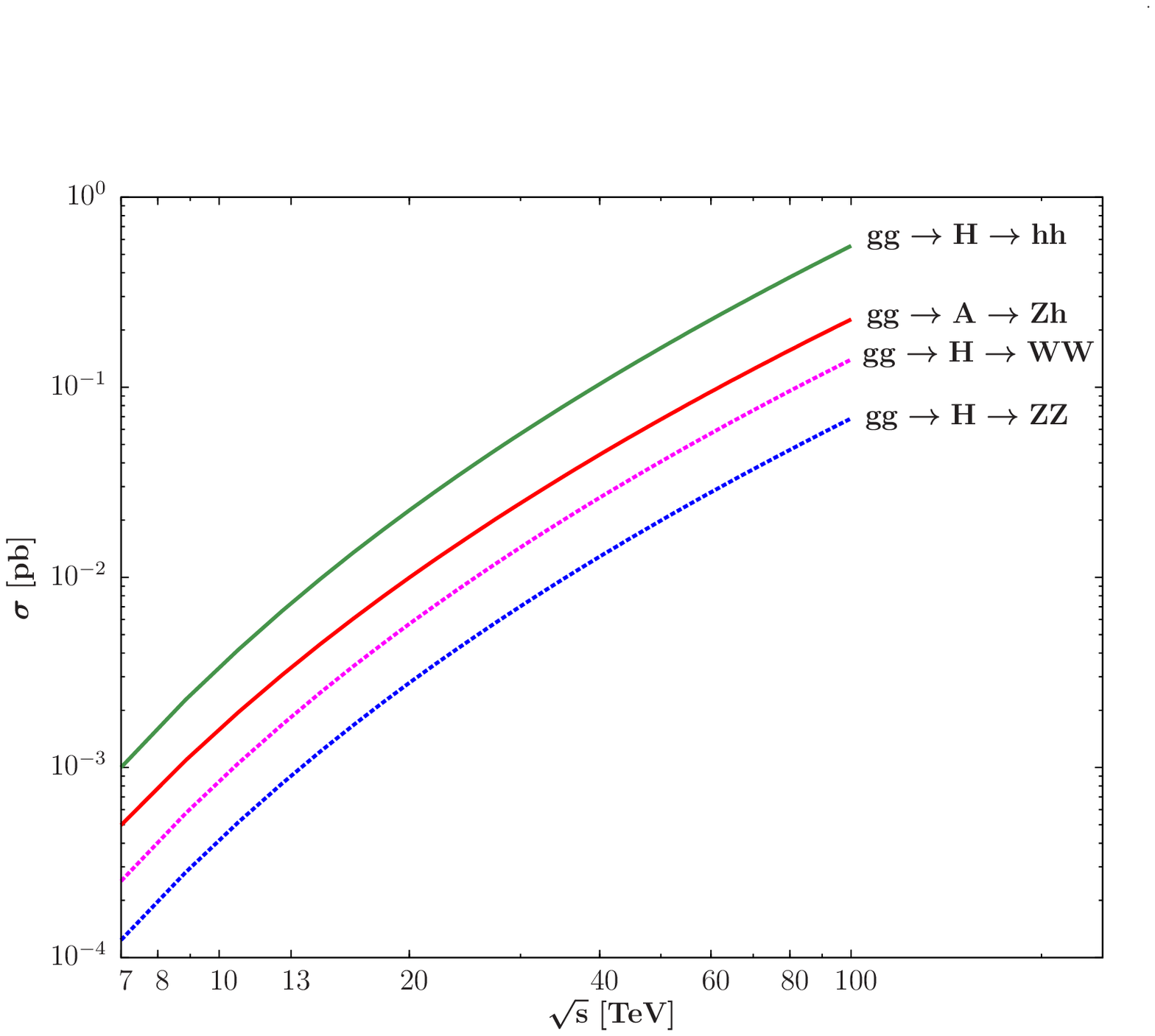} }
\caption{\it Left panel: Subdominant processes for the single production of the CP--even $H$ boson in the $q\bar q \to HW, HZ$ and $qq \to qqH$ channels, and for the pair production of two Higgs bosons in the channels $q\bar q + gg \to  hA, gg \to Hh$, assuming $\cos^2 (\beta-\alpha)=10^{-2}$.
Right panel: The rates for $\Phi$ production in $gg$ fusion followed by decays into the $Zh$ and $WW,ZZ,hh$ final states. The production rates are again functions of $\sqrt s$ from 7 to 100 TeV with $M_\Phi=750$ GeV, $M_h=125$ GeV and $\tan\beta =1$, assuming the couplings found in the hMSSM (2HDM of Type II).}
\label{Fig:subl-pp}
\vspace*{-1mm}
\end{figure}

Finally,  also outside the alignment or decoupling limit, the $H/A$ states that are produced in $gg$ fusion with extremely large rates can have other decay modes than the dominant $\Phi \to t\bar t$. Indeed, since the couplings $\hat g_{HVV}= \hat g_{ZhA}= \cos^2(\beta-\alpha)$ are non--zero in this case, and  would enable additional decays to occur. In the case of the CP--even $H$ boson, important possibilities would be $H\to ZZ$ and $H\to WW$, which would grow like $M_H^3$, to be contrasted with $M_H$ for the $H\to t\bar t$ decay, as a consequence of the growth of the longitudinal gauge boson wave functions with energy, and would dominate at high mass if not for the $\hat g_{HVV}$ coupling suppression.  Another interesting decay for a pseudoscalar Higgs boson,   $A\to hZ$, would also have a non--zero branching ratio if $\hat g_{ZhA}$ were not small. This would also be the case of the other very interesting decay of the CP--even $H$ state, $H\to hh$, which involves the triple coupling $g_{Hhh}$ that is somewhat model-dependent~\footnote{This coupling may also appear in the MSSM, where the coupling $g_{Hhh}$ depends on the supersymmetric spectrum that enters the radiative corrections in the Higgs sector \cite{Review2}. However, if only the dominant radiative correction that also enters the lightest $h$ boson mass~\cite{RC-1loop} is taken into account, the coupling is fixed in terms of $\tb$ and $M_A$ for $M_h=125$ GeV \cite{hMSSM-fully}. This coupling is also small in the decoupling limit that applies for $M_\Phi=750$ GeV.} in a 2HDM, but is in general small in the alignment limit. 

The cross sections for the processes $gg\to H \to WW,ZZ,hh$ and $gg\to A \to hZ$ are shown in the right panel of Fig.~\ref{Fig:subl-pp} as functions of $\sqrt s$, assuming the couplings found in the hMSSM (2HDM of Type II).  The decay rates have been evaluated using the program {\tt HDECAY}  \cite{hdecay}, which includes the relevant higher-order effects.  As can be seen, even for the very small value $\cos^2(\beta-\alpha) \approx 8 \times 10^{-4}$ used
for illustration (and which leads to very tiny branching ratios of the order of  a few times 
$10^{-3}$ for the $H\to ZZ,WW$ and $A \to hZ$ modes and of the order of $10^{-2}$ for the $H\to hh$ decay),  the rates are small but not negligible at the high luminosities planned for high energies. 

\subsection{$\mathbf{\Phi = H, A}$ production at $\mathbf{e^+ e^-}$, $\gamma\gamma$ and $\mathbf{\mu^+ \mu^-}$ colliders}

We study now the production of the $\Phi = H, A$ states at high--energy $e^+ e^-$ colliders. For at least two decades, these machines have been discussed as possible follow-ups for the LHC. Two options for high--energy $e^+ e^-$ linear colliders~\footnote{We do  not discuss here the  FCC-ee~\cite{Fcc-ee} and CEPC~\cite{CEPC-ee} circular machines, which are planned for a maximum centre-of-mass energy of $\sqrt s= 350$ GeV, far below the kinematical threshold for probing directly the 750 GeV $\Phi$  states and the associated matter particles. However, these machines would be able to probe these scenarios indirectly via precise measurements of the standard--like $h$ boson couplings, which would be affected by the presence of the new particles.  We note, however, that a very precise  determination of the $h\gamma\gamma$ coupling could be performed already at the HL--LHC by measuring the ratio of $h\to \gamma\gamma$ and $h \to ZZ^* \to 4\ell^\pm$  signal strengths, as a precision of 
${\cal O}(1\%)$  can be achieved~\cite{golden}.} have been discussed. One is the ILC~\cite{ee-ILC}, which could ultimately reach energies of the order of 1 TeV, as required for $\Phi$ production. The other is CLIC~\cite{ee-CLIC}, which is planned to  cover the multi--TeV scale and can certainly reach a centre-of-mass energy $\gsim 1.5$~TeV that is favoured in the context of this analysis. Both colliders are designed to reach an integrated luminosity of 1 ab$^{-1}$. 


\begin{figure}[!h]
\vspace*{-1.cm}
\centerline{\hspace*{-.2cm} 
\includegraphics[scale=0.8]{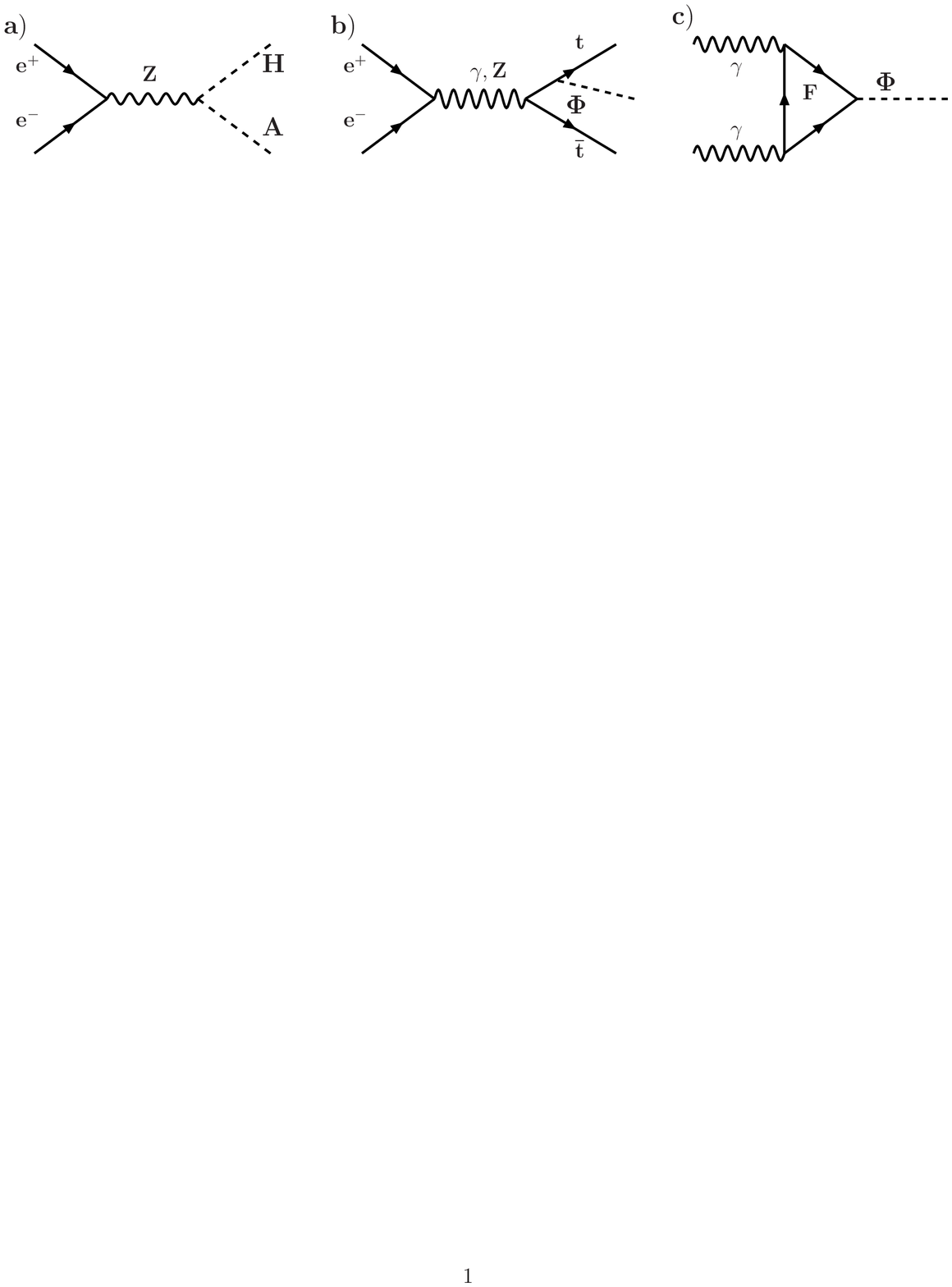} }
\vspace*{-20cm}
\caption{\it Feynman diagrams for $\Phi$ production at $e^+e^-$ colliders.}
\label{diag3}
\vspace*{-2mm}
\end{figure}

In the exact alignment or decoupling limit, the most important channel for $\Phi$
production in  $e^+e^-$ collisions \cite{synopsis} is associated production of the $HA$ states through the  $s$-channel exchange of a virtual  $Z$ boson, $e^+ e^- \to HA$
in Fig.~\ref{diag3}a, as the coupling is then maximal, $\hat g_{ZHA}=\sin(\beta-\alpha) \to 1$. The cross section is simply given by
\beq 
\sigma(\ee \to HA) = \frac{G_\mu^2 M_Z^4}{96 \pi s} (\hat v_e^2+\hat a_e^2) \hat g_{ZHA}^2 
\frac{ \lambda^3}{(1-M_Z^2/s)^2} \, ,
\eeq
where, as usual, the reduced vector and axial--vector couplings of the electrons to the $Z$ boson are given by the $Z$ charges of the electron, $\hat a_e=-1$ and $\hat v_e=-1+4s_W^2$, and $\lambda$ is the usual two--particle phase--space function that, in the case of two equal
particle masses, reduces to the velocity of the $H/A$ bosons: $ \lambda \to \beta= \sqrt {1-4M_ \Phi^2/s}$.

The production rate is shown in the left panel of Fig.~\ref{Fig:ee-phi} for $M_H=M_A=750$ GeV
as a function of the centre-of-mass energy $\sqrt s$. As it scales like $1/s$, the cross section is not that large, namely ${\cal O}(1)$~ fb above the $2M_\Phi$ threshold, leading to a thousand events that can be fully reconstructed for the anticipated luminosity of 1 ab$^{-1}$. The main detection channel would be the four-top final state but, the enhanced $t\bar tgg$ and possibly $t\bar t\gamma \gamma$ final states could also be observed with very high luminosities. 

The other important $\Phi$ production processes are associated production with top quark pairs \cite{ee-ttH}, $e^+ e^- \to t \bar t \Phi$ and Fig.~\ref{diag3}b, for which the combined cross sections are at the level of 0.1 fb at high enough energy, i.e., sufficiently far above the kinematical threshold $\sqrt s \approx 1.1$ TeV. The cross sections are shown  for $H$ and $A$ as functions of $\sqrt s$ in the left panel of Fig.~\ref{Fig:ee-phi}, again for $M_A=M_H= 750$ GeV and $\tb=1$. The signature would be four top quarks in the final state, which should have little background, except from the process $e^+ e^- \to HA$ before the two $H,A$ states are reconstructed. We note that the cross sections for $H$ and $A$ production are slightly different because, at energies below $\sqrt s \approx 3$ TeV, one is not yet in the chiral limit in which top quark mass effects are negligible. In fact, the threshold rise of the cross sections is completely different for $e^+ e^- \to t \bar t A$ and $e^+ e^- \to t\bar t H$, as has been shown in~Ref.~\cite{Bhupal}, and a scan around the $2m_t+ M_\Phi$ threshold could allow for a  distinction between the CP--even and CP--odd Higgs cases.

\begin{figure}[!h]
\vspace*{-2.6cm}
\centerline{\hspace*{-1cm} \includegraphics[scale=0.8]{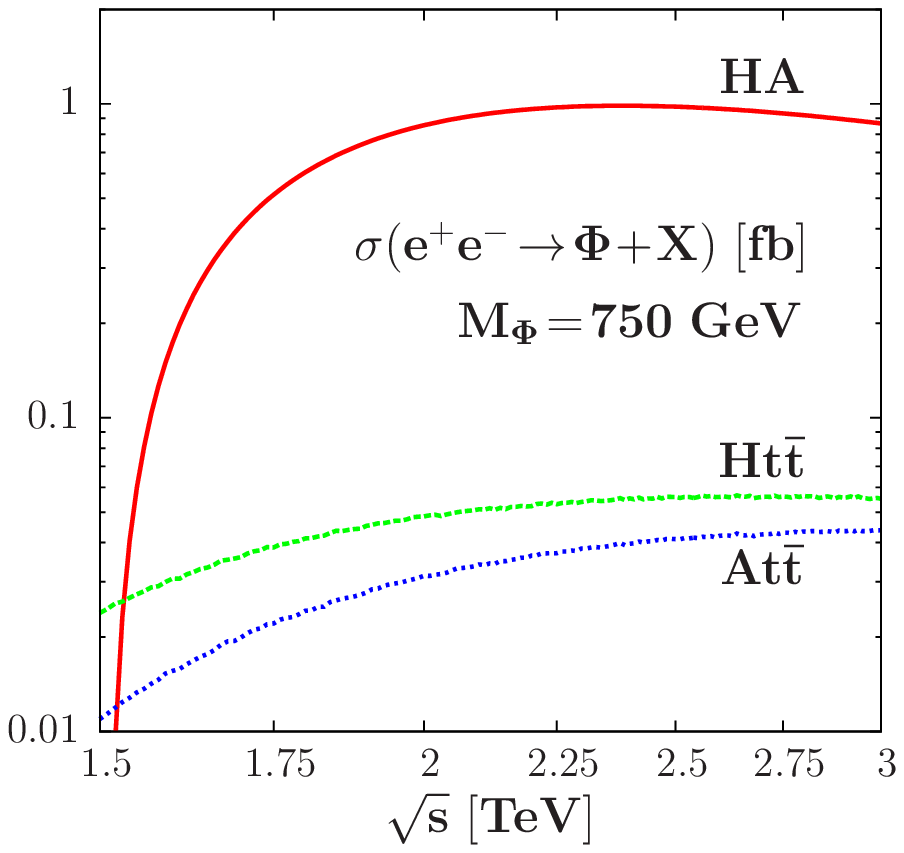}\hspace*{-9mm}
            \hspace*{-8cm} \includegraphics[scale=0.8]{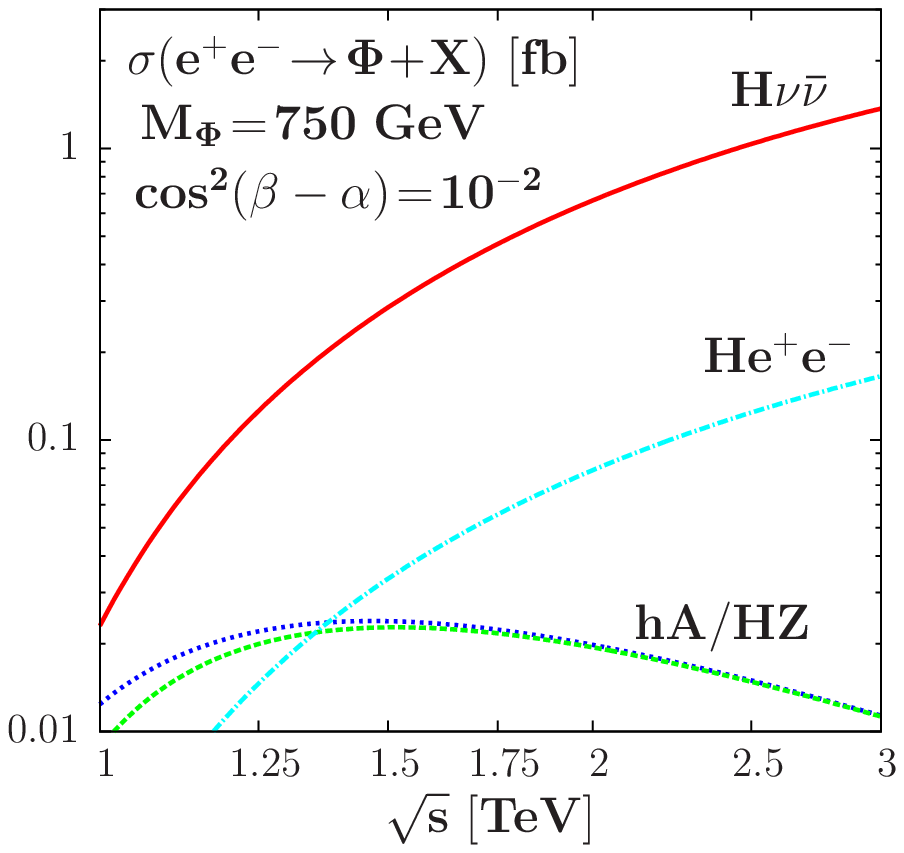}
}
\vspace*{-14.2cm}
\caption{\it
Production cross sections for the $\Phi=H,A$ states at an $e^+ e^-$ collider as a function of the centre-of-mass energy $\sqrt s$,  assuming $M_\Phi= 750 $ GeV and  $\tan\beta=1$. The processes considered are:  $e^+ e^- \to HA$ and  $e^+ e^- \to t\bar t \Phi$ with $\Phi=A$ or $H$ in the alignment limit $\cos^2(\beta-\alpha)=0$ (left panel) and  $e^+ e^- \to H \nu \bar \nu , H e^+ e^-$ and $\ee \to hA,HZ$ for $\cos^2(\beta-\alpha)=10^{-2}$ (right panel).} 
\label{Fig:ee-phi}
\vspace*{-3mm}
\end{figure}

Finally, there are also processes for the single production of the CP--even $H$ boson, 
such as vector boson fusion, $e^+ e^- \to  H \nu \bar \nu$ and $e^+ e^- \to  H e^+ e^-$, 
and Higgs--strahlung, $e^+ e^- \to HZ$, as well as the  associated production of the 
pseudoscalar $A$ and the standard--like $h$ state, $e^+ e^- \to hA$. The rates are, however, suppressed by the small couplings $\hat g_{VVH}=\hat g_{ZhA}= \cos(\beta-\alpha) \to 0$ in the alignment/decoupling limits of the 2HDM and MSSM scenarios, respectively. The cross sections for these processes are shown in the right panel of Fig.~\ref{Fig:ee-phi} as functions of $\sqrt s$ for $\cos(\beta-\alpha) = 0.1$.  

The rates for the $s$--channel processes $e^+ e^- \to Z^* \to hA/HZ$,  which have comparable cross sections at  centre-of-mass energies sufficiently above the kinematical thresholds, as $M_A=M_H$ and $M_h \approx M_Z$, scale like $1/s$ and are therefore small. In turn, as it grows like log$(s/M_W^2)$, the rate for the $WW$ fusion process $e^+ e^- \to  H \nu \bar \nu$ is very large and, at high enough energies, it could dominate all other $\Phi$ production processes, despite the $\cos^2 (\beta - \alpha)$ suppression factor.  The $ZZ$ fusion process $e^+ e^- \to  H e^+ e^-$, which has an order of magnitude smaller rate compared to $WW$ fusion as can be inferred from the $W/Z$ couplings to electrons, could also be observable  for not too tiny $\hat g_{HVV}$ couplings.  

In addition to the conventional $e^+e^-$ mode, future high--energy $e^+e^-$ linear colliders
can be made to run in the $\gamma \gamma$ mode  by using Compton back-scattering of laser light off the high--energy electron beams \cite{ginzburg}.  As discussed earlier, these colliders could have up to $\sim 80$\% of the energy of the $e^+e^-$  collider, with a luminosity that is quite similar.  In the context of the $\Phi = H, A$ states,  the motivation for such a $\gamma\gamma$ machine would again clearly be the direct and precise measurement of the $\Phi \gamma\gamma$ coupling, since the $s$--channel production of resonances, Fig.~\ref{diag3}c, is possible in such a mode \cite{gamma-gamma}.

  As discussed previously, the production of a spin--zero particle at such a collider occurs through the $J_Z=0$ channel.  In the 2HDM studied in this Section, in contrast to the singlet model studied earlier, the total decay widths of the 750 GeV $\Phi$ states are significant, namely about $\Gamma_\Phi \approx 30$~GeV. For polarized initial-state photons, taking into account the total decay width $\Gamma_\Phi$ and  the partial widths for decays into two photons $\Gamma(\Phi  \to \gamma \gamma)$  and into a given final state $X$, $\Gamma( \Phi \to X)$, the cross section for the  process $\gamma \gamma \to \Phi \to X$ is given by eq.~(\ref{gmgmcsec}). As explained earlier, with appropriate choices of the helicities of the $e^{-},e^{+}$ as well as the two laser beams, one can arrange that the two back-scattered photons dominantly have identical helicities: $\lambda_1 \lambda_2=1$, so as to project out the $J_Z=0$ component and therefore favour the resonant Higgs signal. In order to maximize the effective cross section for Higgs production, the $\gamma \gamma$ energy should be tuned so that the peak of the luminosity function at $\sim 0.8 \sqrt{s}_{\ee}$ (for a perfect photon spectrum) occurs at $M_\Phi$. As in our present case the $\Phi$ particles decay almost exclusively into $t\bar t$ final states with BR$(\Phi \to t\bar t) \approx 1$,  the main background is the $\gamma\gamma \to t\bar t$ process whose cross section in the $J_Z\!=\!0$ mode is significant as there is no mass suppression like for light fermions.  
One can impose a polar cut in the centre of mass of the two--photon system to eliminate part of the background events at high invariant masses, which are peaked in the forward and backward directions, with only a moderate loss of the signal.

Fig.~\ref{Fig:gamma} displays the prospects for measuring the doublet $\Phi$ signal in $\gamma \gamma$ collisions, taking into account the interference between signal and background. The cross section $\hat \sigma$ is for the process $\gamma \gamma \rightarrow t \bar t$,  taking into account both the QED process and the resonance production $\gamma \gamma \rightarrow \Phi \rightarrow t \bar t$ and including  the interference. We again make the analysis choosing $\lambda_{e^{-}} \lambda_{l_{1}} = \lambda_{e^{+}} \lambda_{l_{2}} = -1$ and $\lambda_{e^{-}}= \lambda_{e^{+}}$. These choices ensure that the photon luminosity $L_{\gamma \gamma}$ peaks at around $80 \%$ of the energy of the parent $\ee$ collider and that the two colliding photons dominantly have the same helicities.
This analysis was made assuming $M_A=750$ GeV, $M_H=770$ GeV, $\Gamma_A=35$ GeV,  $\Gamma_H=32$ GeV and $\tan\beta=1$ for a 1 TeV parent $\ee$ collider. In the left panel, we show for illustration the case where only the top loop contributions to the $\Phi \gamma\gamma$ couplings are taken into account, whereas in the right panel additional contributions that enhance the previous $H \to \gamma\gamma$  and $A \to \gamma\gamma$ amplitudes by factors 10 and 15, respectively, are also included. Like Fig.~\ref{Fig:Aboost} these figures, too,  display how two closely spaced states would look like a single wide resonance.

\begin{figure}[!h]
\vspace*{-2.6cm}
\centerline{\hspace*{-1cm} \includegraphics[scale=0.8]{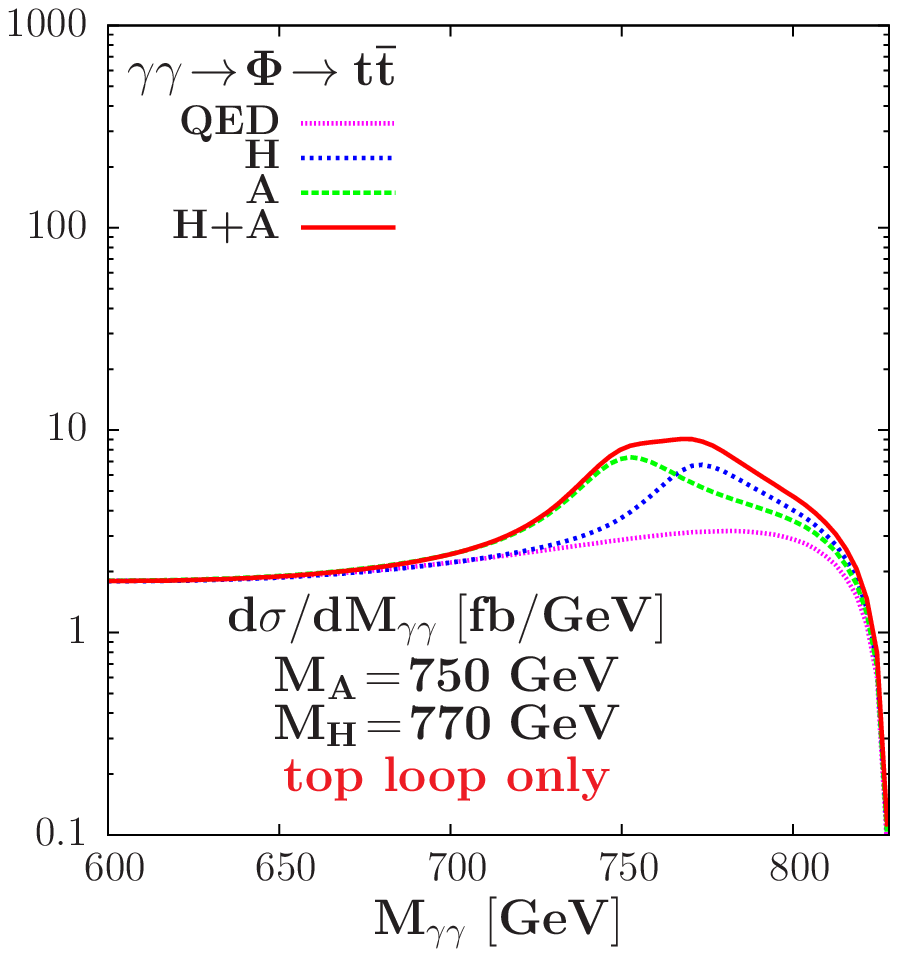}\hspace*{-9mm}
            \hspace*{-8cm} \includegraphics[scale=0.8]{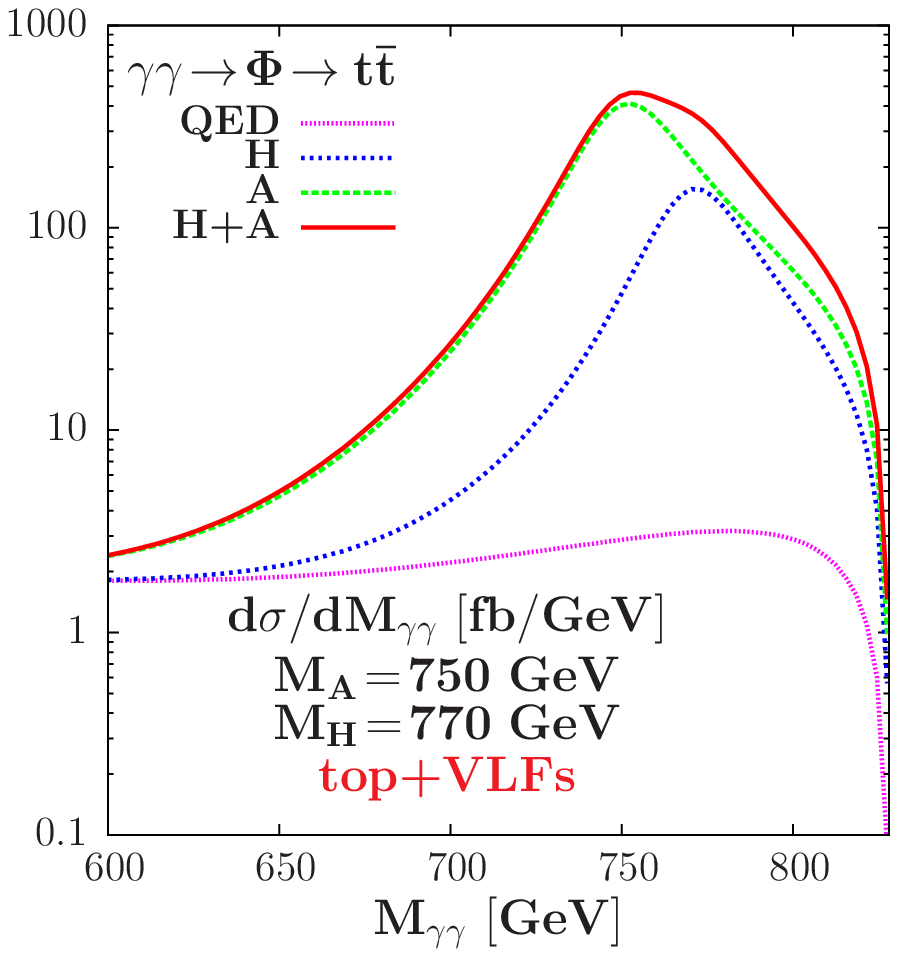}
}
\vspace*{-13.6cm}
\caption{\it The invariant mass distribution ${\rm d\sigma/d}M_{\gamma \gamma}$ in
fb/GeV for the process $\gamma\gamma \to t\bar t$ in the $\gamma\gamma$ mode of a
linear $e^+e^-$ collider. Shown are the pure continuum QED contribution, the additional
separate contributions due to $s$--channel exchanges of the $H$ and $A$ states, and 
the full set of contributions QED$+H+A$. We assume $M_A=750$ GeV, $M_H=770$ GeV, $\Gamma_A=35$ GeV,  $\Gamma_H=32$ GeV and $\tan\beta=1$. In the left panel, only
the top quark loops are taken into account in the $\Phi \gamma\gamma$ couplings.
In the  right panel, additional vector-like fermion contributions that increase $H \to \gamma\gamma$ and $A \to \gamma\gamma$ amplitudes by factors 10 and 15, respectively, are included.}
\label{Fig:gamma}
\vspace*{-2mm}
\end{figure}

For completeness, we close this section with a few words about high--energy muon colliders. 
In $\mu^+ \mu^-$ collisions, the production cross section for a Higgs state 
decaying into a final state $X$ is given, in terms of the partial decay widths, by
\cite{mu-mu} 
\beq
\hspace{-0.3cm}
\sigma (\mu^+ \mu^- \! \to \! \Phi \! \to \! X) \!=\! \frac{ 4\pi \Gamma(\Phi \! \to \! \mu^+ \mu^-) \Gamma (\Phi \! \to \! X)} {(s-M_\Phi^2)^2 + M_\Phi^2 \Gamma_\Phi^2} \! \simeq \! \frac{4\pi}{M_\Phi^2} {\rm BR}(\Phi \! \to \! \mu^+\mu^-) {\rm BR}(\Phi \! \to \! X),
\eeq
where the second term, which gives the effective cross section, is obtained assuming that the 
energy spread of the $\mu^+ \mu^-$ machine is much smaller than the Higgs total decay width. 
In our case, the relevant final state to be considered is $\Phi \to t \bar t$,
which has a branching ration of order 1. As the $\Phi$ mass that we consider here, 
$M_\Phi = M_{H, A} \approx 750$ GeV, is large and the branching fraction
BR$(\Phi \to \mu^+\mu^-)$ low, BR$(\Phi \to \mu^+\mu^-) \approx 1.5 \times 10^{-7}$ 
for $\tan\beta \approx 1$, the production rate would, in this case, be extremely small.  
We will therefore not pursue further this  $\mu^+ \mu^-$ option. 

\subsection{Production of charged Higgs bosons}

We now turn to the discussion of the charged Higgs bosons, which is complicated by
the fact that we do not know their mass.  Indeed, the only available information on $M_{H^\pm}$ is that it should be heavier than about 160 GeV, as a result of the negative searches in top decays $t \to b H^+ \to b \tau \nu$ at LHC Run~1 with $\sqrt s=8$ TeV and about 20 fb$^{-1}$ data \cite{LHC-MH+}.  Therefore, we consider the production cross sections at $pp$ and $\ee$ colliders as functions of  $M_{H^\pm}$ for different centre-of-mass energies. For the other model parameters, we continue with $\tan\beta =1$, $\cos^2 (\beta-\alpha)=0$ and $M_\Phi= M_{H, A} = 750$ GeV. 

We first discuss $H^\pm$ production at $pp$ colliders, for which our results are shown in Fig.~\ref{Fig:H+-pp} for the two centre-of-mass energies $\sqrt s=14$ TeV and  $\sqrt s=100$ TeV, borrowing some results from the recent analysis of Ref.~\cite{Higgs-100TeV}. The dominant process by far is the associated $gb \to tH^\pm$ mechanism, which for $\sqrt s=14$ TeV and $\tan\beta \approx 1$  has a cross section above the pb level for $M_{H^\pm} \lsim 400$ GeV, dropping to 30 fb for a mass  $M_{H^\pm} \approx 1$ TeV.  For low $H^\pm$ masses,  it is followed by the $q\bar q \to \gamma^*, Z^* \to H^+ H^-$ process, which has a rate that is two orders of magnitude lower. A third possibility would be associated  $HH^\pm$ and $AH^\pm$ production via $W$ exchange, $q\bar q' \to W^* \to \Phi W^\pm$,  with a rate comparable to that of pair production for $M_{H^\pm} \approx 400$ GeV  and larger beyond. The cross sections are nevertheless small, below the fb level. Moving to $\sqrt s= 100$ TeV, all cross sections increase by two to three orders of magnitude.  Hence, production of the $H^\pm$ states is copious at these  high energies.

\begin{figure}[!h]
\vspace*{.4cm}
\centerline{\hspace*{-3mm}
\includegraphics[width=7.5cm,height=7.5cm]{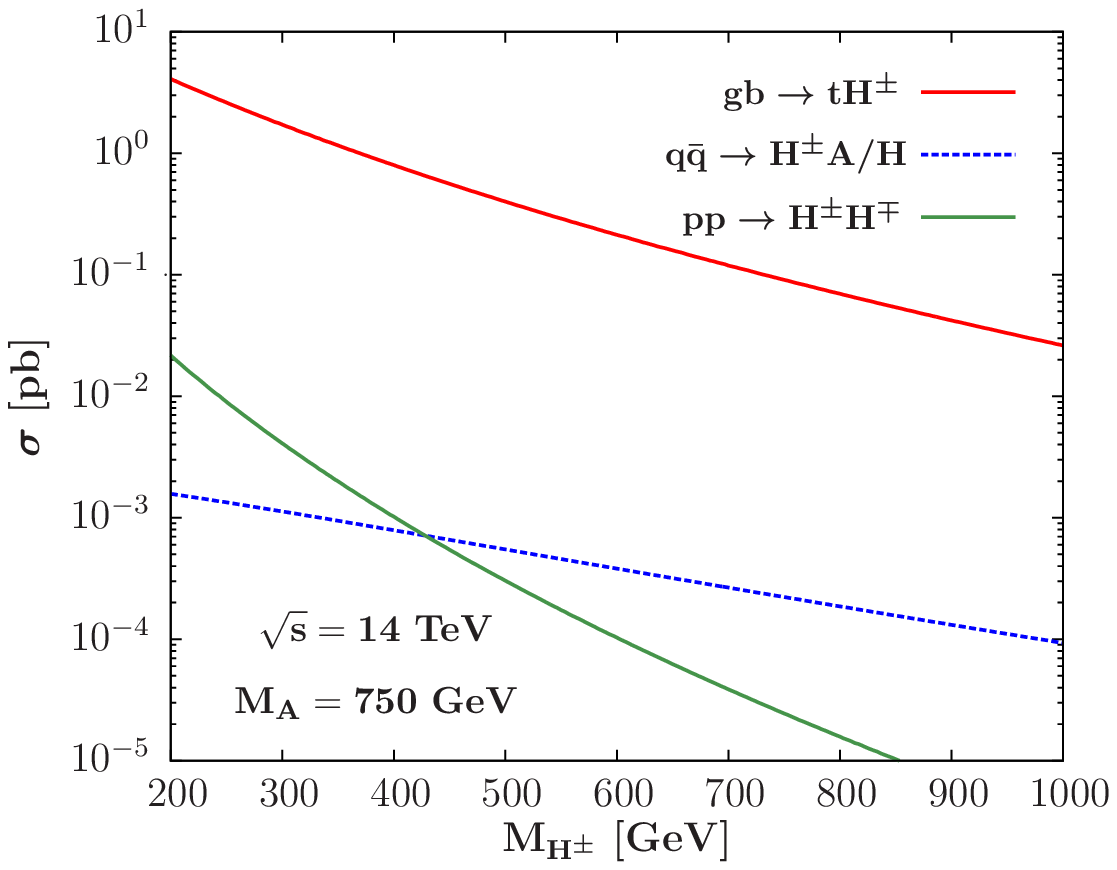}\hspace*{-2mm}
\includegraphics[width=7.5cm,height=7.5cm]{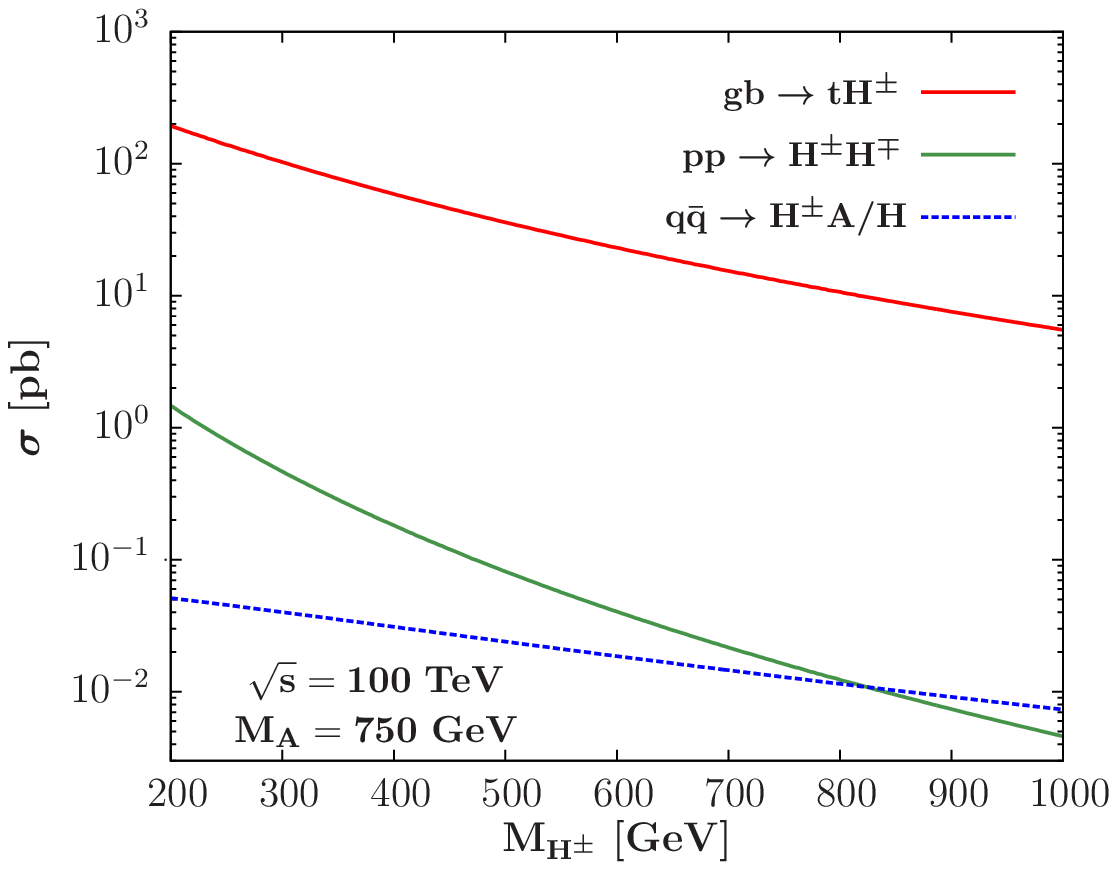}}
\caption{\it Cross sections for production of the charged Higgs boson as functions of its mass for $\tan\beta =1$, $\cos^2 (\beta-\alpha)=0$ and assuming $M_{\Phi}=750$ GeV. We show results for $pp$ colliders at two energies $\sqrt s=14$ TeV (left panel) and 
$\sqrt s=100$ TeV (right panel).}
\label{Fig:H+-pp}
\end{figure}

Turning to  high--energy  $e^+ e^-$ colliders, the most important process for producing charged Higgs states is the pair production process \cite{synopsis}, which proceeds via $\gamma^*, Z^*$ exchange, and for which the cross section depends only on the centre-of-mass energy and $M_{H^\pm}$. It is shown as a function of $M_{H^\pm}$ in Fig.~\ref{Fig:ee-H+} for a fixed centre-of-mass energy of $\sqrt s= 2$ TeV. As can be seen, it drops from about 10 fb at low masses $M_{H ^\pm} \approx 160$ GeV to about 2 fb at $M_{H ^\pm} \approx 750$ GeV. 
Thus, the cross section is approximately a factor of two larger than for $HA$ production,
as a result of the additional photon exchange. For masses above 750~GeV, the cross section drops quickly as a consequence of the velocity  suppression near the kinematical threshold, 
$\sigma (\ee \to H^+ H^-) \propto  \beta_{H^\pm}^3$. Another process in the context of $\ee$ colliders would be  associated production with heavy quarks, $e^+ e^-  \to tbH^\pm$. 
Similarly to associated $\Phi$ production with top quarks, the cross sections may
reach the 0.1 fb level for $\tb=1$ and $H^\pm$ states that are not too heavy \cite{ee-ttH}. 

Another possibility for producing pairs of charged Higgs bosons is in
$\gamma\gamma$ collisions, $\gamma \gamma \to H^+ H^-$.
Again, the cross section depends only on  $M_{H ^\pm}$ at a given energy. 
It is also illustrated in Fig.~\ref{Fig:ee-H+} where, for simplicity, we have fixed the
centre-of-mass energy to $\sqrt s_{\gamma\gamma}= 80\% \sqrt s_{ e^+ e^-} =1.6$ TeV,
and did not fold the cross section of the subprocess with realistic photon spectra. 
As it is mediated by $t$--channel exchange, the cross section has a completely different 
behaviour compared to the $e^+ e^- \to H^+ H^-$ case. It is much larger at low $H^\pm$ masses,
and drops sharply  close to the kinematical threshold. 
 
\begin{figure}[!h]
\vspace*{-2.3cm}
\centerline{\hspace*{-5mm} \includegraphics[scale=0.8]{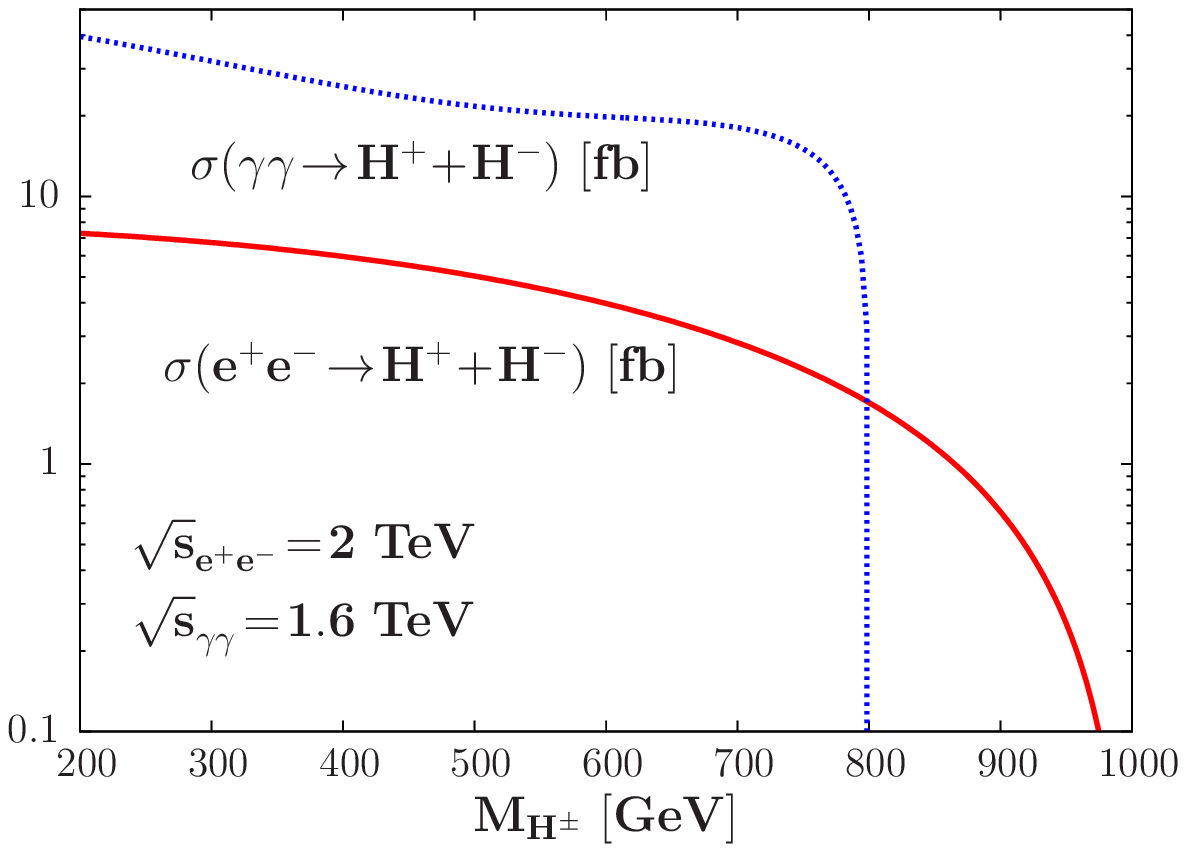}}
\vspace*{-14cm}
\caption{\it Production cross sections for pairs of $H^\pm$ bosons as functions of $M_{H ^\pm}$ at $e^+ e^-$ and $\gamma\gamma$ colliders with a c.m. energy of $\sqrt s=2 (1.6$ TeV) in the $e^+ e^-$ ($\gamma\gamma)$ mode.} 
\label{Fig:ee-H+}
\vspace*{-2mm}
\end{figure}

For $M_{H^\pm} \gsim 180$ GeV, the dominant $H^\pm$ decay mode is by far $H^+ \to t \bar b$. 
In the alignment limit and for small $\tb$ values, all other decays such as $H^\pm \to W^\pm h$ and $H^\pm \to\tau \nu$ are suppressed. There is, however, one other
possibility~\cite{H+3body}: if $M_{H^\pm} \gsim M_\Phi +M_W$, the decays $H^\pm  \to W^\pm \Phi$ take place as the coupling $g_{H^\pm W^\mp \Phi}$ is ${\cal O}(1)$ in the alignment limit of a 2HDM~(in the MSSM, this channel is kinematically closed at the two--body level since we have $M_{H^\pm} \approx M_{\Phi}$ and it is strongly suppressed at the three-body level~\cite{H+3body}). Nevertheless,  BR($H^\pm \to \Phi W)$ is small and does not exceed 50\% even for $M_{H^\pm}=1$ TeV. Hence the main topology in the search for the $H^\pm$ states at proton colliders would be $gb \to tH^- \to tt b$, which would be similar to $t\bar t$ plus jet production, rendering its detection not very easy. In  $\ee$ collisions, one should focus on the $t\bar t b\bar b$ final-state topology. 

Another possibility that should be considered is the decay of the $\Phi = H, A$ states into $H^\pm$  bosons.  Indeed, if we are in the opposite situation to that considered above, i.e., with  $M_{\Phi} \gsim M_{H^\pm} +M_W$, the decays $\Phi \to H^\pm W^\mp$ will take place~(although in an MSSM-like scenario with $M_{H^\pm} \approx M_{\Phi}$, these features do not occur). For our baseline choice $\tb=1$, the two dominant branching ratios are shown in Fig.~\ref{Fig:BRs}, where one can see that, when the decays $\Phi \to H^\pm W^\pm$ are accessible, 
they tend to dominate over the $t\bar t$ channels. At the same time, the total decay widths of the  $\Phi$ states would become much larger than the value $\Gamma_\Phi \approx 30$ GeV that seem to be favoured by the ATLAS search. Beyond $M_{H^\pm} \approx 650$ GeV, these special decays are closed, and one has BR($\Phi \to t\bar t) \approx 1$. These decays are hence disfavoured if we want to stay in a minimal scenario for the observed production rate and the total width  of the diphoton enhancement, with only a few  ingredients. 

\begin{figure}[!h]
\vspace*{-2.2cm}
\centerline{\hspace*{-1cm} \includegraphics[scale=0.8]{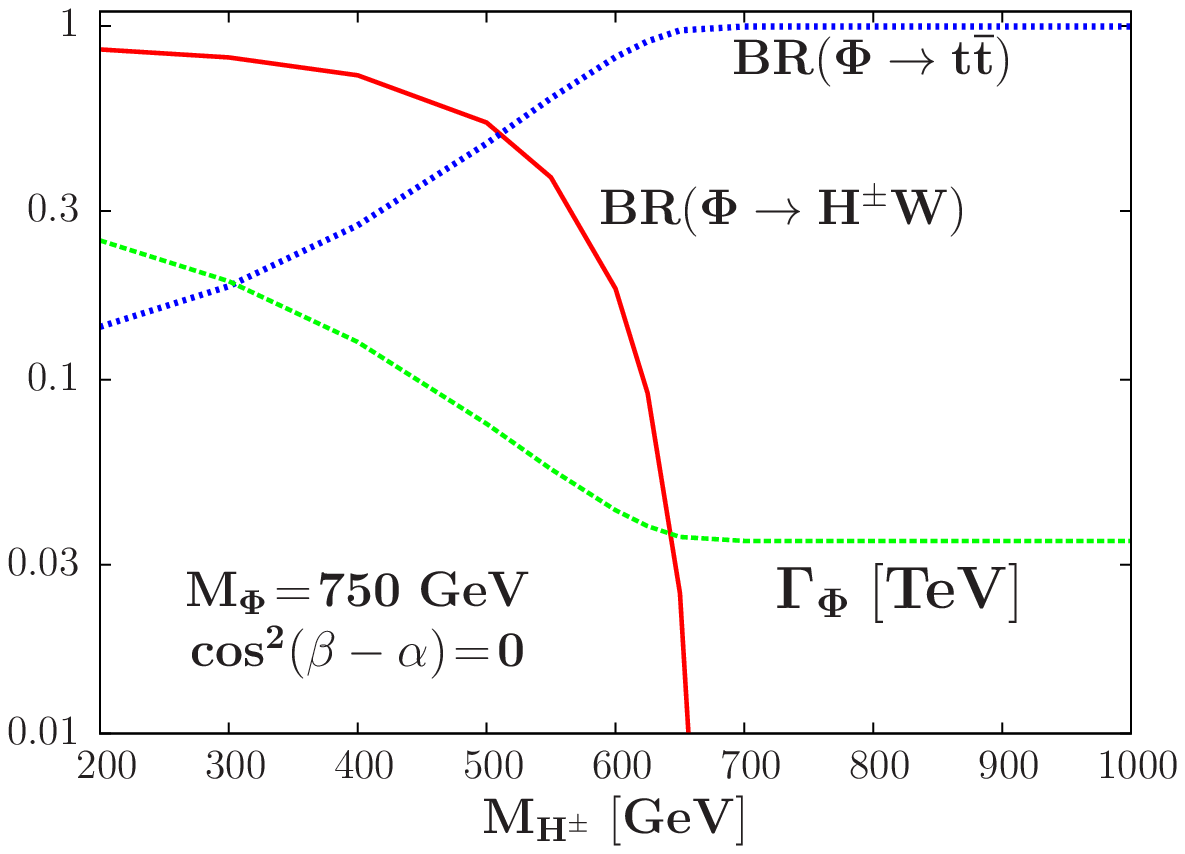}}
\vspace*{-14cm}
\caption{\it Branching ratios for the decays of the $\Phi=H/A$ states and their
total decay widths (in TeV) as functions of $M_{H^\pm}$ for $\tan\beta =1$, $\cos^2 (\beta-\alpha)=0$
and $M_{\Phi}=750$ GeV.}
\label{Fig:BRs}
\vspace*{-.2mm}
\end{figure}


\section{Vector--Like Fermions}

The existence of massive vector--like fermions is a common feature of both the singlet models discussed in Section~2 and the 2HDM scenarios discussed in Section~3. Therefore, in addition to further studies of the $\Phi$ particle itself and searches for possible bosonic partners at the LHC and future colliders, a useful way to probe different models would be via direct production of vector-like fermions. In this Section, we focus on this important aspect of model testing, considering both pair and single production of these fermions. We start by summarising the couplings of these particles, which have also  been discussed frequently elsewhere, see, e.g.,  Refs.~\cite{VLQs,VLQ-india,VLL-recent,lepton1,lepton2}.

\subsection{Couplings, mixing and decays of vector--like fermions} 

Except for singlet neutrinos, which have no electromagnetic and weak charges, all the other vector-like fermions couple to the photon and to the electroweak gauge bosons $W/Z$ 
with typical electroweak strength.  These couplings allow for the pair production processes
at colliders. To obtain the cross sections,  one needs in addition to the electric charge $e_F$ in units of the proton charge, the vector couplings of the vector-like fermion $F$ to the photon and $Z$ boson, which are given by
\beq
v_F^\gamma = e_F \ , \ \ v_F^Z \equiv v_F = \frac{2I_{3L}^F+2I_{3R}^F- 4e_F s_W^2}{4s_W c_W} \, , \ \ v_F^W  = \frac{I_{3L}^F+I_{3R}^F}{\sqrt 2 s_W} \, ,
\label{eq:ZFF}
\eeq
where  $I^{F}_{3L}, I_{3R}^F$ are the third components of the left-- and  right--handed weak isospins of the fermions, and $s_W^2=1-c_W^2 \equiv\sin^2 \theta_W$. The axial--vector couplings of the fermions to the  $Z$ boson, $a_F= (2I_{3L}^F - 2I_{3R}^F)/ (4s_W c_W)$, are zero by construction for vector-like fermions, unlike for Standard Model quarks and leptons. In addition, there is a charged-current coupling between the two components of each fermion doublet and the $W$ boson. As seen in eq.~(\ref{eq:ZFF}), this charged coupling is twice as large in the case of the vector fermions compared to standard ones, because of the vectorial nature of the fermion $F$.   

The vector--like leptons and quarks can mix with the Standard Model fermions that have the same U(1)$_{\rm Q}$ and SU(3)$_{\rm C}$ assignments. This mixing would give rise to new interactions that determine to a large extent the decay properties of the vector-like fermions, and allow for a new production mechanism, namely single production in association with a light fermion  partner. Generic Feynman diagrams for these processes
are shown in Fig.~\ref{diag}. The mixing pattern depends sensitively on the model considered and is, in general,  rather complicated, especially if one includes the mixing between different families. However, this  intergenerational mixing should be very small as it could induce flavour-changing neutral currents  that are severely constrained by existing  data \cite{PDG}. 

In the present analysis, we neglect the intergenerational mixing and treat the  remaining mixing angles as phenomenological parameters. The interactions of the electron and its partner neutrino with vector--like charged and neutral leptons $N,E$ may be written as  
\begin{eqnarray}
\hspace{-0.2cm}
{\cal L}_V = g_W [ \zeta_{W \nu E} \bar{\nu_e}  \gamma_\mu  E 
+ \zeta_{WeN} \bar{e} \gamma_\mu N ]  W^\mu 
+ g_Z [ \zeta_{ZeE} \bar{e} \gamma_\mu E  
+ \zeta_{Z \nu N} \bar{\nu_e} \gamma_\mu N]  Z^\mu  + {\rm h.c.} \, ,
\end{eqnarray}
where $g_W=e/\sqrt{2}s_W$ and $g_Z= e/2s_Wc_W$, and we have anticipated a small mixing 
angle~\cite{F-mixing} so that we have written $\sin \xi_i \simeq \xi_i$. The generalization to the other lepton families 
and to vector--like quarks is straightforward. At least in the case of couplings to third-generation quarks and leptons, 
one should also consider the mixing via the scalar sector~\footnote{One could assume that the 
magnitude of the mixing through the Higgs boson is proportional to the fermion masses and hence is negligible for the first two generations of fermions, compared to the mixing through the vector bosons.  Note, however, that such an intergenerational mixing might provide an explanation for the flavour--changing $h\to \mu^\pm \tau^\mp$ decay of the standard--like Higgs boson hinted during Run 1 of the LHC~\cite{Phi-JLK}.}, including the Standard Model--like $h$ state as well as the $\Phi$ and (in doublet models) the $H^\pm$ states. 
Normalizing the  mixing through the $Z$ and Higgs bosons in the same way by also defining $g_S= e/2s_Wc_W$,
the Lagrangian describing this mixing, where the sum is over all scalars and fermions, is given by 
\begin{eqnarray}
{\cal L}_S &=& g_S \sum_{S,f}  \zeta_{S f F } \bar{f} F S +  {\rm h.c.} \, ,
\end{eqnarray}
which we now use in analysis of some possible phenomenological signatures.

The heavy fermions can decay through mixing into massive gauge bosons plus their ordinary light partners,  $F \to Vf$ with $V=W,Z$ or $F \to hf$ (and eventually even $F\to \Phi f$ 
for very heavy fermions); see Fig.~\ref{diag}a (left). Using the scaled masses $v_X=M_X^2/m_F^2$ and neglecting the ordinary fermion masses (which should be an excellent approximation even in the case of the top quark), the partial decay widths are given by \cite{VLQ-india,lepton2}
\begin{eqnarray}
\Gamma (F \to Wf') &=& \frac{\alpha} {16 s_W^2 c_W^2} \zeta_{W Ff'}^2 \ 
\frac{m_F^3}{M_W^2}  ( 1-v_W)^2 (1+2v_W) \, , \nonumber \\
\Gamma (F \to Zf) &=& \frac{\alpha} {32 s_W^2 c_W^2} \zeta_{ZFf}^2 \ 
\frac{m_F^3}{M_Z^2}  ( 1-v_Z)^2 (1+2v_Z) \, , \nonumber \\
\Gamma (F \to S f) &=& \frac{\alpha} {32 s_W^2 c_W^2} \zeta_{SFf}^2 \ 
\frac{m_F^3}{M_S^2}  ( 1-v_S)^3 \, ,
\end{eqnarray}
When there is no Higgs channel, the charged-current decay mode is always dominant compared
to the neutral one and, for fermion  masses much larger than $M_V$, the branching ratios are 1/3 and 2/3 for the $F\to fZ$  and $F \to f'W$ modes, respectively.  Note that for Majorana neutrinos,  both the $N\to l^-W^+ /\nu_l Z$ and $N \to l^+ W^- / \bar{\nu_l} Z$ decays are possible. This gives nice like-sign lepton search signatures, and makes the Majorana neutrino total decay widths twice as large as for Dirac neutrinos.

The decay pattern above assumes at least an approximate mass degeneracy between the members of the same weak isodoublet, for instance $m_L \approx m_N$ in the case of vector--like leptons. However, if one allows for a mass difference between the two states,  one would have charged current decays such as $L^\pm \to W^\pm N$  or $N \to W^\pm L^\mp$; see
Fig.~\ref{diag}a (right). If $m_L \gsim m_N+M_W$, the $L^\pm$ decay is at the two--body level and, since it has a partial width that is not suppressed by any mixing angle, it is the dominant one.  In turn, $N$ decays through mixing to ordinary leptons and gauge bosons. An interesting situation is when $ 0< m_L - m_N < M_W$, so that the decay occurs at the three--body level with the virtual $W$ boson decaying into almost massless fermions, $L \to N W^* \to Nf \bar f$. In this case, depending in the virtuality of the $W$ boson and the mixing with light fermions, the $L \to Nf \bar f$ and $L \to \ell Z, \nu W$ modes might compete with each other rendering the situation quite model--dependent. 
 
\begin{figure}[!h]
\vspace*{-2.cm}
\centerline{\hspace*{-1cm} 
\includegraphics[scale=0.89]{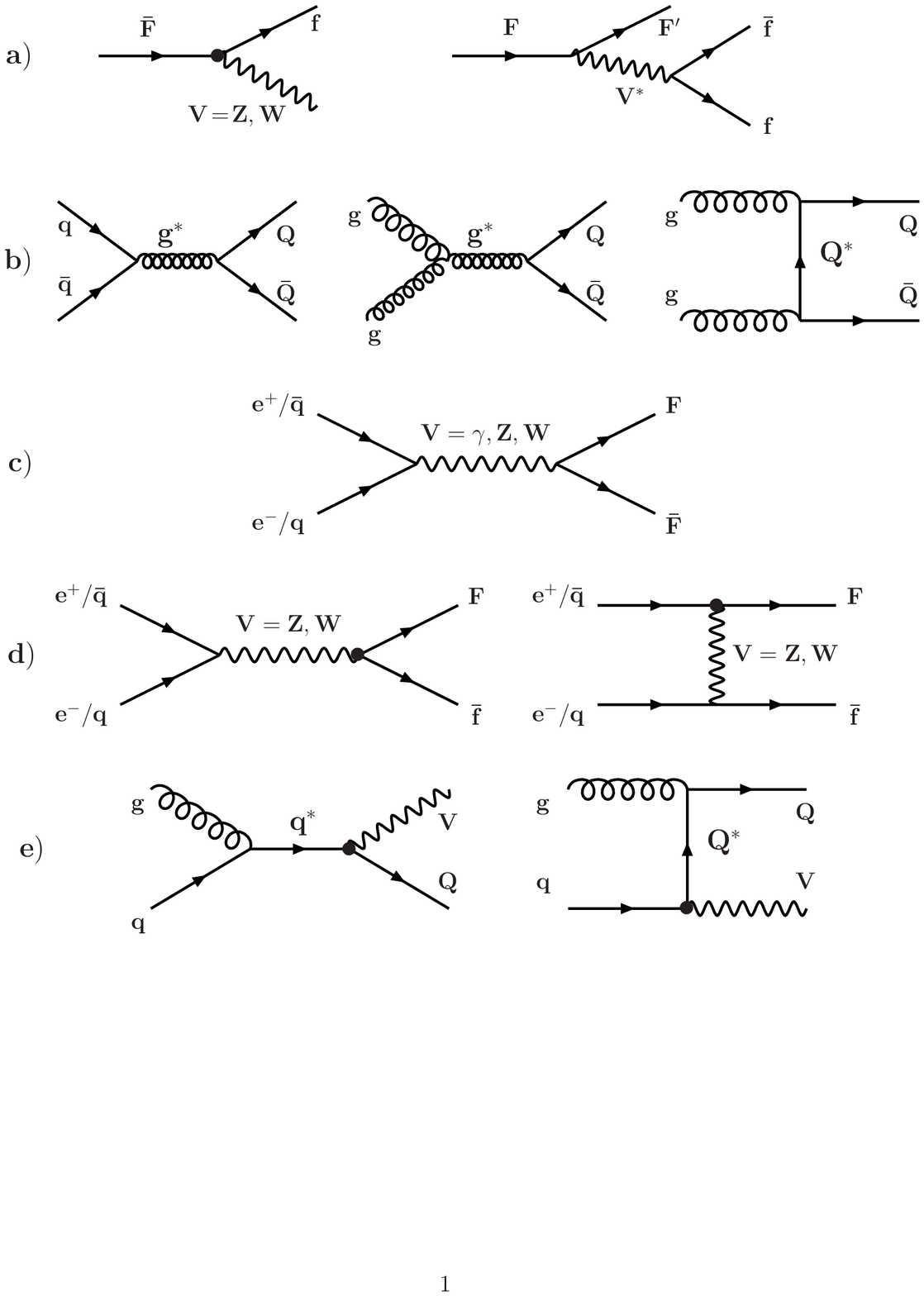} }
\vspace*{-8.5cm}
\caption{\it Generic Feynman diagrams for the processes involving vector--like fermions.
a) Two-body decays through mixing and two-- or three--body decays through gauge interactions.
b) Pair production of heavy quarks in $pp$ collisions.
c) Pair production in $\ee$ and $pp$ collisions via electroweak gauge couplings. 
d) Single production of heavy fermions in $\ee$ and $pp$ collisions through mixing. 
e) Production of one heavy quark in association with a gauge boson $V=W,Z$ in the
$gq$ subprocess at hadron colliders. The symbol $\bullet$ at a vertex indicates mixing between heavy and ordinary fermions.}
\label{diag}
\vspace*{-5mm}
\end{figure}

Note finally that for small mixing angles (and also if only three--body decays into lighter
vector--like fermions are kinematically accessible), the vector--like fermions have rather narrow widths if they are not too heavy, but the widths increase rapidly with the vector-like fermion mass, $\Gamma \sim m_F^3$. However,
for mixing angles  of order $\zeta \lsim 0.1$, the total widths do not exceed the 100 GeV range even for $m_F \sim  {\cal O} (1$~TeV). 

\subsection{Experimental constraints} 

We now summarize briefly the present experimental constraints on the masses of the
vector-like fermions and their mixings with the ordinary fermions. The mixing will, in general, alter the couplings of the electroweak gauge bosons to light quarks and leptons from their Standard Model values. The $Z f\bar f $ couplings, which are reduced by mixing factors 
$\cos^2\xi \approx 1- \xi^2$, have been very accurately determined at LEP through the measurements of total, partial and invisible decay widths, as well as forward--backward and polarization asymmetries. As they have been found to agree with the Standard Model predictions to the level of at least one percent, the mixing angles are constrained to be smaller than ${\cal O}(10^{-1})$ \cite{F-mixing}. The vector-like quark mixing with the top quark can presumably be slightly larger \cite{VLQs,VLQ-india}, as the Tevatron and LHC data on the top quark production cross sections and asymmetries are at the level of 10\% accuracy only \cite{PDG}. 

Note that in the case of vector--like leptons, the precise measurements of the 
anomalous magnetic moments of the electron and muon lead to even more stringent constraints:
$\zeta_i < {\cal O}(10^{-2})$ for light vector-like leptons with  masses of the order of 100 GeV \cite{Fg-2}. Indeed, because there is no chiral protection, the contribution of heavy lepton loops to $(g-2)_{e,\mu}$ is proportional to $m_e/m_{L}$, in contrast to $m_{e,\mu}^2/m_{L}^2$ for chiral couplings, and very small $\zeta_i$ and/or large $m_L$ are needed to accommodate the data.

Turning to the masses, the constraints in the vector-like lepton case are rather weak. From the null results of searches for new states and from the measurement of $Z$ decay widths at LEP, one can infer a bound $m_L \gsim {\cal O}(100)$ GeV independently of the mixing \cite{VLL-pair-LEP}, except for singlet neutrinos that have no weak couplings to the $Z$ boson. 
For small mixing angles, one could push these bounds up to $m_L \gsim {\cal O}(200)$ GeV
for single production at LEP2 in favourable cases \cite{VLL-sing-LEP}. For these single 
production processes, the limits from the Tevatron are not competitive with the LEP2 limits while those of the previous LHC run are comparable.  For instance, CMS searched for Majorana neutrinos \cite{Majorana} that couple to muons in the same--sign muon plus two jet final states and, for mixing angles $\zeta_{WN\mu} \lsim 0.1$, they obtain a bound of $m_N \gsim 200$ GeV on the Majorana neutrino masses \cite{CMS-Majorana}. 

The most constraining searches on vector--like leptons have been conducted by the ATLAS collaboration \cite{ATLAS-VLL1,ATLAS-VLL2}~\footnote{There are similar CMS analyses \cite{CMS-VLL}, but only at $\sqrt s=7$ TeV and  hence with reduced sensitivity.}. A first search \cite{ATLAS-VLL1} has been performed in the context of the Type-III seesaw model originally introduced in order to explain the smallness of the neutrino masses \cite{TypeIII}, which contains triplets with a neutral and two charged $L^\pm$ leptons that are assumed to be degenerate in mass.  Their decays into ordinary third-generation leptons are suppressed, so that only the easier final states with $e/\mu$ that have higher sensitivity are considered. The dominant search channel is $pp \to L^\pm N \to W^\pm \nu W^\mp \ell^\pm$,  where one $W$ boson decays leptonically and the other hadronically, resulting in same-sign or opposite-sign lepton pairs in the final state.  In a subsequent analysis \cite{ATLAS-VLL2}, the additional $pp \to L^+ L^-$  channel has been considered as well as the decay mode $L^\pm \to Z \ell^\pm$, which leads to three-lepton final states. In addition, a different model was considered  with a single charged vector--lepton state that we denote by one--VLL.  

For heavy leptons with masses $m_L=200$ GeV, the cross sections assumed in the ATLAS analyses at $\sqrt s=8$ TeV were 34 fb and 844 fb for the one--VLL and Type-III seesaw models, the huge difference being  mainly due to the fact that Type--III has both charged and neutral leptons than can be produced in the $pp\to L^\pm N$ process. The 95\% confidence level exclusion limits obtained by ATLAS are then $m_L \gsim 170$ GeV in the one-VLL model and $m_L \gsim 400$ GeV in the Type-III model. 

In our analysis, we will assume a lower bound on the charged vector--like lepton mass of $m_E=400$ GeV and sometimes we will fix $m_E$ to this specific value in order to enhance their loop  contributions to the $\Phi \gamma \gamma$ couplings, as can be seen from Fig.~\ref{fig:enhancement}. For the neutral vector--like lepton, a comparable mass $m_N \approx m_E$ will be assumed in general, though sometimes one can consider a relatively smaller mass in order to allow for the (possibly invisible) decay $\Phi \to N\bar N$. We note that, if $m_N \lsim m_E +M_W$, the $E$ decay 
will be more involved  and the experimental constraint on $m_E$ would become much weaker. 

The case of vector-like quarks is completely different, since the bounds on  their masses from
negative searches at Run 1 of the LHC are much more severe. This is particularly true in the case of vector--like partners of the top and bottom quarks. Indeed, depending on the decay branching fractions, one has for instance the bounds   $m_{T} > 950$~GeV for a branching ratio $\mathrm{BR} (T \to h t)=1$  and $m_{B} > 813$~GeV~for if $\mathrm{BR} (B \to W t) = 1$ \cite{LHC-VLQ}. (Similar limits can be set on quarks with exotic charges that we do not discuss here.) Although these limits are model--dependent, we assume as a general rule that vector-like quarks are heavier than about 800 GeV to 1 TeV, in order to evade these bounds. 


\subsection{Production of vector-like fermions in pp collisions}

We discuss now the production of the new fermions at hadron colliders. 
First, because they couple to gluons like ordinary quarks, vector--like quarks can be pair produced in the strong interaction process $pp \to Q \bar Q$ with rates that depend only on the heavy quark mass $m_Q$ and the strong coupling constant $\alpha_s = g_s^2/4\pi$.  As
can be seen from Fig.~\ref{diag}b, two processes are in play, $q\bar q $ annihilation and $gg$ fusion, with the latter largely dominating at higher-energy colliders for relatively small quark masses. The partonic cross sections at leading order (LO) are given in term of the velocity  $\beta_Q = \sqrt{1-4m_Q^2/\hat s}$ by \cite{pp-QQ}
\beq
\hat \sigma  (q\bar q \to Q\bar Q) &=& \frac{4 \pi \alpha_s^2} {27 \hat s} 
\beta_Q (3- \beta_Q^2) \, , \nonumber \\
\hat \sigma  (gg \to Q\bar Q) &=& \frac{\pi \alpha_s^2} {48 \hat s} \bigg[
\frac14 \beta_Q (\beta_Q^4 - 2\beta_Q^2 +143) + (33- 18 \beta_Q^2 + \beta_Q^4) 
\log \frac{1+ \beta_Q}{1- \beta_Q} \bigg] \, .~~
\eeq
The total hadronic cross section, i.e., after folding with the parton luminosities which
are chosen here to be those of the MSTW2008 fit~\cite{MSTW},  is shown in Fig.~\ref{Fig:pp-VLQ-pair} as a function of the heavy quark mass for several $pp$ collider centre-of-mass energies $\sqrt s=8,13, 14,33$ and 100 TeV. For the value $m_Q=1$ TeV that is close to the experimental lower bound,  the cross section, which is at the few fb level at $\sqrt s=8$ TeV, increases by one order of magnitude at $\sqrt s=13$ or 14 TeV and by four orders of magnitude at  $\sqrt s=100$ TeV. For higher quark masses, the increase of the rate with energy is
even steeper. For instance,  the production rate of about 100 fb at $\sqrt s=100$ TeV for $m_Q=3$ TeV is five orders of magnitude larger than at $\sqrt s=14$ TeV.
This clearly shows the advantage of a higher energy proton-proton collider. 
Note that we have evaluated the rate only at the leading order. At NLO in QCD, 
supplemented by the next-to-leading threshold logarithmic corrections (NLO+NLL), one
has to multiply the production rate above by a factor $K\approx 1.5$ at the LHC \cite{pp-QQ-NLO}  for heavy quark masses between 1 and 2 TeV. 

\begin{figure}[!h]
\vspace*{-2.4cm}
\centerline{\hspace*{-1cm} 
\includegraphics[scale=0.75]{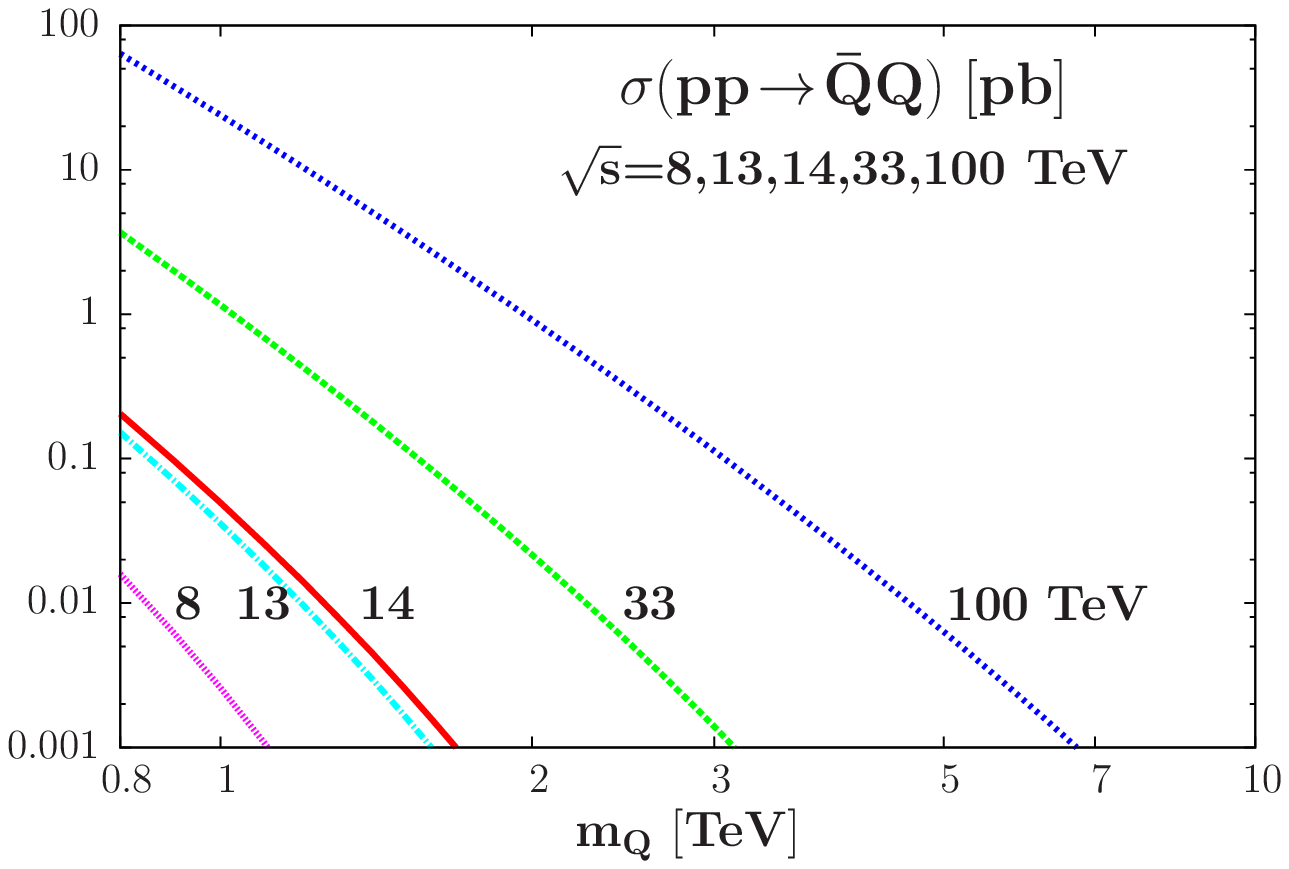} }
\vspace*{-13.2cm}
\caption{\it
The production cross sections in $pp$ collisions of vector--like quark pairs  as functions of
the  mass for several collider energies. Only the LO contributions are included and 
the MSTW parton distributions have been adopted.}
\label{Fig:pp-VLQ-pair}
\vspace*{-2mm}
\end{figure}

Vector-like leptons can also be pair produced in proton--proton  collisions via their electroweak couplings to $V=\gamma,W/Z$ bosons in the  Drell--Yan process $q \bar q \to V^* \to L\bar L$~\cite{Drell-Yan}, Fig.~\ref{diag}c.  The cross section for the  $q\bar q \to L \bar L$ subprocess where $L$ can be either a charged $E$ or a neutral $N$ lepton with a velocity  $\beta_L= (1-4m_L^2/ \hat s)^{1/2}$ reads
\beq
\hat \sigma (q\bar q \to L \bar L) = \frac{2 \pi \alpha^2}{9 \hat s} \beta_L (3-\beta_L^2) 
\left[e_q^2 e_L^2 + \frac{2 e_q e_L v_q v_L}{1-M_Z^2/\hat s} + \frac{(a_q^2+ v_q^2)
v_L^2}{(1-M_Z^2/ \hat s)^2} \right] \, .
\eeq
In the case of an electrically-charged $E$ state, both the $\gamma$ and $Z$ boson channels and their interference have to be considered, while only the channel with $Z$ boson exchange has to be considered for a non-singlet neutral lepton $N$ (an iso-singlet would not couple through the usual gauge interactions).  In addition, one could produce pairs of charged and neutral leptons via the exchange of a virtual $W$ boson, $q\bar q ' \to W^{*\pm} \to E^\pm N$. For comparable masses, $m_E \approx m_N$, the cross section  is simply given by
\beq
\hat \sigma (q\bar q \to L \bar N + \bar L N ) = \frac{4 \pi \alpha^2}{9 \hat s} \frac 
{\beta_L (3-\beta_L^2) }{(1-M_W^2/ \hat s)^2} \times \frac{1}{8 s_W^4} \, ,
\eeq
where we assume unit CKM--like matrix elements for simplicity.

\begin{figure}[!h]
\vspace*{-2cm}
\centerline{\hspace*{-1cm} 
\includegraphics[scale=0.8]{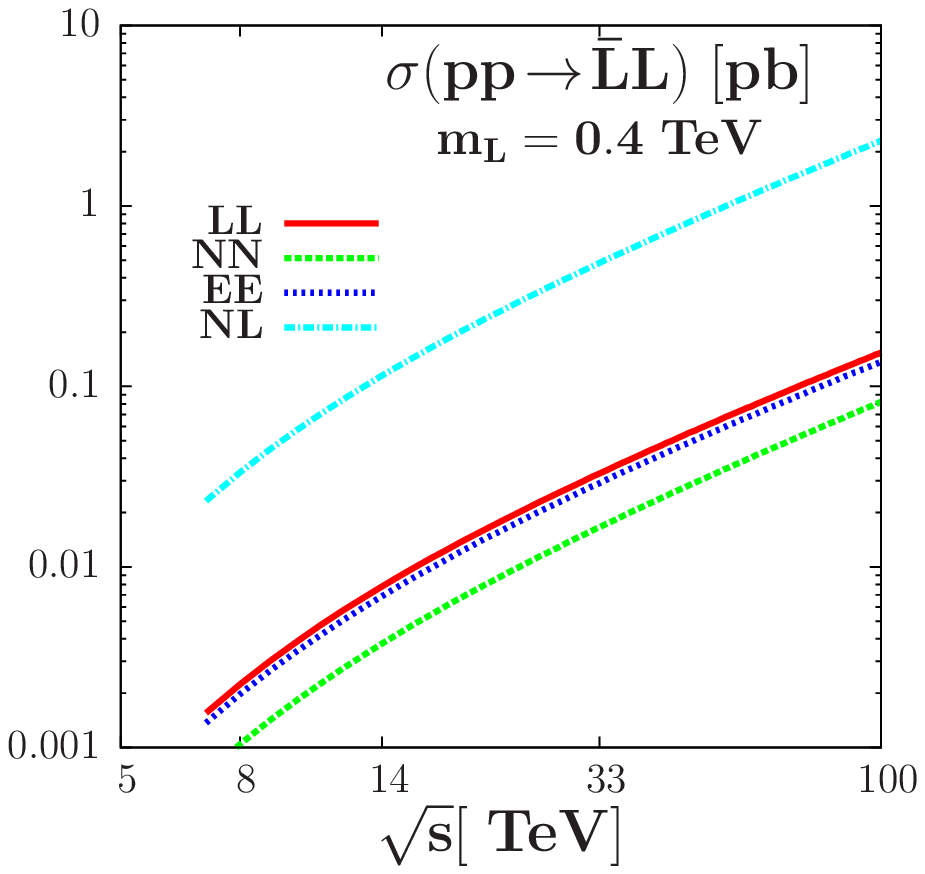}
\hspace*{-9cm} 
\includegraphics[scale=0.8]{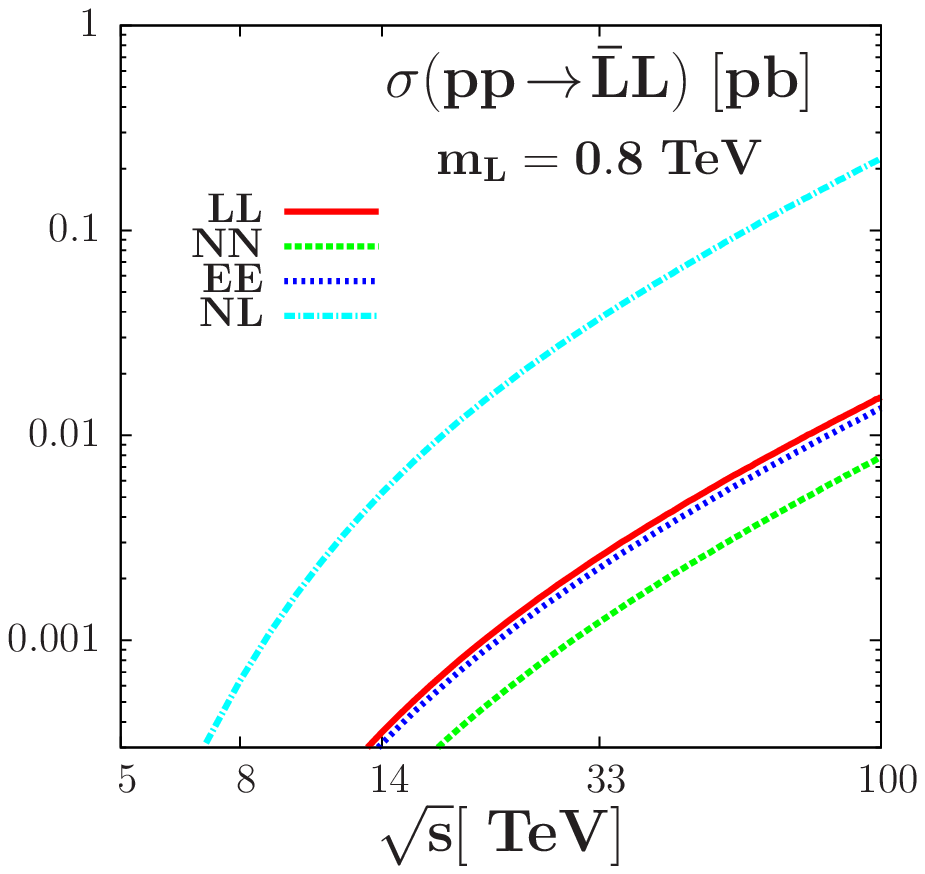} }
\vspace*{-14.2cm}
\caption{\it
Cross sections in $pp$ collisions for the pair production of vector--like lepton pairs as functions of
the centre-of-mass energy  for two values of the  lepton  mass: $m_L=400$ GeV (left) and $m_L=800$ GeV (right). 
The produced particles have the following electric charge and weak isospin
assignments: $L (-1, -\frac12)$, $N (0, +\frac12)$ and  $E (-1, 0)$.}
\label{Fig:pp-VLL-pair}
\vspace*{-2mm}
\end{figure}

The cross sections for producing pairs of vector--like charged and neutral leptons of masses $m_L=400$ GeV and $m_L=800$ GeV are shown in Fig.~\ref{Fig:pp-VLL-pair} as functions of $\sqrt s$. (Here also, we omitted the enhancement factor $K\!\approx\! 1.5$ due to QCD corrections
\cite{DY-NNLO}). The heavy leptons considered are those of Model 4 discussed in Section~2, which have the following assignments for electric charge and weak isospin $L (-1, 
-\frac12)$, $N (0, +\frac12)$ and  $E (-1, 0)$. One notices that, because of the smaller electroweak gauge coupling compared to the QCD coupling, the cross sections for vector--like leptons are three to four orders of magnitude smaller than those for vector--like quarks. In addition, the rates are much smaller for the neutral-current processes with the photon and/or $Z$ exchanges than for the charged-current $W$  exchange.
While the cross sections for $NN,LL,EE$ production are comparable and barely reach the  
fb level for $m_L=400$ GeV at the 8 TeV LHC, they are a factor 20 larger for $NL$ production.
The latter process is thus the best probe of vector leptons in pair production. Note again that the cross sections increase by about two orders of magnitude when moving from $\sqrt s=8$ TeV to $\sqrt s=100$ TeV. 

Let us finally mention that, for $m_L=200$ GeV, the cross section for single $N\ell$ and associated $NL$ pair production at the LHC with $\sqrt s=8$ TeV are, respectively, $\sigma(pp\to N\bar l + N \bar l ) \approx 15$ fb and $\sigma(pp \to L^\pm N) \approx 600$ fb. These values will be needed when we will discuss the sensitivity of future searches on the vector--like fermions.     

The other important set of processes for vector--like fermion production is provided by single production through mixing. In the case of vector--like leptons, the situation is rather simple, as there is only one relevant process which is $q \bar q \to V^* \to L \ell$
where the intermediate state  can be either a $W$ or $Z$ boson, see Fig.~\ref{diag}d (left).
For instance, the partonic cross section for the production of a heavy neutrino in association with a charged ordinary lepton is simply given by 
\beq
\hat \sigma (q\bar q \to \bar N \ell + \bar \ell  N) = \frac {\pi \alpha^2}{18 \hat s} \frac 
{ (1- \mu_N^2)^2 (1+ \frac12 \mu_N^2) } { (1-M_W^2/ \hat s)^2} \frac{ \xi^2_{ W N \ell} }  {s_W^4} \, ,
\label{eq:pp-singL}
\eeq
where $\mu_N \equiv m_N/\sqrt{\hat s}$. This is the cross section that led to the constraints $m_N \gsim 200$ GeV for $\xi^2_{WN\ell}  \approx  0.01$ derived by the CMS collaboration in the search for Majorana neutrinos decaying into two same-sign muons and jets  \cite{CMS-Majorana}. The cross sections for other leptonic final states can be obtained by simply adapting the couplings. 

The production rates in the process $pp  \to \bar N \ell + \bar \ell  N$  are shown in the left--hand side of Fig.~\ref{Fig:pp-sing1} for various collider energies as a function  of the heavy lepton mass for a mixing angle $\xi^2= 10^{-2}$. They are not too small and, 
for $m_N=400$ GeV, they increase from the level of 1 fb at $\sqrt s=8$ TeV to
about 100 fb at $\sqrt s=100$ TeV. The chances of observing such a process are therefore not
negligible at very high energy $pp$ colliders. 

\begin{figure}[!h]
\vspace*{-2.6cm}
\centerline{\hspace*{-9mm} \includegraphics[scale=0.8]{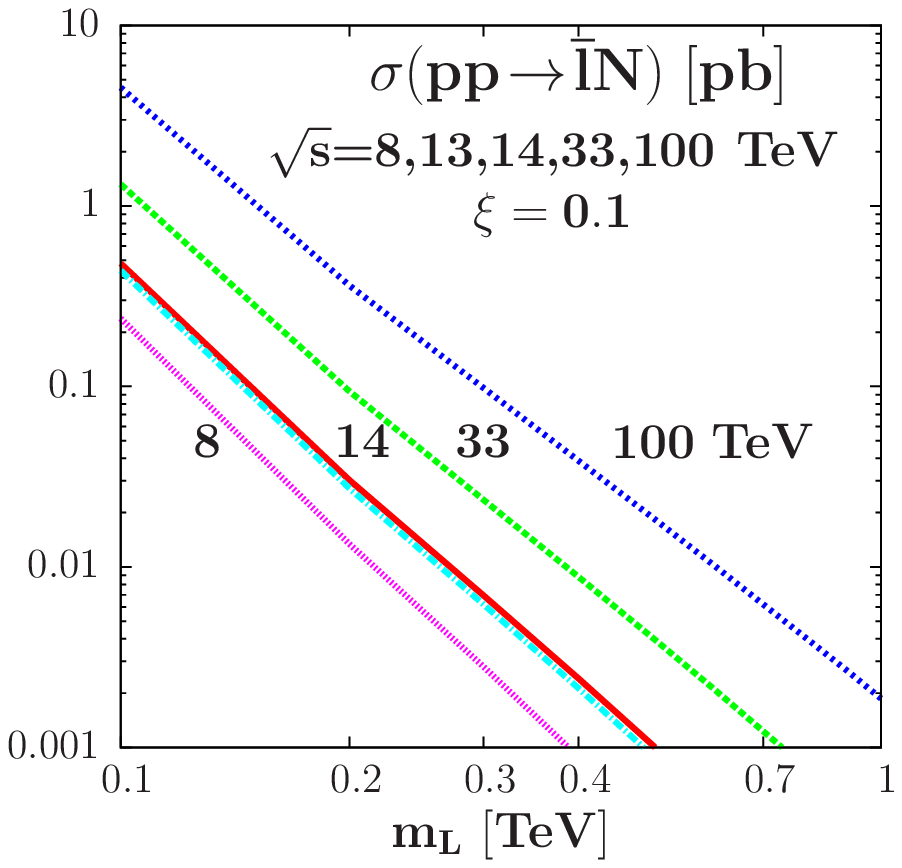}
\hspace*{-9.2cm} \includegraphics[scale=0.8]{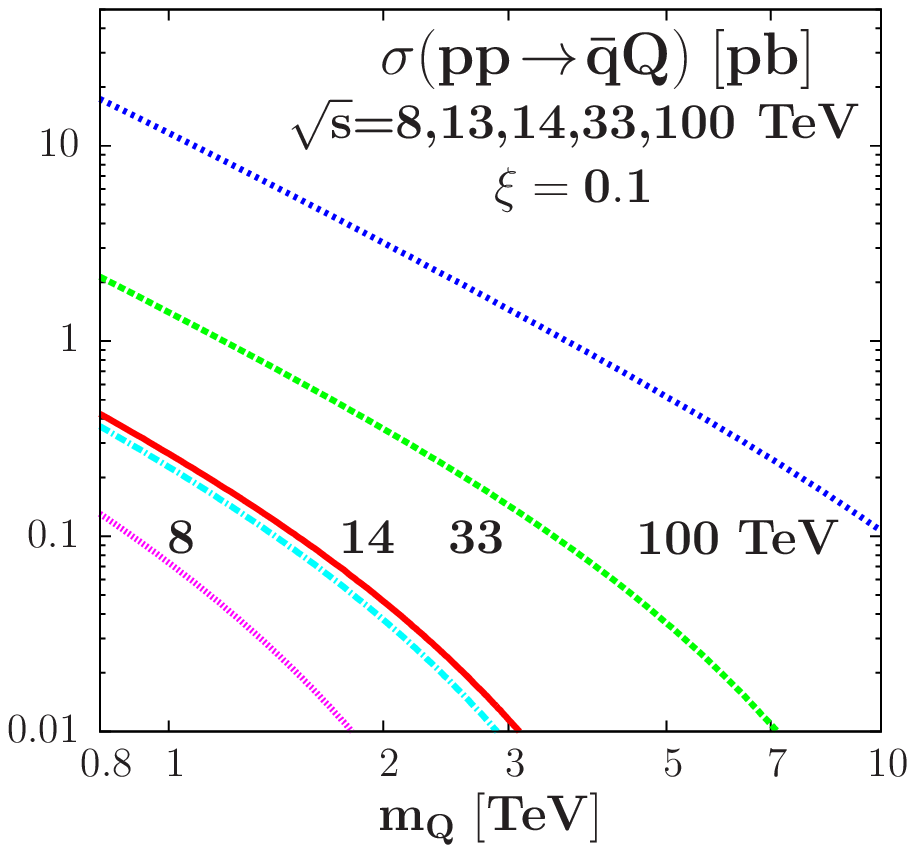} }
\vspace*{-14.cm}
\caption{\it
Cross sections for the associated production of the heavy neutral vector--like leptons with Standard Model leptons
(left) and for the associated production of heavy vector--like quarks and Standard Model quarks (right) 
as functions of the heavy fermion masses. In both cases, the mixing angles have been taken generically to be $\xi^2=10^{-2}$ and the rates are shown at various collider energies.} 
\label{Fig:pp-sing1}
\vspace*{-3mm}
\end{figure}

There are also processes for the single production of the vector-like quarks in association with their light quark partners through mixing as seen above for leptons. The rates are indeed suppressed by the small mixing angles but, for  high masses, some compensation is provided by the more favourable phase space, compared to the pair production channels.  Several single production processes exist but we will consider the two which  should, in principle, lead to the largest cross sections. 

First are the processes in which the production occurs through the virtual exchange  
of the $V=W,Z$ vector bosons. There is the possibility of $s$--channel $Z$ boson exchange 
in $q\bar q \to Q\bar q$ but also $W$ exchange in $q\bar q \to Q \bar q'$ as already mentioned  for vector--like leptons. These diagrams lead in general to contributions $\propto 
1/\hat s$, given by an expression similar to eq.~(\ref{eq:pp-singL}), that are small. There are also contributions that are generated by the exchange of the $V=W,Z$ bosons in the $t$--channel and which involve the mixing between the heavy $Q$ and the initial state light $q$ parton; see Fig.~\ref{diag}c (right). This gives rise to an extremely large enhancement of the cross section, which can be approximated in this case by
\begin{eqnarray}
\hat \sigma (q q  \to q Q ) =  \frac{ 2\pi \alpha^2}{3 \hat s} \, {\zeta_{VqQ}^2 } \,  
(v_q^2 + a_q^2)  (1- \mu^2)  {\cal F} (v)  \, ,
\label{eq:qqqQ}
\end{eqnarray}
where $v= M_V^2/ \hat s$, $\mu^2= m_Q^2/\hat s$ and $a_q, v_q$ are the full axial and vector 
$V q\bar q$ couplings: $a_q= 2I_q^3/(4c_W s_W)$ and $v_q=  (2I_q^3-4e_q s_W^2)/(4c_W s_W)$
for the $Z$ boson and $a_q= v_q =1/(2 \sqrt 2 s_W)$ for the $W$ boson. The function ${\cal F} (x)$, obtained when only the leading contributions at high energies are included, is given by  (the complete expressions can be found in Ref.~\cite{lepton2})
\beq 
{\cal F} (x) \approx  \frac{1}{x} + 2 \log \frac{1}{x (1-\mu^2)} \, .
\label{eq:Fsinglet}
\eeq
Note that in  eq.~(\ref{eq:qqqQ}) above the generic parton $q$ stands for both quarks and antiquarks. 

The cross sections of the $pp  \to qQ$  mechanism (they have been multiplied by  a factor of two to obtain the rates for the charged conjugate process with $\bar Q$), where $Q$ is a ``first generation" vector--like quark that couples to the $u$ quark with a mixing angle $\xi= 10^{-1}$, are displayed in the right--hand side of Fig.~\ref{Fig:pp-sing1} as a function of $m_Q$ for the usual collider energies.  As can be seen, they are quite substantial for the assumed mixing and at the LHC with $\sqrt s=14$ TeV, one obtains cross sections of the order of 100 fb (10 fb) for $m_Q=1.5$~TeV (2 TeV). For such masses, the rates are much larger than for pair production, despite the mixing and the smaller electroweak coupling. The rates increase more rapidly with centre-of-mass energy and for $m_Q=5$ TeV the cross section is still at the pb level at a 100 TeV collider for the chosen mixing angle  $\xi= 10^{-1}$.

A second process that is relevant for single vector--like production is the associated 
production with a vector boson, $qg \to VQ$ with $V=W,Z$. There are two channels involved 
in this mechanism: one in which the $qg$ pairs annihilates through the $s$--channel exchange
of $q^*$ which splits into the $VQ$ pair, and another in which the heavy quark $Q$  is
exchanged in the $t$--channel; see Fig.~\ref{diag}e. The total production cross section, in the limit $\hat s , m_Q \gg M_V$ which is appropriate in our case, is given  by the very simple expression\footnote{All these processes for single vector--like quark production are similar to those involved in single top quark \cite{single-top}. Surprisingly enough, we did not find in the literature the simple formula of eq.~(\ref{eq:sing-VQ}) for the total cross section in this specific single top production case.}
 \beq
\hat \sigma( qg \to VQ)= \frac{\alpha_s G_\mu}{12 \sqrt 2} \hat g_V^2 \xi^2_{VqQ} \mu_Q^2 \bigg[
 (1+2\mu_Q^2+2\mu_Q^4) \log \frac{1}{\mu_Q^2} - \frac12 (1-\mu_Q^2)(3+7 \mu_Q^2)
\bigg]
\label{eq:sing-VQ}
\eeq
with $G_\mu$ the Fermi coupling constant, $\hat g_W= \sqrt 2$ and $\hat g_Z = 1/ c_W$ 
and again $\mu_Q=m_Q/ \sqrt {\hat s}$. We have evaluated the cross sections in the case of 
a heavy vector--like quark $Q$ that couples to the first generation $u,d$ quarks with final
states involving both the $W$ and $Z$ bosons with masses that have been neglected (when folding with the parton distributions, we have thus summed over all first generation quark and antiquark contributions). The results for the production cross sections  are displayed in Fig.~\ref{Fig:pp-sing2} as a function of the mass $m_Q$ for several energies of the $pp$ 
colliders and a mixing angle fixed to $\xi=0.1$.

Despite of the $\xi^2$ suppression, the cross sections are extremely large. Already for 
a $1$ TeV quark, the rates at the $\sqrt s=14$ TeV  LHC  are an order of magnitude
larger than in the QCD pair production process. At $\sqrt s=100$ TeV, the rates
relative to $\sqrt s=14$ TeV are of the same order,  $\sigma \approx 25$ pb, for a mass 
$m_Q=1$ TeV. But for higher $Q$ masses,  there can be a huge difference: the rates for $pp\to VQ$ are larger than those of $pp\to Q\bar Q$ by an order of magnitude for $m_Q=3$ TeV and by three orders for $m_Q=10$ TeV. 

Hence, these single production processes are the most copious sources of vector--like quarks and they allow to probe mixing angles that are very small (more than an order of magnitude smaller than $\xi=0.1$). With such large  rates, one could even probe second and third generation heavy quark if inter-family mixing is not prohibitively  small. 

\begin{figure}[!h]
\vspace*{-2.4cm}
\centerline{\hspace*{-1cm} 
\includegraphics[scale=0.75]{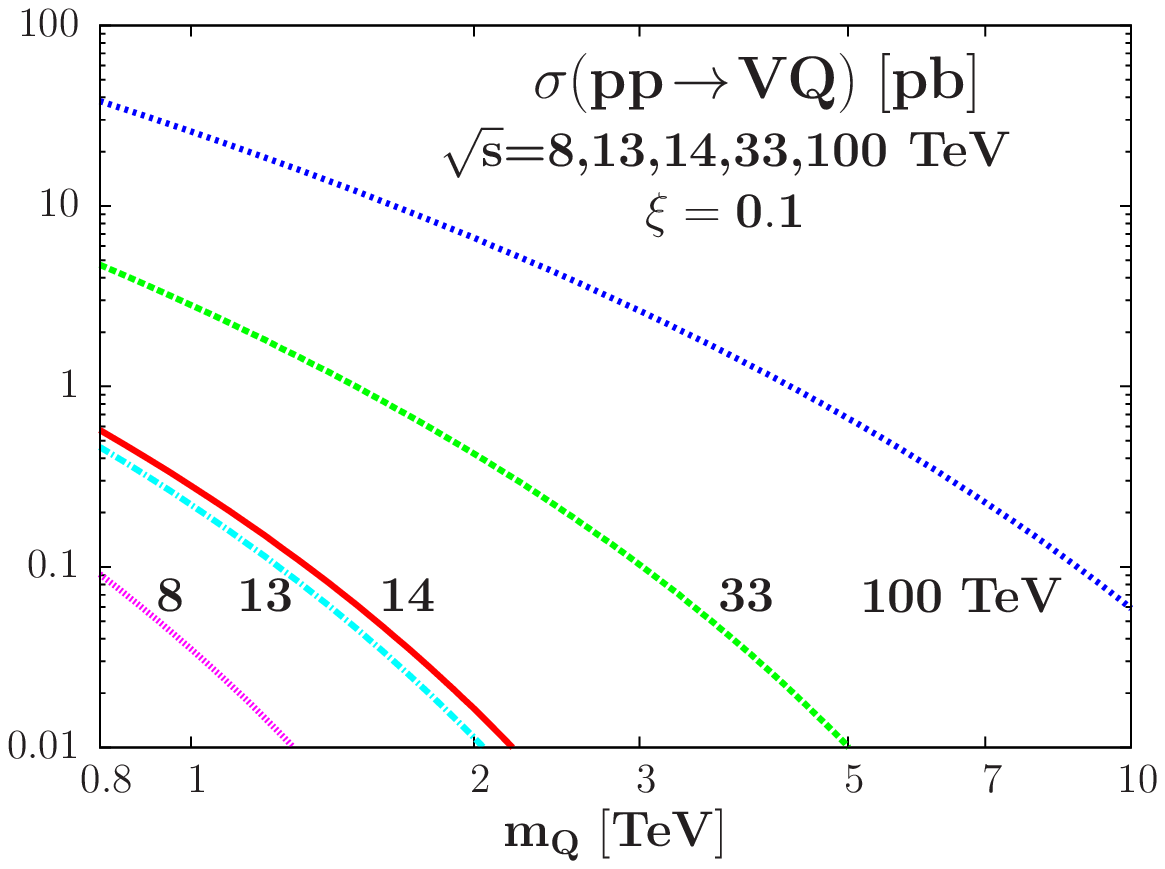} }
\vspace*{-13.2cm}
\caption{\it
The production cross sections in $pp$ collisions of single vector--like quarks in the
process $qg \to QW$ and $qg \to QZ$    as a function of the  mass for several collider energies; a mixing angle $\xi=0.1$ is assumed.}
\label{Fig:pp-sing2}
\vspace*{-2mm}
\end{figure}

We close this Section by discussing the sensitivity of future searches at the LHC and at higher-energy $pp$ colliders to the masses and couplings of the vector--like fermions. The LHC experiments have already searched for the coloured heavy quarks during  Run~1, with exclusion sensitivity to masses above $800$~GeV \cite{LHC-VLQ} from pair production as discussed above (we will not discuss single production which is more model-dependent). In order to extrapolate this sensitivity to higher energies and luminosities, we assume that a similar number of events would be required at another centre-of-mass energy, while noting that more efficient multivariate discrimination and boosted techniques might lead lead to improved sensitivity at the future LHC runs and higher-energy $pp$ colliders.

Neglecting this possibility of improved analyses techniques and simply considering the cross sections for the pair production processes discussed in the previous subsection, we have evaluated the prospective sensitivities to heavy vector-like quarks at the LHC with various energies and luminosities and at higher-energy $pp$ colliders. Our results are  summarised in  Table~\ref{tab:heavyF} in the four models with vector--like fermion content  discussed in Section~2. We have used the facts that Models 2, 3 and 4 have, respectively, twice, four times and four times as many heavy quark degrees of freedom as Model 1.

As can be seen, depending on the considered model, sensitivities to vector--like quark masses between 1.7 TeV and 2.0 TeV can be reached at the LHC with $\sqrt s=14$ TeV and 300 fb$^{-1}$ of data.  Thanks to the huge increase in the production cross sections when moving to higher centre-of-mass energies, the sensitivity improves by a factor of approximately 8 at a $pp$ collider with $\sqrt s=100$ TeV energy and 20 ab$^{-1}$ of data, allowing one to probe vector--like masses higher than the  10 TeV level; see also Ref.~\cite{100TeV}. 

\begin{table}[!h] 
\centering
\renewcommand{\arraystretch}{1.2}
\begin{tabular}{| c | c | c |}
\hline
& Vector-like quark mass sensitivity & Vector-like lepton mass sensitivity\\
{model} & 100fb$^{-1}$ ~300fb$^{-1}$ ~300fb$^{-1}$ ~20ab$^{-1}$ & 100fb$^{-1}$ 
~300fb$^{-1}$ ~300fb$^{-1} ~$20ab$^{-1}$\\
& 13~TeV ~~14~TeV ~~33 TeV ~100~TeV & 13~TeV ~~14~TeV ~~33$\;$TeV ~100$\;$TeV\\
\hline
\textbf{1} \! &\! ~~~~1.4~~~~  ~~~~1.7~~~~ ~~~~3.1~~~~ ~~~11.7~~~ & -\\
\hline
\textbf{2} \! &\! ~~~~1.5~~~~  ~~~~1.8~~~~ ~~~~3.4~~~~ ~~~12.7~~~ & -\\
\hline
\textbf{3} \! &\! ~~~~1.6~~~~  ~~~~2.0~~~~ ~~~~3.7~~~~ ~~~13.7~~~ & -\\
\hline
\textbf{4} \! &\! ~~~~1.6~~~~  ~~~~2.0~~~~ ~~~~3.7~~~~ ~~~13.7~~~ & \hspace*{-3mm}
             0.56~~  ~~0.73~~~~ ~~~1.7~~~ ~~5.3\\
\hline
\end{tabular}
\caption{\it Prospective model sensitivities to massive vector-like quarks (left) and leptons (right) [with the particle masses in TeV] in the indicated $pp$ collider and scenario.}
\label{tab:heavyF}
\vspace*{-4mm}
\end{table}

Turning to vector--like leptons,  future sensitivities may be derived from the ones 
obtained by the ATLAS collaboration in searches at the 8 TeV LHC with approximately 20 fb$^{-1}$ of data in the Type-III seesaw model discussed previously \cite{ATLAS-VLL1,ATLAS-VLL2}
and which led to a bound $m_L \gsim 400$ GeV in this model (assuming of course the specific multi--lepton decay  pattern which simplified the search and led to the quite stringent bound).

In the singlet Model 4 of Section~2, which includes a family of vector--like fermions,  there are SU(2) lepton doublets with charged and neutral components and an SU(2) singlet charged lepton. The cross sections at the LHC are $\sigma(pp\to N\bar l) \approx 13$ fb (for 
$\xi^2 \approx 0.01$) and $\sigma(pp \to L^\pm N) \approx 600$ fb, i.e., $ \approx 30\%$ smaller than those assumed by ATLAS to set the limits that were discussed previously. The present sensitivity in Model 4 will be thus slightly worse  and, if  the vector-like leptons are degenerate, may be assumed to be $m_L \approx 300$ GeV from pair production (the limit 
from single production depends on the mixing angle).  
Using the same equal-sensitivity assumption as for vector quarks, we find the prospective future sensitivities to vector-like leptons of Model 4 that are shown in the last column of Table~\ref{tab:heavyF} for the LHC and for higher-energy $pp$ colliders.  

We conclude that hadron colliders have indeed some sensitivity to heavy leptons; see 
also Ref.~\cite{100TeV}.  Nevertheless, the sensitivities above should be taken only as indications, as they are quite model--dependent. For instance, they are weaker if decays into third generation leptons are allowed or even dominant. The sensitivity also weakens if there is a mass difference between the neutral leptons that would lead to cascade decays. Finally, if the lightest vector--like lepton is the dark matter particle and is thus stable, all these searches become extremely difficult because the $N$ lepton escapes  detection.  
 
\subsection{Production of vector-like fermions in $\mathbf{\ee}$ collisions}

We turn now to $\ee$ colliders. Thinking that the vector-like leptons might be the lightest, we focus first on lepton pair production: $e^+ e^- \to L^+ L^-$, which is kinematically possible for $m_{L} \lsim \frac12 \sqrt s$. The total cross section for fermion pair 
depends on the electric charge and the vector--like $ZF \bar F$ coupling given in eq.~(\ref{eq:ZFF})
\beq
\sigma (e^+ e^- \to F \bar F) = \sigma_0 N_c \frac{1}{2} \beta_F (3-\beta_F^2) 
\left[e_e^2 e_F^2 + \frac{2 e_ee_F v_e v_F}{1-M_Z^2/s} + \frac{(a_e^2+ v_e^2)
v_F^2}{(1-M_Z^2/s)^2} \right] \, ,
\eeq
where $N_c$ is the colour number, $\sigma_0=4\pi \alpha^2/3s$ is the point--like QED 
cross section for muon pair production,  $\beta_F= (1-4m_F^2/s)^{1/2}$ the velocity 
 and the reduced couplings of the electron to the $Z$ boson are $v_e=
(-1+4s_W^2)/(4s_W c_W)$ and $a_e=-1/(4s_W c_W)$. 

The cross sections for lepton pair production are shown in the left panel of Fig.~\ref{Fig:eeVLL} as functions of $\sqrt s$ for the masses $m_L = m_N = 400$ GeV.  We note that in the case of neutral vector-like leptons the cross section for $e^+ e^- \to NN$ proceeds 
only through $Z$--boson exchange, Fig.~\ref{diag}c, and the rate is in general much smaller than for charged leptons. For Dirac particles the cross section rises as $\beta_N$ close to threshold, whereas Majorana fermions have a $\beta_N^3$ behaviour, like scalars, which could be a useful diagnostic tool. 

We note that, since they have no axial couplings to the $Z$ boson, there is no forward--backward asymmetry in $e^+ e^- \to L^+ L^-$, $A_{FB} \propto a_F = 0$. Furthermore, the polarization vectors of the heavy leptons have vanishing longitudinal components with respect to the flight direction, whereas the transverse components are small and positive \cite{lepton2}.  

The right panel of Fig.~\ref{Fig:eeVLL} shows the pair production cross sections at $\sqrt{s} = 3$~TeV for the vector-like fermions $U, D, E$ and $N$ as functions of $m_F$. It is clear that a 3 TeV $\ee$ collider such as CLIC would be able to produce copiously and study in detail any of these vector-like fermions, even if they have masses close to the beam energy.

\begin{figure}[!h]
\vspace*{-2.5cm}
\centerline{\hspace*{-1cm} 
\includegraphics[scale=0.75]{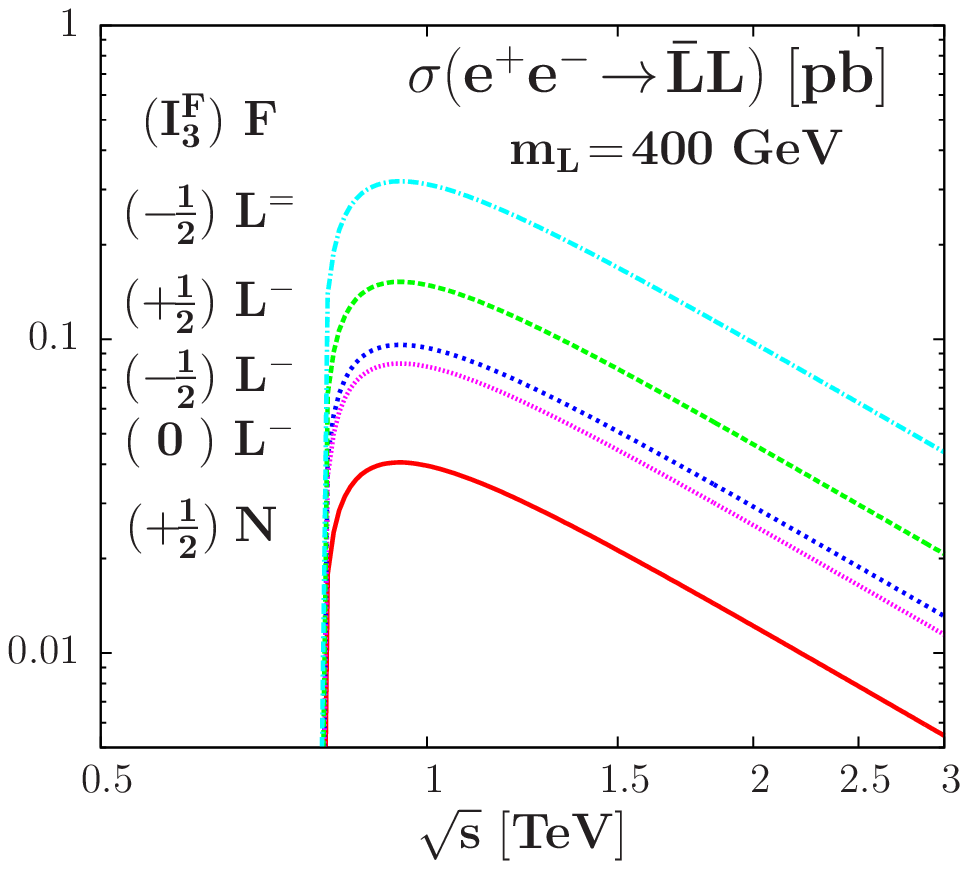}
\hspace*{-8cm} 
\includegraphics[scale=0.75]{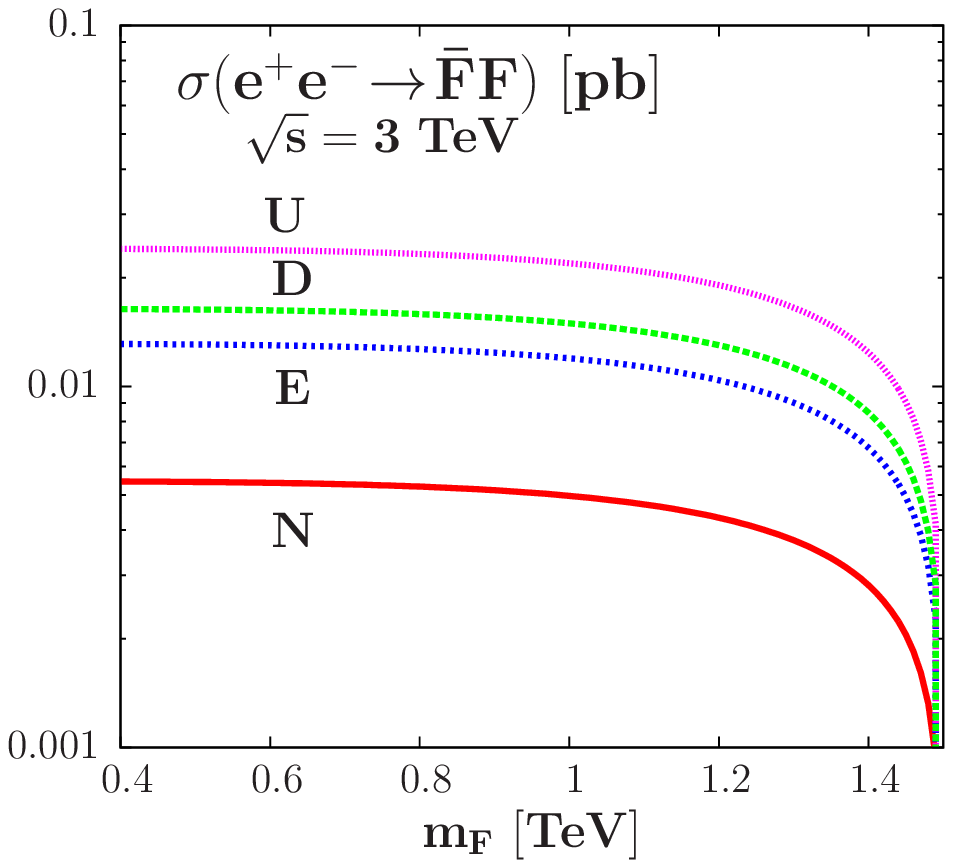} }
\vspace*{-13.2cm}
\caption{\it
Left panel: Cross sections for pair-production of the heavy vector--like charged and neutral
leptons at an $e^+ e^-$ collider as functions of the centre-of-mass energy $\sqrt s$ 
for the mass values $m_L = m_N = 400$ GeV, ordered with decreasing cross sections.
In addition to the pair production  processes $e^+ e^- \to N \bar N, L^+ L^-$ with the isospin of the particles given in brackets, we also include the pair-production of doubly-charged leptons, $e^+ e^- \to L^{++}  L^{- - }$. Right panel: Cross sections at $\sqrt{s} = 3$~TeV for the pair production of vector-like fermions $U, D, E$ and $N$ as functions of $m_F$.
}
\label{Fig:eeVLL}
\end{figure}

Single production of vector-like leptons is possible in $\ee$ collisions in association with the light Standard Model leptons,  $e^+ e^- \to L_i^\pm l_i^\mp$ and $e^+ e^- \to N_i \nu_i$, via the mixing angles $\xi_i$ \cite{lepton2,lepton4}. 
These processes are, in contrast to pair production, not democratic. One would have,
in terms of the reduced electron couplings, $\hat a_e=-1$ and $\hat v_e= -1+ 4s_W^2$,  
\begin{eqnarray}
\sigma(\ee \to F\bar f) = \sigma_0 N_c \frac{ \zeta_{ZFf}^2 } {128 c_W^4s_W^4} (\hat a_e^2 + \hat v_e^2)  \frac{(1-\mu^2)^2(1+\frac12 \mu^2)}{(1-z)^2} \, ,
\end{eqnarray}
with $\mu^2\!=\!m_F^2/s$ and $z\!=\!M_Z^2/s$. In addition, for vector-like leptons with couplings to the electron,
one needs to include additional $t$--channel vector boson exchanges: 
$W$--exchange for the neutral leptons and $Z$--exchange for the charged leptons;
Fig.~\ref{diag}d.   
Complete expressions for the angular distributions and the total cross sections for these processes in the 
general case can be found in Refs.~\cite{lepton1,lepton2}. Here we simply give the dominant piece 
in the limit where $M_W, M_Z \ll \sqrt s $, with only the leading logarithms being retained:
\begin{eqnarray}
\sigma (\ee \to \bar Ee ) & = & \sigma_0 \frac{ 3 \zeta_{ZeE}^2 } {64 c_W^4s_W^4 } (\hat v_e^2 + \hat a_e^2)  (1- \mu^2)  {\cal F} (z)  \, , \nonumber \\
\sigma(\ee \to \bar N_e \nu_e) &=& \frac{ 3\sigma_0  }{16s_W^4} (\hat v_e^2 + \hat a_e^2)  
(1- \mu^2)  {\cal F} (w) \, ,
\end{eqnarray}
where the function ${\cal F}$ is given in eq.~(\ref{eq:Fsinglet}). 
The cross sections are shown in  Fig.~\ref{Fig:eeVLell} as functions of $\sqrt s$ for 
$m_L = m_N = 400$ GeV, assuming $\xi_i^2=10^{-2}$ for all mixing angles. 
They are much smaller in the case of neutral and charged leptons with couplings only to the second- and third-generation
Standard Model leptons, since the cross sections for vector-like leptons coupled to electrons
are enhanced by some two to three orders of magnitude thanks to the 
$t$--channel exchanges. This effect is particularly marked  for neutral vector-like leptons,
where rates of the order of 1 pb for $\bar N \nu_e + \bar \nu_e N$ production can be 
obtained for mixing angles $\zeta^2=10^{-2}$, larger than in pair production,
which is not mixing suppressed. The reason is that there is a contribution that grows like $1/w = s^2/M_W^2$ and another like log$(1/w)$. The cross section for charged vector-like leptons is an order of magnitude less as a result of the smaller $Z$ couplings, is 
nevertheless also significant.

\begin{figure}[!h]
\vspace*{-2.5cm}
\centerline{\includegraphics[scale=0.75]{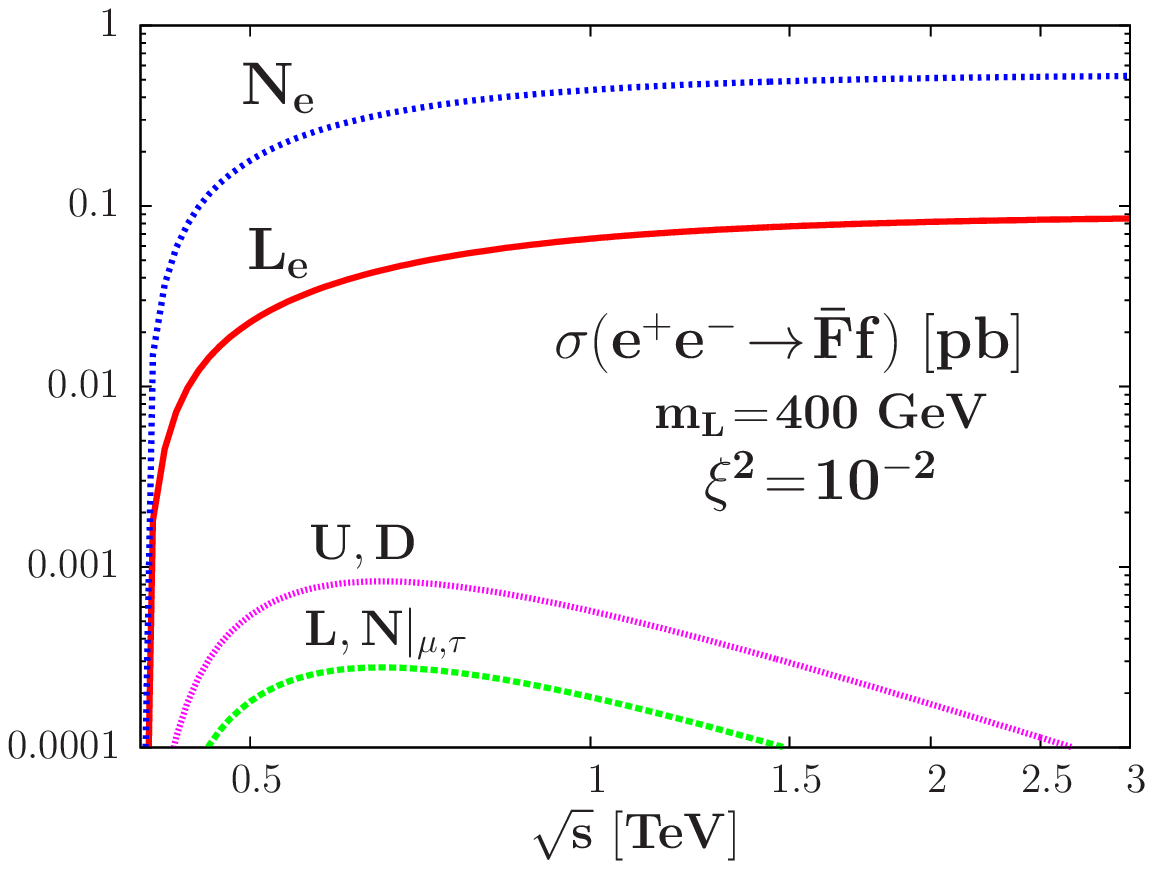} }
\vspace*{-13.5cm}
\caption{\it Cross sections for single production of the heavy vector--like charged and neutral leptons in association with ordinary leptons at an $e^+ e^-$ collider,
as functions of the centre-of-mass energy $\sqrt s$  for the mass values $m_L = m_N = 400$ GeV, assuming mixing $\xi^2 = 10^{-2}$ with ordinary leptons $e$ or $\mu, \tau$. The cross
section in the case of vector--like quarks with the same mass is also shown. 
}
\label{Fig:eeVLell}
\vspace*{-2mm}
\end{figure}

Regarding their detection, as mentioned previously, these fermions decay through mixing into gauge bosons and ordinary leptons, $L^\pm  \to \ell^\pm Z, \nu W^\pm$  and $N \to\nu Z, e^\pm W^\mp$,  with rates that are in general twice as large for the charged decays as for the neutral ones. There are potentially also  decays involving the light $h$ boson in the final state, which have magnitudes that are strongly dependent on the mixing pattern.  If only these decays are considered, detecting the vector-like leptons in an $e^+ e^-$ environment is not a problem, provided that the  production rates are large enough (such decays involving leptons and missing energy could be relatively easily detectable even at $pp$ colliders).

However, there is a special and more complicated situation that could be of interest.
If the lightest vector-like fermion is a neutral lepton, it might be (meta-)stable and a candidate for the dark matter particle. This vector-like lepton would be invisible and escape detection, except in a `neutrino counting' experiment where the heavy neutrino is 
pair-produced in association with a photon radiated from the initial state,
$\ee \to N\bar N \gamma$. However, the charged vector lepton $E$ could
decay into the neutral one and a $W$ boson, which would be off--shell if the mass difference 
$m_E - m_N$ is small, $E \! \to \! N W^* \! \to \! N f\bar f$, with the $N$ state again escaping detection. These channels with soft leptons and a large amount of missing energy will be more difficult to probe (in particular at hadron colliders) since they have signatures similar to supersymmetry with a compressed spectrum.

Finally, a very interesting set of processes for the vector-like leptons would be production in association with a Higgs boson~\footnote{In addition, vector-like lepton production in association with a vector boson might be important. As the cross sections for $\ee \to W^+W^-$ and $ZZ$ production are rather large,  the emission of heavy plus light leptons by virtual $V$ bosons through mixing, $e^+ e^- \to VV^* \to V L\bar \ell$, might lead to rates that are not negligible compared to the single production processes, in particular for heavy leptons that do no mix with the electrons and whose rates are thus not enhanced.},  $e^+ e^- \to L \bar L$ + Higgs, taking advantage of the large Yukawa couplings  $\propto m_L/v$. Most favoured by phase space is the process $e^+ e^- \to h L \bar L$ with $h$ the standard--like Higgs boson with a mass $M_h=125$ GeV. This is similar to the associated $t\bar t h$ processes  discussed in Ref.~\cite{ee-ttH}, and the cross sections are shown in the left panel of Fig.~\ref{Fig:LLHiggs} as functions of $\sqrt s$, again for a mass
$m_L=400$ GeV. 

We consider the cases of the $N$ state with isospin  $+\frac12$ and  the $L^-, L^{--}$ states with isospin $-\frac12$, and assume that the Yukawa couplings are simply proportional to the vector-like lepton masses $Y_{L_i}= Y_{N_i}  = m_{i}/v$. Here, the dominant contribution to the cross section comes from vector-like lepton pair production through $s$--channel $\gamma$ and $Z$ exchange, with one of them emitting the $h$ boson,  $\ee \to \bar L L^* \to h \bar LL$, and this is the only contribution retained in the figure. As expected, simply because of the electric charge, the largest cross section is obtained for $L^{--}$, followed by that for $L^-$ and then that for $N$. One has rates of a few fb to a few tens of fb at centre-of-mass energies $\sqrt s \approx 1.5$ TeV, where the phase space and the $1/s$  behaviours are not too penalizing. 

\begin{figure}[!h]
\vspace*{-2.6cm}
\centerline{\hspace*{-9mm} \includegraphics[scale=0.75]{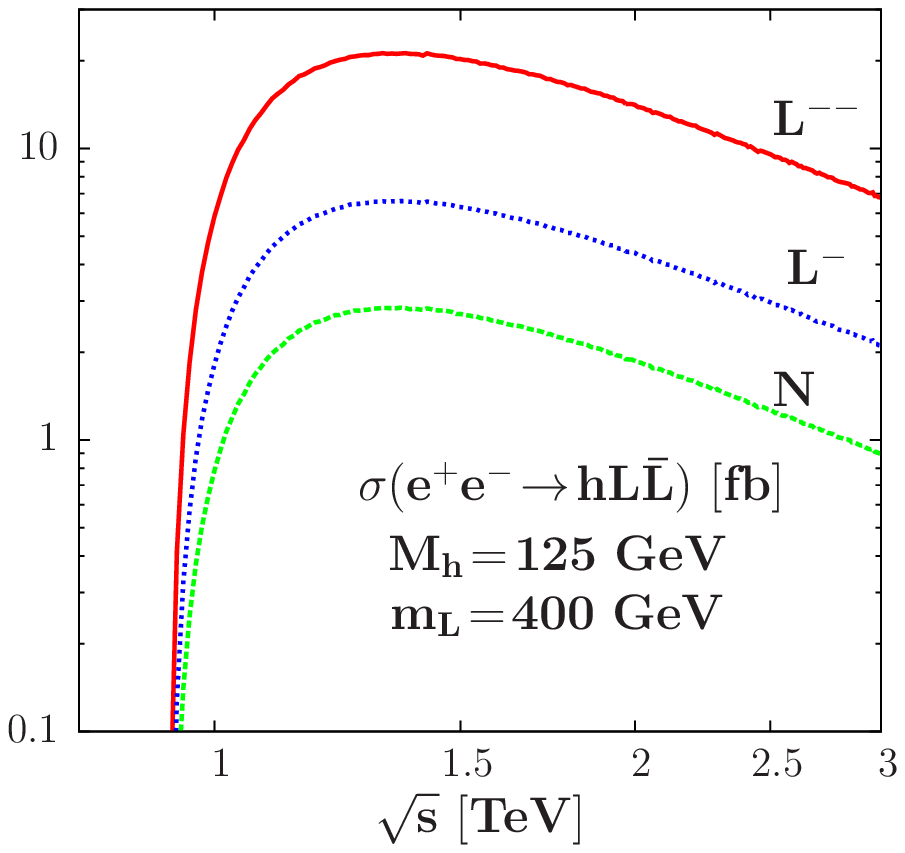}
\hspace*{-8.7cm} \includegraphics[scale=0.75]{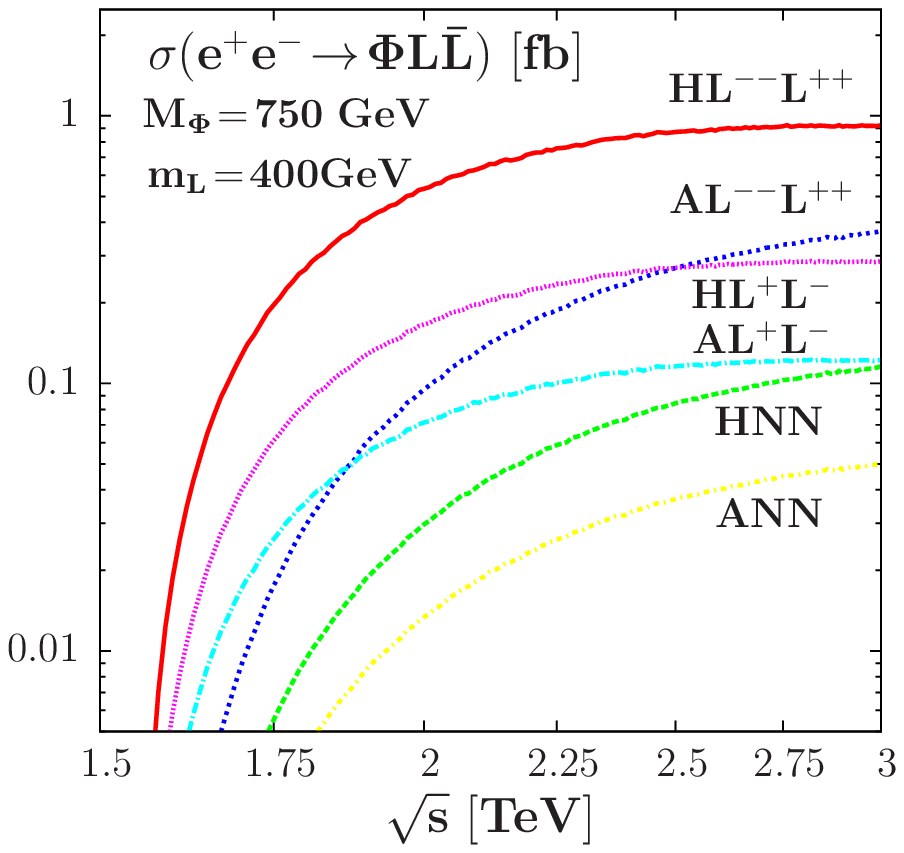} }
\vspace*{-13.2cm}
\caption{\it
Cross sections for production of the heavy charged and neutral vector--like leptons
in association with the Standard Model--like $h$ boson (left panel) and the $\Phi=H/A$ states (right panel). We assume $m_L \! = \! m_N \! = \! 400$ GeV and consider the  $L^{--}, L^-, N$ cases with standard--like Yukawa couplings $\propto m_{i}/v$.} 
\label{Fig:LLHiggs}
\end{figure}

Another mechanism is $\Phi = H, A$ production in association with a pair of heavy leptons, $e^+ e^- \to \bar L L \Phi$ (there is also single production in the channels  $e^+ e^- \to \bar L \ell \Phi$, but this is suppressed by the mixing in addition to phase space) and the rates are shown in the right panel of Fig.~\ref{Fig:LLHiggs} for the $L^{--}, L^-, N$ lepton and $H/A$~ cases. Except for the phase-space suppression, the production rates follow those of the $h\bar LL$ processes, dominated by the $\ee \to L^{++}L^{--}\Phi$ channel, for which the cross section can reach the fb level. This is the case for production with the $H$ 
state, in particular, as the rate is twice as large as in the case of $A$ for energies below 3 TeV where the chiral limit is not yet attained. 

We close this Section by noting that the charged vector--like fermions can also be pair
produced at photon colliders in the processes $\gamma \gamma \to F \bar F$, which would be particularly interesting for doubly--charged leptons since the production rates are proportional to $e_F^4$; for singly charged leptons, the rate is a factor of four higher than for the scalar case shown in Fig.~\ref{Fig:ee-H+}. Heavy leptons having couplings to the electron can also be singly produced in the $e\gamma$ option of linear $\ee$ colliders and the relevant processes, $e\gamma \to N W $ and $e\gamma \to E Z$ (similar to Fig.~\ref{diag}e), might have large rates for not too tiny mixing angles.  

\section{Conclusions}

The first question for the continuation of LHC Run~2 is whether the initial hint of
the 750 GeV $\Phi$ resonance decaying into two photons will be confirmed. If so, measurements of its strength and width
would provide information on the parameters of the state (or states) that could discriminate
between different models. For example, the 2HDM benchmark models studied here suggest
that there may be two near-degenerate $\Phi$ states, with natural widths greater than the experimental resolution,
yielding an asymmetric lineshape. As follow-ups, more detailed measurements would be possible at a higher-energy $pp$ collider, and it would be possible to make a precision measurement of $\Gamma(\Phi \to \gamma\gamma)$ via $\gamma \gamma$ collisions at an $\ee$ collider with centre-of-mass energy $\simeq 1$~TeV~\cite{ee-ILC}, as discussed in Sections 2 and 3.

In the case of the 2HDM benchmarks, the decay $\Phi \to t \bar t$ decay would be dominant,
and its detection may lie within reach of the LHC experiments. One should also search for the other diboson  decays of the $\Phi$ state, $ \Phi \to Z \gamma$, $ZZ$ and $W^+ W^-$, which would help distinguish between different singlet models as seen in Table~1,  as well as other fermionic decay modes such as $\bar b b$ and $\tau^+ \tau^-$ that could show up in the 2HDM benchmarks, along possibly with the interesting $H\to hh$ and $A\to hZ$ channels. 
Thorough searches for these decay modes will require high--luminosity running for the LHC~\cite{HL-ATLAS,HL-CMS} followed by a higher-energy $pp$ collider such as HE-LHC~\cite{LHC-33}, SPPC~\cite{SPPC} or FCC-hh~\cite{FCC-hh}. 

A common feature of the benchmark scenarios considered here is the need for additional vector-like matter particles, which we assume to be fermions.  As discussed in Section~4, future LHC searches for these particles are very promising, with the capability to explore all the parameter spaces of some singlet models. Measurements of pair production of vector-like fermions at $\ee$ machines are also promising, particularly for vector-like leptons. These may well be lighter than vector-like quarks,
which could help explain the magnitude of the $\Phi \to \gamma\gamma$ signal: see
Figs.~\ref{fig:enhancement} and \ref{Fig:Aboost}. Single production of vector-like fermions
in association with Standard Model fermions
would have lower thresholds than pair production, and hence may be more accessible to the
LHC and/or an $\ee$ collider, depending on the magnitudes of their mixing with their Standard Model counterparts. Single production processes would be very interesting for measuring their mixing, and thereby constraining models of the vector-like fermions.

The 2HDM benchmark scenarios have the distinctive feature that they predict the
existence of two neutral Higgs bosons $H, A$ contributing to the $\Phi$ signal, as well as
a pair of accompanying charged Higgs bosons $H^\pm$. The two $H, A$ states would not
be exactly degenerate. A typical separation is $\sim 15$~GeV, which might also help explain
the hint of a non-negligible width for $\Phi$ reported by ATLAS. The LHC and
other experiments might be able to resolve the $\Phi$ structure into two peaks, each with
a measurable natural width. As discussed in Section 3, there are many interesting opportunities to search for pair production of the $H, A$ and $H^\pm$ states in $pp$ collisions, as well as their production in association with $W$, $Z$ or $h$ bosons, or $\bar t t$ pairs. Similar final states may also be accessible in $\ee$ collisions, as may various processes for $H^\pm$ production. CLIC~\cite{ee-CLIC}, with its centre-of-mass energy 
up to $3$~TeV would be particularly well placed for such studies.

We have in this paper barely skimmed the surface of the physics possibilities that would be
opened up if the existence of the $\Phi$ state is confirmed. Such a discovery would open a new
era in particle physics, with a new layer of degrees of freedom at the TeV scale. If the $\Phi$
discovery is confirmed, it will shine a new light on options for possible future colliders, placing a
premium on those with sufficient energy to produce the new particles, while also suggesting a
new motivation for precision low-energy experiments. We await with interest the verdicts of
the ATLAS and CMS Collaborations.

\section*{Acknowledgements}

Discussions with and help from our collaborators S.A.R. Ellis, V. Sanz, R. Singh 
and T. You, and comments from H. Ito and M. McCullough are gratefully acknowledged. 
The work of AD is supported by the ERC Advanced Investigator Grant Higgs@LHC, and he
thanks the CERN Theory Department for its hospitality.
The work of JE was supported partly by the London Centre for Terauniverse Studies (LCTS), using funding from the European Research Council via the Advanced Investigator Grant 26732, and partly by the STFC Grant ST/L000326/1.
The work of JQ was supported by the STFC Grant ST/L000326/1.
The work of R.M.G. is supported by the Department of Science and Technology, India under Grant No. SR/S2/JCB-64/2007 under the J.C. Bose Fellowship scheme and the Indo-French
LIA-IFTHEP.

\end{document}